\documentclass{svmult}
\usepackage{latexsym}
\usepackage{amsmath}
\usepackage{amssymb}
\usepackage{amsfonts}

\newcounter{smallarabics}
\newenvironment{arabicenumerate}
{\begin{list}{{\normalfont\textrm{\arabic{smallarabics})}}}
  {\usecounter{smallarabics}\setlength{\itemindent}{0cm}
   \setlength{\leftmargin}{5ex}\setlength{\labelwidth}{4ex}
   \setlength{\topsep}{0.75\parsep}\setlength{\partopsep}{0ex}
   \setlength{\itemsep}{0ex}}}
{\end{list}}

\newcounter{smallroman}

\newcommand{\ben}{\begin{arabicenumerate}}  
\newcommand{\een}{\end{arabicenumerate}}

\newcommand{\beq}{\begin{equation}}
\newcommand{\eeq}{\end{equation}}
\newcommand{\bea}{\begin{eqnarray}}
\newcommand{\eea}{\end{eqnarray}}
\newcommand{\bex}{\begin{example}}
\newcommand{\eex}{\end{example}}
\def\bel{\begin{lemma}}
\def\eel{\end{lemma}}
\def\bet{\begin{theorem}}
\def\eet{\end{theorem}}
\def\bed{\begin{definition}}
\def\eed{\end{definition}}
\def\ber{\begin{remark}}
\def\eer{\end{remark}}

\def\rr{{\mathbb R}}
\def\zz{{\mathbb Z}}
\def\cc{{\mathbb C}}

\def\Conv{{\rm Conv}}
\def\AW{{\rm AW}}
\def\PF{{}}
\def\part{{\rm par}}

\def\Im{{\rm Im}}
\def\semi{{\rm semi}}
\def\Re{{\rm Re}}

\def\cl{{\rm cl}}

\def\cpl{{\rm cpl}}
\def\bar{\overline}

\def\l{{\rm l}}

\def\perpr{{\rm perp}}

\def\r{{\rm r}}

\def\c0inf{C_0^\infty}

\def\s{{\rm s}}
\def\sa{{\rm s/a}}
\def\12{{\frac12}}

\def\AW{{\rm AW}}

\def\sp{{\rm sp}}

\def\proof{
\noindent{\bf Proof.}\ \ }

\def\h{{\rm h}}

\def\cZ{{\cal Z}}
\def\cY{{\cal Y}}
\def\cB{{ B}}
\def\cV{{\cal V}}

\def\cU{{\cal U}}

\def\cE{{\cal E}}

\def\cW{{\cal W}}

\def\fM{{\mathfrak M}}  
\def\fA{{\mathfrak A}}  

\def\fin{{\rm fin}}

\def\a{{\rm a}}

\def\i{{\rm i}}

\def\Span{{\rm Span}}
\def\Dom{{\rm Dom}}

\def\loplus{\mathop{\oplus}\limits}

\def\sgn{{\rm sgn}}

\def\Tr{{\rm Tr}}

\def\Ker{{\rm Ker}}

\def\qed{$\Box$\medskip}

\def\12{\frac{1}{2}}

\def\e{{\rm e}}

\def\d{{\rm d}}
\def\Ran{{\rm Ran}}

\def\one{{\bf 1}}
\def\cH{{\cal H}}

\def\ii{ j }

\def\l{{\rm l}}
\def\r{{\rm r}}

\def\loplus{\mathop{\oplus}\limits}

\def\sgn{{\rm sgn}}

\def\Tr{{\rm Tr}}

\def\Ker{{\rm Ker}}

\def\cX{{\cal X}}
\def\I{{\cal I}}

\def\cK{{\cal K}}

\def\12{\frac{1}{2}}

\def\e{{\rm e}}

\def\d{{\rm d}}
\def\Ran{{\rm Ran}}

\def\one{{ 1}}
\def\cH{{\cal H}}

\def\W{{\cal W}}
\def\bep{\begin{proposition}}
\def\eep{\end{proposition}}

\def\fr{{\rm fr}}

\def\s{{\rm s}}

\def\otimesal{\vskip 0.2 em\mathop{ \hbox{\raise 1.45 ex
  \hbox{$\scriptscriptstyle\circ$}
\kern -1.0 em \hbox{$\otimes$}}}}
\def\oplusal{\vskip 0.2 em \mathop{\vski\hbox{\raise 1.45 ex
  \hbox{$\scriptscriptstyle\circ$}
\kern -1.0 em \hbox{$\oplus$}}}}
\def\Gammal{\hbox{\raise 1.68 ex 
\hbox{$\scriptscriptstyle\circ$}\kern -0.50 em $\Gamma$}}
\def\Bal{\hbox{\raise 1.68 ex 
\hbox{$\scriptscriptstyle\circ$}\kern -0.50 em $B$}}

\def\CARal{{\rm C\hskip 0.25 em \hbox{\raise 1.72 ex 
\hbox{$\scriptscriptstyle\circ$}\kern -0.57 em A}R}}
\def\t{{\scriptscriptstyle\#}}

\def\otimesal{\mathop{\hbox{\raise 1.45 ex
  \hbox{$\scriptscriptstyle\circ$}
\kern -0.92 em \hbox{$\otimes$}}}}
\def\oplusal{\mathop{\hbox{\raise 1.45 ex
  \hbox{$\scriptscriptstyle\circ$}
\kern -0.92 em \hbox{$\oplus$}}}}
\def\Gammal{\hbox{\raise 1.68 ex 
\hbox{$\scriptscriptstyle\circ$}\kern -0.50 em $\Gamma$}}

\begin{document}

\title*{Introduction to Representations \\
of the Canonical Commutation \\
and Anticommutation Relations}
\titlerunning{Introduction to Representations of the CCR and CAR}
\author{Jan Derezi\'{n}ski}
\institute{
Department of Mathematical Methods in Physics \\ 
Warsaw University \\ Ho\.{z}a 74, 00-682, Warszawa, Poland\\
e-mail { jan.derezinski@fuw.edu.pl}}

\maketitle
\setcounter{minitocdepth}{2}

\section{Introduction}

Since the early days of quantum mechanics it has been noted that the position
operator $x$ and the momentum operator $D:=-\i\nabla$ satisfy
the following commutation relation:
\beq \ [x,D]=\i.\label{cccc}\eeq
Similar commutation relation hold in the context of the second
quantization.  The bosonic creation
operator $a^*$ and the annihilation operator
$a$ satisfy
\beq [a,a^*]=1.\label{ccc1}\eeq
If we set $a^*=\frac1{\sqrt2}(x-\i D)$, $a=\frac1{\sqrt2}(x+\i D)$, then
 (\ref{cccc}) implies
 (\ref{ccc1}), so we see that both kinds of commutation relations are closely
related.

Strictly speaking the formulas (\ref{cccc}) and (\ref{ccc1}) are ill defined because it is not
clear how to interpret the commutator of unbounded operators. Weyl proposed to
replace (\ref{cccc}) by
\beq
\e^{\i \eta x}\e^{\i q D}=\e^{-\i q\eta}\e^{\i q D}\e^{\i \eta x},\ \eta,q\in\rr,
\label{wewe}\eeq
which has a clear mathematical meaning \cite{Weyl}.  (\ref{cccc})
is often called the canonical commutation relation (CCR)
 in the Heisenberg form and (\ref{wewe}) in the Weyl form.

It is natural to ask whether the commutation relations fix the operators $x$
and $D$ uniquely up to the
 unitary equivalence. If we assume that we are given  two self-adjoint
operators $x$ and $D$  acting
irreducibly on a Hilbert space and
satisfying (\ref{wewe}), then the answer is positive,
as proven by Stone and von  Neumann  \cite{vN}, see also \cite{Su}.

It is useful to generalize
the relations  (\ref{cccc}) and (\ref{ccc1}) to
systems with many degrees of freedom.
They were used by Dirac to describe quantized electromagnetic field in 
\cite{Di1}.

In systems with many degrees of freedom it is often useful to use a
more abstract setting for the CCR.
One can  consider $n$ self-adjoint operators $\phi_1,\dots,\phi_n$
satisfying relations
\beq [\phi_j,\phi_k]=\i\omega_{jk},\label{phi}\eeq
where $\omega_{jk}$ is an antisymmetric matrix.
Alternatively one can consider the Weyl (exponentiated) form of  (\ref{phi})
satisfied by the so-called Weyl operators $\e^{\i(y_1\phi_1+\cdots+y_n\phi_n
)}$, where $(y_1,\dots,y_n)\in\rr^n$.

The  Stone-von Neumann Theorem about the uniqueness 
can be extended to the case of regular
representations of the CCR in the Weyl form  if $\omega_{jk}$ is
a finite dimensional symplectic
matrix. 
Note that in this case  the relations (\ref{phi}) are invariant with respect
to the symplectic group. This invariance is implemented by a projective 
 unitary representation of the symplectic group, as has been noted by Segal
 \cite{Seg1}.
 It can be expressed in terms
 of a 
representation
 of the two-fold covering of the symplectic group --- the so-called
 metaplectic representation, \cite{Weil,Sh}.

The symplectic invariance of the 
CCR plays an important role in many problems of
quantum theory and of partial differential equations. An interesting and
historically perhaps  first nontrivial 
application is due to Bogolubov, who used it in
the theory of superfluidity of the Bose gas 
\cite{Bog}. Since then, the idea of using symplectic
transformations in bosonic systems often goes in the physics literature under
the name of the Bogolubov method, see e.g. \cite{FW}.

The Canonical Anticommutation Relations (CAR) appeared in mathematics before
quantum theory in the context  of Clifford algebras \cite{Cl}.
 A Clifford algebra is the
associative algebra generated by elements $\phi_1,\dots,\phi_n$ satisfying the
relations
\beq
[\phi_i,\phi_j]_+=2\delta_{ij},
\label{ccar}
\eeq
where $[\cdot,\cdot]_+$ denotes the anticommutator. It is natural to assume 
 that the $\phi_i$ are self-adjoint.
 It is not
difficult to show that if the representation (\ref{ccar})
is irreducible, then it is
 unique up to the unitary equivalence for $n$ even
 and  that there are two inequivalent representations for  an odd $n$.

In quantum physics, the CAR appeared in the description of fermions \cite{JW}.
 If
$a_1^*,\dots,a_m^*$ are fermionic creation and
$a_1,\dots,a_m$ fermionic annihilation operators, then they satisfy
\[
[a_i^*,a_j^*]_+=0,\ \ [a_i,a_j]_+=0,\ \ [a_i^*,a_j]_+=\delta_{ij}.\]
If we set $\phi_{2j-1}:=a_j^*+a_j$, $\phi_{2j}:=-\i(a_j^*-a_j)$, then we
see that they satisfy the relations (\ref{ccar}) with $n=2m$.

Another application of the CAR in quantum physics are
Pauli  \cite{Pa} and Dirac  \cite{Di2} matrices
used in the description
 of spin $\frac12$ particles.

Clearly, the relations (\ref{ccar}) are preserved by orthogonal
transformations applied to $(\phi_1,\dots,\phi_n)$.
The orthogonal invariance of the CAR is implemented by 
  a projective unitary representation.
 It can be also expressed in terms of a representation of the
double covering of the orthogonal group, called the Pin group. 
The so-called spinor representations of orthogonal groups
 were studied by Cartan \cite{Ca}, and Brauer and Weyl \cite{BW}

The orthogonal invariance of the CAR relations appears in many disguises
in algebra,
differential geometry and quantum physics. 
In quantum physics it is again often called the method of Bogolubov 
transformations. A particularly interesting application of this method
can be found
in the theory of superfluidity (a version of 
the BCS theory that can be found e.g. in \cite{FW}).

The  notion of a representation of the
 CCR and CAR gives a convenient framework to
describe Bogolubov transformations and their unitary implementations. Analysis
of Bogolubov transformations becomes especially interesting in the case of
an infinite number of degrees of freedom. In this case  there exist many
inequivalent representations of the CCR and CAR, as  noticed in the 50's,
e.g. by Segal \cite{Seg2} and
Gaarding and Wightman \cite{GaW}.

The most commonly used representations of the CCR/CAR  are the so-called Fock
representations, defined in 
bosonic/fermionic Fock spaces. These spaces have a distinguished vector
$\Omega$ called the vacuum killed by the annihilation operators and
cyclic with respect to 
 creation operators. They were introduced in quantum physics by
Fock \cite{Fo} to describe systems of many particle systems with the
Bose/Fermi statistics. Their mathematical structure, and also the essential
self-adjointness of bosonic field operators,
 was established by Cook \cite{Coo}. 

The passage from a  one particle system to a system with an arbitrary number of
 particles subject to the Bose/Fermi statistics is usually called second
 quantization. Early mathematical research on abstract aspects
 of second quantization was done by Friedrichs
\cite{Fr} and Segal \cite{Seg1,Seg2}.

 In the case of an infinite number of degrees of
freedom, 
the symplectic/orthogonal invariance of representations of the CCR/CAR
becomes much more subtle. The unitary implementability of symplectic/orthogonal transformations
in the Fock space is described by the Shale/Shale-Stinespring Theorem. These
theorems say 
 that implementable symplectic/orthogonal transformation belong to a
relatively small group $Sp_2(\cY)$/$O_2(\cY)$, \cite{Sh}/\cite{ShSt}.
In the case
of an infinite number 
of degrees of freedom 
there also exists  an analogue of the
 metaplectic/Pin representation. This seems to have been first noted
by Lundberg \cite{Lu}.

Among early works describing these results 
let us mention the book by Berezin \cite{Ber}. It gives
 concrete formulas for the implementation of Bogolubov
transformations in bosonic and fermionic Fock spaces. Related problems were
discussed, often independently, by other researchers, such as Ruijsenaars \cite{Ru1,Ru2}.


As stressed by Segal \cite{Seg2}, it is natural to apply
the language of $C^*$-algebras in the description of  the CCR and CAR.
 This is easily done in
the case of the
 CAR, where there exists an obvious candidate for the $C^*$-algebra
of the CAR over a given Euclidean space \cite{BR2}. If  this Euclidean space is of countably 
infinite 
dimension, the $C^*$-algebra of the CAR is isomorphic to the so called
$UHF(2^\infty)$ algebra studied by Glimm.
 Using representations of this $C^*$-algebra one can construct
 various non-isomorphic kinds of
factors ($W^*$-algebras with a trivial center),
 studied in particular by Powers \cite{Po} and Powers and St{\oe}rmer
\cite{PoSt}. 

In the case of the CCR, the choice of the corresponding
 $C^*$-algebra is less obvious. The most
popular choice is the $C^*$-algebra generated by the Weyl operators, studied
in particular by Slawny \cite{Sla}. One can, however, argue that the ``Weyl
algebra'' is not very physical and that there are other more natural choices
of the $C^*$-algebra of the CCR.
Partly to avoid discussing such (quite academic) topics, 
 in our lecture notes
we avoid the language of $C^*$-algebras. On the other hand, we will use the
language of $W^*$-algebras, which seems  natural in this context.

 One class of representations of the CCR and CAR -- the quasi-free
representations -- is especially widely used in quantum physics.
In mathematical literature they have been first identified  by 
Robinson \cite{Rob} and Shale and Stinespring \cite{ShSt}.
Quasi-free representations were extensively studied,
especially by Araki \cite{Ar6,Ar2,Ar3,ArShi,AY} and van Daele \cite{vD}.

 A concrete
realization of quasi-free representations is furnished by the so-called
Araki-Woods representations  \cite{AW} in the bosonic 
and  Araki-Wyss representations  \cite{AWy} in
the fermionic case. We describe these representations in detail. 
 From the physical point of view, they can be viewed as
a kind of a thermodynamical limit of representations for a finite number of
degrees of freedom.  From the mathematical point of view,
they provide interesting and physically well motivated
examples of factors of type II and III. 
 It is very instructive to use the
Araki-Woods and Araki-Wyss  representations as  illustrations for the
Tomita-Takesaki theory and for the so-called standard form of a $W^*$-algebra
\cite{Ha}
(see also \cite{Ar2,Co,BR1,St, DJP}). They are quite often used in recent
works on quantum statistical physics, see e.g. \cite{DJ2,JP}

It is interesting to note that the famous paper of Haag, Hugenholtz and 
Winnink \cite{HHW} about the KMS condition was directly inspired by 
the Araki-Woods representation.

Araki-Woods/Araki-Wyss representations can be considered also in the case of
a finite number of degrees of freedom. In this case, they are equivalent to a
multiple of the usual representations of the CCR/CAR. This equivalence can be
described by the GNS representation with respect to
 a quasi-free state composed
with an appropriate unitarily implemented Bogolubov transformation. We
discuss this topic in the section devoted to ``confined'' Bose/Fermi gas.

It is easy to  see that real subspaces of a complex Hilbert space form a
complete complemented lattice, where the complementation is given by the
symplectic orthogonal complement.
It is also clear that von Neumann algebras on a given Hilbert space form a
complete complemented lattice with the commutant as the complementation. It
was proven by Araki \cite{Ar4} (see also \cite{EO})
 that von Neumann algebras on a bosonic Fock
space associated to real subspaces of the classical phase space also form a
complemented complete lattice isomorphic to the corresponding lattice of real
subspaces. We present this result, used often in algebraic quantum field
theory. We also describe the fermionic analog of this result (which seems to  
 have been overlooked 
in the literature).

In the last section we describe a certain class of operators that we call
Pauli-Fierz operators, which are used to describe a small quantum system
interacting with a bosonic reservoir,
 see  \cite{DG,DJ,DJ2,BFS} and references therein.
 These operators have interesting
mathematical and physical properties, which have been studied in recent years
by various authors. Pauli-Fierz operators provide a good opportunity to
illustrate the use of various representations of the CCR.

The concepts discussed in these lectures, in particular 
representations of the 
CCR and CAR,   constitute, in one form or another, a part
of the 
standard language of mathematical physics. More or less explicitly they are
used in any textbook on quantum field theory. Usually the authors  first 
discuss quantum fields ``classically''---just the relations they satisfy
without specifying their representation.
 Only then one introduces their
representation in a Hilbert space. In the zero temperature, it is usually the
Fock representation determined by the requirement that the Hamiltonian  
should be
bounded from below,see e.g. \cite{DB}.
 In positive temperatures one usually  chooses the GNS
representation given by an appropriate thermal state.

The literature devoted to topics contained in our 
lecture notes is quite large. Let
us mention some of the monographs.
The  exposition of the $C^*$-algebraic approach to the 
CCR and CAR can be found in
\cite{BR2}. This monograph provides also extensive historical remarks.
One could also consult  an older monograph \cite{Em}.
Modern exposition of the mathematical formalism of second quantization can be
also found e.g.  in \cite{GJ,BSZ}.
We would also like to mention the book by Neretin \cite{Ne}, which describes infinite
dimensional metaplectic and Pin groups, and review articles by Varilly and
Gracia-Bondia \cite{VGB1,VGB2}. 
A very comprehensive article devoted to the CAR $C^*$-algebras
 was written by Araki \cite{Ar5}. Introductions to Clifford algebras can be
 found in \cite{LM,Tr,Lo}. In this collection of lecture notes
 De Bi\`{e}vre discusses the localizability for bosonic fields \cite{DB}.

The theory of the CCR and CAR involves a large number 
of concepts coming from algebra,
analysis and physics. This is why
 the literature about this subject is 
very scattered and uses various conventions, notations
and  terminology.
 Even the meaning of the expressions ``a representation of
the CCR'' and ``a representation of the CAR'' depends
slightly on the author.

In our lectures
we want to stress
 close analogies
between the CCR and CAR. Therefore, we tried
to present both formalisms in a possibly parallel way.

 We also
want to draw the  reader's attention to 
  $W^*$-algebraic aspects of the theory.
 They
shed a lot of light onto some aspects of
mathematical physics. 
The CAR and CCR are also
 a rich source of illustrations for various concepts of
the  theory of $W^*$-algebras.

We often refrain from giving proofs. Most of them are quite elementary and
can be easily provided by
the interested reader.

\medskip

\noindent{\bf Acknowledgments.} 
The research of the author  was  partly supported by the
Postdoctoral Training Program HPRN-CT-2002-0277 and the Polish KBN grants
 SPUB127 and 2 P03A 027 25. Part of the work was done  when the author
 visited the
 Erwin Schr\"odinger Institute as the Senior Research Fellow.

The author profited from
 discussions with V. Jak\v{s}i\'{c} and S. De Bi\`{e}vre. He would also like
 to express his gratitude to H. Araki for reading a previous version of the
 manuscript and pointing out some  errors.

\section{Preliminaries}
\label{2}

In this section we review our basic notation, mostly about vector spaces and
linear operators.

\subsection{Bilinear forms}


Let $\alpha$ be a bilinear  form on a vector space $\cY$.
The action of $\alpha$  on
 a pair of vectors $y_1,y_2\in\cY$ will be written 
as
$y_1\alpha y_2$. 
We say that a linear map $r$ on $\cY$  preserves $\alpha$ iff
\[(ry_1)\alpha(ry_2)=y_1\alpha y_2,\ \ \ y_1,y_2\in\cY.\ \ \ \]
We say that $\alpha$ is nondegenerate,
 if
for any non-zero 
$y_1\in \cY$ there exists $y_2\in\cY$ such that $y_2\alpha y_1\neq0$.

An antisymmetric nondegenerate form is called symplectic.
A symmetric nondegenerate form is called a scalar product.

\subsection{Operators in Hilbert spaces}

The
  scalar
 product of two vectors $\Phi,\Psi$ in a Hilbert space will be denoted by
$(\Phi|\Psi)$. It will be antilinear in the first argument
 and linear in the
  second.

If $\cH_1,\cH_2$ are Hilbert spaces, then $B(\cH_1,\cH_2)$, resp.
 $U(\cH_1,\cH_2)$
 denotes bounded, resp. unitary operators from $\cH_1$ to $\cH_2$.

$A^*$ denotes the hermitian adjoint of the operator $A$.

An operator $U:\cH_1\to\cH_2$ is called antiunitary iff it is antilinear,
bijective and $(U\Phi|U\Psi)=\bar{(\Phi|\Psi)}$.

$B^2(\cH_1,\cH_2)$ denotes Hilbert-Schmidt operators from $\cH_1$ to $\cH_2$,
 that is $A\in B^2(\cH_1,\cH_2)$ iff $\Tr A^*A<\infty$. Note that
 $B^2(\cH_1,\cH_2)$ has a natural structure of the  Hilbert space with the
 scalar product
\[(A|B):=\Tr A^*B.\]

$B^1(\cH_1,\cH_2)$ denotes trace  class operators from $\cH_1$ to $\cH_2$,
 that is $A\in B^1(\cH_1,\cH_2)$ iff $\Tr (A^*A)^{1/2}<\infty$.

 For a single
 space $\cH$, we will write $B(\cH)=B(\cH,\cH)$, etc. 
 $B_\h(\cH)$ will denote bounded self-adjoint operators on $\cH$
(the subscript $\h$ stands for ``hermitian'').
$B_+(\cH)$ denotes positive bounded operators on $\cH$. 
Similarly,  $B_+^2(\cH)$ and 
$B_+^1(\cH)$ stand for positive Hilbert-Schmidt and
 trace class operators on $\cH$ respectively.

By saying that $A$ is an operator from $\cH_1$ to $\cH_2$ we will mean 
that the domain  of $A$, denoted by $\Dom A$ is a subspace of $\cH_1$
and $A$ is a linear map from $\Dom A$ to $\cH_2$.

The spectrum of an operator $A$ is denoted by $\sp A$.

\subsection{Tensor product}
\label{s5.1}
We assume that the reader is familiar with the concept of the algebraic tensor
product of two vector spaces. The tensor product of a vector $z\in\cZ$ and a
vector $w\in\cW$ will be, as usual,  denoted by $z\otimes w$.

Let 
$\cZ$ and $\cW$ be two Hilbert spaces. 
The notation $\cZ\otimes\cW$ will be used to denote the tensor product of
$\cZ$ and $\cW$ in the
sense of Hilbert spaces. 
Thus $\cZ\otimes\cW$
is a Hilbert space equipped with the operation
\[ \cZ\times\cW
\ni (z,w)\mapsto z\otimes w\in\cZ\otimes \cW,\]
and with a scalar product satisfying
\beq
(z_1\otimes w_1|z_2\otimes w_2)
=(z_1|z_2)(w_1|w_2).\label{sca}\eeq
$\cZ\otimes\cW$ is the
completion of the algebraic tensor product of $\cZ$ and $\cW$ in the norm
given by (\ref{sca}).

\subsection{Operators in a tensor product}
Let $a$ and $b$  be  (not necessarily everywhere defined) operators on $\cZ$
and $\cW$. Then we can define a linear
 operator $a\otimes b$ with the domain equal
to the algebraic tensor product of $\Dom\, a$ and  $\Dom\, b$,
satisfying
\[(a\otimes b) (z\otimes w):=(az)\otimes(bw).\]

If
$a$ and $b$ are densely defined, then so is $a\otimes b$.

If $a$ and $b$ are closed, then $a\otimes b$ is closable. To see this, note
that the algebraic tensor product of $\Dom\, a^*$ and $\Dom\, b^*$ is contained in
the domain of $(a\otimes b)^*$. Hence  $(a\otimes b)^*$ has a dense domain.

We will often denote the closure of $a\otimes b$ with the same symbol.

\subsection{Conjugate Hilbert spaces}
\label{s3.6}

Let $\cH$ be a complex  vector space.
The space $\bar\cH$ conjugate to $\cH$ is a complex vector
together  with a distinguished  antilinear bijection
\begin{equation}\cH\ni \Psi\mapsto\bar \Psi\in\bar\cH.\label{bar}\end{equation}
The map (\ref{bar}) is called a conjugation on $\cH$. 
It is convenient to denote the inverse of the map (\ref{bar}) by the same
symbol. Thus $\bar{\bar\Psi}=\Psi$.

Assume in addition that $\cH$ is a Hilbert space.
Then we assume that $\bar\cH$ is also a Hilbert space and (\ref{bar}) is
antiunitary, so that  the scalar product on $\bar \cH$ satisfies
\[(\bar\Phi|\bar\Psi=\bar{(\Phi|\Psi)}.\]

For $\Psi\in\cH$, let $(\Psi|$ denote the operator in $B(\cH,\cc)$ given by
\[
\cH\ni\Phi\mapsto(\Psi|\Phi)\in\cc.\]
We will write $|\Psi):=(\Psi|^*$.

By the Riesz lemma,  the map 
\[\bar\cH\ni\bar\Psi\mapsto (\Psi|\in \cB(\cH,\cc) \]
is an  isomorphism between $\bar \cH$ and 
the dual of $\cH$, that is $\cB(\cH,\cc)$.

If $A\in \cB(\cH)$, then $\bar A\in \cB(\bar\cH)$ is defined by
\[\bar\cH\ni\bar\Psi\mapsto
\bar A\ \bar\Psi:=\bar{A\Psi}\in\bar\cH.\]
We will identify $\bar{B(\cH)}$ with $B(\bar\cH)$.

If $\cH$ is a real vector space, we always take $\Psi\mapsto\bar\Psi$ to be
the identity.

Let $\cZ,\cW$ be Hilbert spaces.
We will often use the 
 identification of the set of Hilbert-Schmidt
operators $B^2(\cZ,\cW)$ with
$\cW\otimes\bar\cZ$, so that $|\Phi)(\Psi|\in B^2(\cZ,\cW)$
corresponds to $\Phi\otimes\bar{\Psi}\in\cW\otimes\bar\cZ.$

We
identify $\bar{\cZ\otimes\cW}$ with $\bar\cW\otimes\bar\cZ$.

If $A\in B(\cZ,\cW)$, then $A^\t\in B(\bar\cW,\bar\cZ)$ is defined as
$A^\t:={\bar A}^*$ (recall that
$*$ denotes the hermitian conjugation). $A^\t$ is
sometimes called the transpose of $A$.

 This is especially useful if $A\in B(\bar\cZ,\cZ)$. Then we
say that $A$ is symmetric iff $A^\t=A$ and antisymmetric if $A^\t=- A$.
In other words, $A$ is symmetric if
\[(z_1|A \bar z_2)=(z_2| A\bar z_1),\ \ z_1,z_2\in\cZ;\]
antisymmetric if
\[(z_1|A \bar z_2)=-(z_2| A\bar z_1),\ \ z_1,z_2\in\cZ.\]
The space of symmetric and antisymmetric bounded operators from $\bar\cZ$ to
$\cZ$ is denoted by $B_\s(\bar\cZ,\cZ)$ and $B_\a(\bar\cZ,\cZ)$ resp.
The space of Hilbert-Schmidt symmetric and antisymmetric operators from $\bar\cZ$
to $\cZ$ is denoted by $B_\s^2(\bar\cZ,\cZ)$ and $B_\a^2(\bar\cZ,\cZ)$ resp.

\ber Note that, unfortunately, in the literature, e.g. in \cite{RS1},
 the word ``symmetric'' is sometimes used in a
different meaning (``hermitian but not necessarily self-adjoint''). \eer

\subsection{Fredholm determinant}

Let $\cY$ be a (real or complex) 
Hilbert space.

Let $\one+B^1(\cY)$ denote the set of operators of the form $\one
+a$ with $a\in B^1(\cY)$.

\bet There exists a unique
 function $\one+B^1(\cY)\ni r\mapsto\det r$ that
satisfies 
\ben\item If $\cY=\cY_1\oplus\cY_2$ with $\dim\cY_1<\infty$ and $r=r_1\oplus
 \one$, then $\det r=\det r_1$, where $\det r_1$ is the usual 
determinant of a finite dimensional operator $r_1$;
\item $\one+B^1(\cY)\ni 
r\mapsto \det r$ is continuous in the trace norm.
\een
\eet

$\det r$ is called the Fredholm determinant of $r$, see e.g.  \cite{RS4}
Sect. XIII.17.



\section{Canonical commutation relations}

In this section we introduce one of the basic concept of our lectures --  a
representation  of the canonical commutation relations (CCR).
 We choose the exponential form of the CCR -- often called
the Weyl form of the CCR.

In the literature the terminology related to CCR often depends on the author,
\cite{Em,BR2,BSZ,DG}.
 What we call  representations
of the CCR is known  also
as  Weyl or a Heisenberg-Weyl systems. 

\subsection{Representations of the CCR}

 Let $\cY$ be a real vector space equipped with an antisymmetric form
 $\omega$.
(Note that $\omega$ does not need to be nondegenerate).
Let $\cH$ be a Hilbert space. Recall that $U(\cH)$ denotes the set of unitary
operators on $\cH$.
 We say that a map
\beq
\cY\ni y\mapsto W^\pi(y)\in U(\cH)
\label{iui}\eeq is a 
{\em  representation of the CCR
 over $\cY$ in $\cH$}
 if
\beq\begin{array}{l}W^\pi(y_1)W^\pi(y_2)=\e^{-\frac{\i}{2} y_1\omega y_2}
W^\pi(y_1+y_2),\ \ \ \ y_1,y_2\in\cY.
\end{array}\label{weyl1}\eeq


Note that (\ref{weyl1}) implies
\bet Let $y,y_1,y_2\in\cY$. Then
\begin{eqnarray}
W^\pi(y_1)W^\pi(y_2)&=&\e^{-\i y_1\omega y_2}
W^\pi(y_2)W^\pi(y_1),\label{wey1}\\[3mm]
W^{\pi*}(y)&=&W^\pi(-y),\ \ \ \ \ W^\pi(0)=1,\label{wey2}\\[3mm]
W^\pi(t_1y)W^\pi(t_2y)&=&W^\pi((t_1+t_2)y),\ \ \ t_1,t_2\in\rr.\label{wey3}
\end{eqnarray}
\label{weyl2}
\eet
(\ref{wey1}) is known as the canonical commutation relation in the Weyl form.

We say that a subset $K\subset\cH$ is cyclic for (\ref{iui})
if \[\Span\{W^\pi(y)\Psi\ :\ 
\Psi\in K,\ y\in\cY\}\]
 is dense in $\cH$. We say that $\Psi_0\in\cH$
is cyclic
if  $\{\Psi_0\}$ is cyclic.

We say that the 
representation (\ref{iui})
 is irreducible if the only closed subspace of $\cH$ preserved by $W^\pi(y)$
 for all $y\in\cY$ is $\{0\}$ and $\cH$.
 Clearly, in the case of an
irreducible representation, all  nonzero vectors in $\cH$ are cyclic.

Suppose we are given two representations of the CCR over the same
 space $(\cY,\omega)$:
\bea
&&\cY\ni y\mapsto W^{\pi_1}(y)\in U(\cH_1),\label{req1}\\[3mm]
&&\cY\ni y\mapsto W^{\pi_2}(y)\in U(\cH_2).\label{req2}
\eea
We say that (\ref{req1}) is unitarily equivalent to (\ref{req2}) iff there
exists a unitary operator $U\in U(\cH_1,\cH_2)$ such that
\[UW^{\pi_1}(y)=W^{\pi_2}(y)U,\ \ \ y\in \cY.\]

Clearly, given a representation of the CCR (\ref{iui}) and  a linear
transformation  $r$ on $\cY$ that
preserves $\omega$,
\[\cY\ni y\mapsto W^\pi(ry)\in U(\cH)\]
is also a representation of the CCR.

If we have two  representations of the CCR 
\bea
&&\cY_1\ni y_1\mapsto W^{\pi_1}(y_1)\in U(\cH_1),\nonumber\\[3mm]
&&\cY_2\ni y_2\mapsto W^{\pi_2}(y_2)\in U(\cH_2),\nonumber
\eea
then
\[\cY_1\oplus\cY_2\ni(y_1,y_2)\mapsto W^{\pi_1}(y_1)\otimes
 W^{\pi_2}(y_2)\in U(\cH_1\otimes\cH_2)\]
 is also a
representation of the CCR.

By  (\ref{wey3}), for any representation of the CCR,
\beq\rr\ni t\mapsto W^\pi(ty)\in U(\cH) \label{gro}\eeq 
is a 1-parameter group. We say that  a representation of the CCR (\ref{iui}) is
{\em  regular} if
(\ref{gro}) is strongly continuous for each $y\in\cY$.
Representations of the CCR that  appear in applications are usually  regular.

\subsection{Schr\"odinger representation of the CCR}

 Let $\cX$ be a finite dimensional real space. Let $\cX^\t$ denote the space
 dual to $\cX$.
Then  the form
\beq(\eta_1,q_1)\omega(\eta_2,q_2)=\eta_1q_2-\eta_2 q_1\label{symp}\eeq
on $\cX^\t\oplus\cX$
is  symplectic.

Let $x$ be the generic name of the variable in $ \cX$, and simultaneously
 the 
 operator of multiplication by the variable $x$
 in $L^2(\cX)$.
More precisely
for any $\eta\in\cX^\t$ the symbol $\eta x$ denotes the
 self-adjoint operator on
 $L^2(\cX)$ acting on its domain as
\[(\eta x\Psi)(x):=(\eta x)\Psi(x).\]

 Let $D:=\frac1\i\nabla_x$ be the momentum operator on
 $L^2(\cX)$. 
More precisely, for any $q\in\cX$ the symbol
 $q D$ denotes the self-adjoint operator on  $L^2(\cX)$ acting
on its domain as 
\[(q D\Psi)(x):=\frac1\i q\nabla_x\Psi(x).\]

It is easy to see that
\bet The map
\beq\cX^\t\oplus\cX\ni(\eta,q)\mapsto\e^{\i(\eta x+q D)}\in
 U(L^2(\cX))\label{sch}\eeq 
is an irreducible  regular representation of the CCR over $\cX^\t\oplus\cX$ in
$L^2(\cX)$.
\eet

(\ref{sch}) is  called the {\em Schr\"odinger representation}.

Conversely, suppose that $\cY$ is a finite dimensional real vector space
equipped with a 
symplectic form $\omega$. Let
\beq \cY\ni y\mapsto W^\pi(y)\in U(\cH)\label{ccc}\eeq
be a  representation of the  CCR.
There exists a real vector space $\cX$ such that the symplectic space
$\cY$ can be identified with
$\cX^\t\oplus\cX$ equipped with the symplectic form (\ref{symp}).
 Thus we can rewrite (\ref{ccc}) as
\[\cX^\t\oplus\cX\ni(\eta,q)\mapsto W^\pi(\eta,q)\in U(\cH)\]
satisfying
\[W^\pi(\eta_1,q_1)W^\pi(\eta_2,q_2)=\e^{-\frac{\i}{2}(\eta_1 q_2-\eta_2 q_1)}
W^\pi(\eta_1+\eta_2,q_1+q_2).\]
In particular,
\beq\cX^\t\ni\eta\mapsto  W^\pi(\eta,0)\in U(\cH),\label{rer1}\eeq
\beq\cX\ni q\mapsto W^\pi(0,q)\in U(\cH)\label{rer2}\eeq
are  unitary representations satisfying
\[W^\pi(\eta,0)W^\pi(0,q)=\e^{-\i\eta q}W^\pi(0,q)W^\pi(\eta,0).\]
If (\ref{ccc}) is regular, then (\ref{rer1}) and (\ref{rer2}) are strongly
continuous.

The following classic result says that a  representation of the   CCR over a
symplectic space with a finite number of degrees of freedom is essentially
unique up to the multiplicity (see e.g. \cite{BR2,Em}):
\bet[The Stone--von Neumann theorem]
Suppose that $(\cY,\omega)$ is a finite dimensional 
symplectic space and (\ref{ccc}) a regular representation of the  CCR.
Suppose that we fix an identification
of $\cY$ with
$\cX^\t\oplus\cX$. 
Then, there exists a Hilbert space
$\cK$ and a unitary operator
$U: L^2(\cX)\otimes\cK\to \cH$
such that
\[W^\pi(\eta,q)U=U\left(\e^{\i(\eta x+q D)}\otimes 1_\cK\right) .\]
The representation of the CCR (\ref{ccc})
is irreducible iff $\cK=\cc$. \label{stvn}\eet

\begin{corollary} Suppose that $\cY$ is a 
finite dimensional symplectic space. Let
$\cY\ni y\mapsto W^{\pi_1}(y)\in U(\cH)$ and
$\cY\ni y\mapsto W^{\pi_2}(y)\in U(\cH)$ be two regular irreducible
representations of the CCR. Then they are unitarily equivalent.
\label{cosa}
\end{corollary}

\subsection{Field operators}

In this subsection
we assume that we are given a regular representation
$\cY\ni y\mapsto W^\pi(y)\in U(\cH)$.
Recall that $\rr\ni t\mapsto W^\pi(ty)$
 is a strongly continuous unitary group.
By the Stone theorem, for any $y\in \cY$, we can define its
self-adjoint generator
\[\phi^\pi(y):=-\i\frac{\d}{\d t}W^\pi(ty)\Big|_{t=0}.\]
$\phi^\pi(y)$ will be called the {\em  field operator}
corresponding to $y\in\cY$.

\begin{remark} Sometimes, the operators $\phi^\pi(y)$ 
are called {\em Segal field operators}. \end{remark}



\bet Let $y,y_1,y_2\in\cY$. 
\ben
\item Let $\Psi\in\Dom \:\phi^\pi(y_1)\cap\Dom\:\phi^\pi(y_2)$,
$c_1,c_2\in\rr$. Then
\[\begin{array}{rl}
\Psi&\in\Dom\:\phi^\pi(c_1y_1+c_2y_2),\\[3mm]
\phi^\pi(c_1y_1+c_2y_2)
\Psi&=c_1\phi^\pi(y_1)\Psi+c_2\phi^\pi(y_2)\Psi.\end{array}\]
\item
 Let $\Psi_1,\Psi_2\in\Dom \ \phi^\pi(y_1)\cap\Dom\ \phi^\pi(y_2)$. Then
\beq
(\phi^\pi(y_1)\Psi_1|\phi^\pi(y_2)\Psi_2)-
(\phi^\pi(y_2)\Psi_1|\phi^\pi(y_1)\Psi_2)
=\i y_1\omega y_2(\Psi_1|\Psi_2).\label{hei}\eeq
\item $\phi^\pi(y_1)+\i\phi^\pi( y_2)$
 is a closed operator on the domain
\[\Dom \ \phi^\pi(y_1)\cap\Dom\ \phi^\pi(y_2).\]
\een\eet

(\ref{hei}) can be written somewhat imprecisely as
\beq[\phi^\pi(y_1),\phi^\pi(y_2)]=\i y_1\omega y_2\label{impr}\eeq
and is called the
 canonical commutation relation in the Heisenberg form.




\subsection{Bosonic 
Bogolubov transformations}

Let $(\cY,\omega)$ be a finite dimensional symplectic space.
Linear maps on $\cY$ preserving $\omega$ are then automatically invertible.
 They  form a group, which will be called the {\em symplectic group of
$\cY$} and denoted $Sp(\cY)$.

Let
$\cY\ni y\mapsto W(y)\in U(\cH)$ be a regular
irreducible representation of canonical commutation relations.
The following theorem is an immediate  consequence of Corollary \ref{cosa}:

\bet For any $r\in Sp(\cY)$ there exists $U\in U(\cH)$, defined uniquely up
to a phase factor (a complex number of absolute value $1$), such that
\beq U W(y)U^*=W(ry).\label{qfa}\eeq
Let $\cU_r$ be  the class of unitary operators satisfying (\ref{qfa}). Then
\[Sp(\cY)\ni r\mapsto \cU_r\in  U(\cH)/U(1)\]
is a group homomorphism, where $U(1)$ denotes the group of unitary scalar
operators on $\cH$.
\label{bogo}\eet

One can ask whether one can 
fix uniquely the phase factor appearing  in the above theorem
and obtain
a group homomorphism of $Sp(\cY)$ into $U(\cH)$ satisfying (\ref{qfa}). 
This is impossible, the best what one can do is
the following improvement of Theorem \ref{bogo}
\bet For any $r\in Sp(\cY)$ there exists a unique pair of operators
$\{U_r,-U_r\}\subset U(\cH)$ such that
\[U_r W(y)U_r^*=W(ry),\]
and such that we have a group homomorphism
\beq Sp(\cY)\ni r\mapsto \pm U_r\in  U(\cH)/\{1,-1\}.\label{bogo2}\eeq
\label{bogo1}\eet

The representation  (\ref{bogo2}) is called the {\em metaplectic representation
of $Sp(\cY)$.}

 Note that the
homotopy group of $Sp(\cY)$ is $\zz$. Hence for any 
$n\in\{1,2,\dots,\infty\}$ we can construct the $n$-fold covering of
$Sp(\cY)$. The image of the metaplectic representation is isomorphic to the
double covering of $Sp(\cY)$.  It is often called the metaplectic group.

In the physics literature the fact that symplectic transformations can be
unitarily implemented is generally associated with the name of Bogolubov, who
successfully applied this idea to
the superfluidity of the Bose gas.
 Therefore, in the physics literature, the transformations
described in Theorems \ref{bogo} and \ref{bogo1} are often called
 {\em  Bogolubov transformations}.

The proofs of Theorems \ref{bogo} and \ref{bogo1} are most conveniently given
in the Fock representation, where one has simple formulas for $U_r$
 (see e.g. \cite{Fol,Ber}). 
We will describe these formulas later on (in the more general context of an
infinite number of degrees of freedom).

\section{Canonical anticommutation relations}

In this section we introduce the second basic concept of our lectures, that of
a representation of the canonical anticommutation relations  (CAR). Again, there is no uniform terminology in this
domain \cite{Em,BR2,BSZ}. What we call a representation of the CAR is often called 
 Clifford relations, which is perhaps more justified
 historically. Our terminology is intended to stress the analogy between the 
CCR
 and CAR.

\subsection{Representations of the CAR}

Let $\cY$ be a real vector space with a positive scalar product
$\alpha$.  Let $\cH$ be a Hilbert space. Recall that $B_\h(\cH)$ denotes the
set of bounded self-adjoint operators on $\cH$.
 We say that a linear
map
\beq\cY\ni y\mapsto\phi^\pi( y)\in B_\h(\cH)\label{car1a}\eeq
is a {\em representation of the CAR
 over $\cY$ in $\cH$ } iff
$\phi^\pi(y)$ satisfy
\beq[\phi^\pi(y_1),\phi^\pi(y_2)]_+= 2 y_1\alpha y_2,\ \ \ y_1,y_2\in\cY
.\label{car2a}\eeq

We will often drop $\pi$, and write $|y|_\alpha:=(y\alpha y)^{1/2}$.

\ber The reason of putting the factor $2$ in (\ref{car2a}) is the identity
$\phi(y)^2=y\alpha y$. Note, however, that in (\ref{impr}) there is no factor
of $2$, and therefore at some places the treatment of the 
CCR and CAR will be not
as parallel as it could be.
\eer



\bet
\ben\item 
 $\sp \ \phi(y)=\{-|y|_\alpha,|y|_\alpha\}$
 \item
Let $t\in\cc$, $y\in\cY$. Then
$\|t+\phi(y)\|=\max\{|t+|y|_\alpha|,|t-|y|_\alpha|\}$.
\item
$\e^{\i \phi(y)}=\cos|y|_\alpha+\i\frac{\sin|y|_\alpha}{|y|_\alpha}\phi(y)$.
\item
Let $\cY^\cpl$ be the completion of $\cY$ in the norm $|\cdot|_\alpha$. Then there exists a unique
extension
of (\ref{car1a}) to a continuous map
\beq\cY^\cpl\ni y\mapsto \phi^{\pi^\cpl}(y)\in B_\h(\cH).\label{car1aa}\eeq
Moreover,  (\ref{car1aa}) is  a representation of the CAR.
\een\label{cca}\eet

Motivated by the last statement, henceforth we will assume that $\cY$
is complete---that is, $\cY$ is a real Hilbert space.

By saying that $(\phi_1,\dots,\phi_n)$ is a representation of the 
CAR on $\cH$ we
 will mean that we have a representation of the CAR
 $\rr^n\ni y\mapsto\phi(y)\in B_\h(\cH)$, where
 $\rr^n$ is equipped with the
 canonical scalar product,   $e_i$ is the canonical basis of
 $\rr^n$ and 
 $\phi_i=\phi^\pi(e_i)$. Clearly, this is equivalent to the relations
$[\phi_i,\phi_j]_+=2\delta_{ij}$.

We say that a subset $K\subset\cH$ is cyclic for (\ref{car1a})
if $\Span\{\phi^\pi(y_1)\cdots\phi^\pi(y_n)\Psi\ :\ 
\Psi\in K,\ y_1,\dots,y_n\in\cY,\ n=1,2,\dots\}$ is dense in $\cH$. We say that $\Psi_0\in\cH$
is cyclic
for (\ref{car1a})
if  $\{\Psi_0\}$ is cyclic.

We say that (\ref{car1a}) is irreducible if 
 the only closed subspace of $\cH$ preserved by $\phi^\pi(y)$
 for all $y\in\cY$ is $\{0\}$ and $\cH$.
Clearly, in the case of an
irreducible representation, all  nonzero vectors in $\cH$ are cyclic.

Suppose we are given two representations of the CAR over the same
 space $(\cY,\alpha)$:
\bea
&&\cY\ni y\mapsto \phi^{\pi_1}(y)\in B_\h(\cH_1),\label{req1a}\\[3mm]
&&\cY\ni y\mapsto \phi^{\pi_2}(y)\in B_\h(\cH_2),\label{req2a}
\eea
then
we say that (\ref{req1a}) is unitarily equivalent to (\ref{req2a}) iff there
exists a unitary operator $U\in U(\cH_1,\cH_2)$ such that
\[U\phi^{\pi_1}(y)=\phi^{\pi_2}(y)U,\ \ \ y\in \cY.\]

Let $\cY_1$, $\cY_2$ be 
two real Hilbert spaces. Suppose that  $I$ is a self-adjoint operator on
$\cH_1$  and
\[\cY_1\oplus\rr\ni(y_1,t)\mapsto \phi^{\pi_1}(y_1)+tI\in B_\h(\cH_1),\]
\[\cY_2\ni y_2\mapsto \phi^{\pi_2}(y_2)\in B_\h(\cH_2)\]
are representations of the CAR. Then
\[\cY_1\oplus\cY_2\ni(y_1,y_2)\mapsto
\phi^{\pi_1}(y_1)\otimes\one+I\otimes \phi^{\pi_2}(y_2)
\in
B(\cH_1\otimes\cH_2) \] is a representation of the CAR.

If $r\in B(\cY)$ preserves  
the scalar product (is  isometric), and we are given a representation of the CAR
(\ref{car1a}), then  
\[\cY\ni y\mapsto \phi^\pi(ry)\in B_\h(\cH)\]
is also  a representation of the CAR.

Most of the above material was very similar to its CCR
counterpart. The following construction, however, has no analog in the
context of the CCR:

\bet Suppose that $\rr^n\ni y\mapsto\phi(y)$ is a representation of the
CAR. Let $y_1,\dots,y_n$ be an orthonormal basis in $\rr^n$. Set
\[Q:=\i^{n(n-1)/2}\phi(y_1)\cdots\phi(y_n).\]
Then the following is true:
\ben\item $Q$ depends only on the orientation of the basis  (it changes the
sign under the change of the orientation).
\item $Q$ is unitary and self-adjoint, moreover, $Q^2=1$.
\item $Q\phi(y)=(-1)^n \phi(y)Q$, for any $y\in\cY$.
\item If $n=2m$, then 
$Q=\i^m\phi(y_1)\cdots\phi(y_{2m})$ and
\[\rr^{2m+1}\ni (y,t)\mapsto\phi(y)\pm tQ\]
are two representations  of the CAR.
\item If $n=2m+1$, then 
$Q=(-\i)^m\phi(y_1)\cdots\phi(y_{2m+1})$ and
$\cH=\Ker(Q-1)\oplus\Ker(Q+1)$ gives a decomposition
of $\cH$  into a direct sum of subspaces preserved by our representation.
\een\label{qu}\eet

\subsection{Representations of the CAR  in
  terms of Pauli matrices}

In the space $\cc^2$ we introduce the usual Pauli spin matrices $\sigma_1$,
$\sigma_2$ and $\sigma_3$. This means
\[\sigma_1=\left[\begin{array}{cc}0&1\\1&0\end{array}\right],\ \ \ \ 
\sigma_2=\left[\begin{array}{cc}0&-\i\\\i&0\end{array}\right],\ \ \ \ 
\sigma_3=\left[\begin{array}{cc}1&0\\0&-1\end{array}\right].\]
Note that $\sigma_i^2=1$, $\sigma_i^*=\sigma_i$,  $i=1,2,3$,
 and\beq\begin{array}{l}
\sigma_1\sigma_2=-\sigma_2\sigma_1=\i\sigma_3,\\[3mm]
\sigma_2\sigma_3=-\sigma_3\sigma_2=\i\sigma_1\\[3mm]
\sigma_3\sigma_1=-\sigma_1\sigma_3=\i\sigma_2.\end{array}
\label{dzd}\eeq
Moreover, $B(\cc^2)$ has a basis $(1,\sigma_1,\sigma_2,\sigma_3)$.
Clearly, $(\sigma_1,\sigma_2,\sigma_3)$ is a representation of the CAR
over $\rr^3$.  

In the algebra $B(\otimes^n\cc^2)$ we introduce the
operators
\[\sigma_i^{(j)}:=1^{\otimes(j-1)}\otimes\sigma_i\otimes1^{\otimes(n-j)},\
\ \
i=1,2,3,\ \ j=1,\dots,n.\]
Note that
 $\sigma_i^{(j)}$ satisfy (\ref{dzd}) for any $j$ and commute for distinct
$j$.
Moreover, $B(\otimes^n\cc^2)$ is generated as an algebra
 by $\{\sigma_i^{(j)}\ :\
j=1,\dots, n,\ i=1,2\}$.
Set $I_j:=\sigma_3^{(1)}\cdots\sigma_3^{(j)}$.
In order to  transform spin matrices into a representation of the CAR we need
to apply the so-called {\em Jordan-Wigner construction}. According to this
construction,
\[\left(\sigma_1^{(1)},\sigma_2^{(1)},I_1\sigma_1^{(2)},I_1\sigma_2^{(2)},\dots,
I_{n-1}\sigma_1^{(n)},I_{n-1}\sigma_2^{(n)}\right)\]
is a representation of the CAR over $\rr^{2n}$. By adding $\pm I_n$ we obtain 
 a representation of the CAR over $\rr^{2n+1}$. 

The  following theorem can be viewed as a fermionic analog of the Stone-von
Neumann Theorem \ref{stvn}.
 It is, however, much easier to prove and belongs to standard results 
 about Clifford algebras \cite{Lo}.

\bet \ben\item
Let $(\phi_1,\phi_2,\dots,\phi_{2n})$ be a representation of the 
CAR over $\rr^{2n}$ in a
Hilbert space $\cH$. Then there exists a Hilbert space $\cK$ and a
unitary operator \[U:\otimes^n\cc^{2}\otimes\cK\to \cH\] 
such that
\[\begin{array}{l}
U\left(I_{j-1}\sigma_1^{(j)}\otimes 1_{\cK}\right) =\phi_{2j-1}U,\\[3mm]
U\left(I_{j-1}\sigma_2^{(j)}\otimes 1_{\cK} \right)=\phi_{2j}U,\ \ \
j=1,\dots,n. 
\end{array}\]
The representation is irreducible iff $\cK=\cc$.
\item Let $(\phi_1,\phi_2,\dots,\phi_{2n+1})$ be a representation of the CAR over $\rr^{2n+1}$ in a
Hilbert space $\cH$. Then there exist Hilbert spaces $\cK_-$ and
$\cK_+$ and a
unitary operator
\[U:\otimes^n\cc^{2}\otimes(\cK_+\oplus\cK_-)
\to \cH\]
such that
\[\begin{array}{l}
U\left(I_{j-1}\sigma_1^{(j)}\otimes 1_{\cK_+\oplus\cK_-}\right) =\phi_{2j-1}U,\\[3mm]
U\left(
I_{j-1}\sigma_2^{(j)}\otimes 1_{\cK_+\oplus\cK_-}\right) =\phi_{2j}U,
\ \ \
j=1,\dots,n,\\[3mm]
U\left(I_n\otimes(1_{\cK_+}\oplus-1_{\cK_-})\right)
=\phi_{2n+1}U.
\end{array}\]\een
\eet

\begin{corollary} Suppose that $\cY$ is an 
even dimensional real Hilbert space. Let
$\cY\ni y\mapsto \phi^{\pi_1}(y)\in B_\h(\cH)$ and
$\cY\ni y\mapsto \phi^{\pi_2}(y)\in B_\h(\cH)$ be two  irreducible
representations of the CAR. Then they are unitarily equivalent.
\label{cosar}
\end{corollary}

\subsection{Fermionic
Bogolubov transformations}

Let $(\cY,\alpha)$ be a finite dimensional real Hilbert space (a Euclidean
  space). 
Linear transformations on $\cY$
 that preserve the scalar product are invertible and 
 form a group, which will be called the orthogonal group of
$\cY$ and denoted $O(\cY)$.

Let
\beq
\cY\ni y\mapsto \phi(y)\in B_\h(\cH)
\label{iui4car}\eeq be an irreducible  
 representation of the CAR.
The following theorem is an immediate  consequence of Corollary \ref{cosar}:

\bet Let $\dim\cY$ be even.
For any $r\in O(\cY)$ there exists $U\in U(\cH)$ such that
\beq U \phi(y)U^*=\phi(ry).\label{fgf1a}\eeq
The unitary operator 
$U$ in (\ref{fgf1a}) is defined uniquely up to a phase factor. Let $\cU_r$
  denote the class of such 
operators. Then
\[O(\cY)\ni r\mapsto \cU_r\in  U(\cH)/U(1)\]
is a group homomorphism.
\label{bogocar}\eet

One can ask whether one can 
fix uniquely the phase factor appearing  in the above theorem
and obtain
a group homomorphism of $O(\cY)$ into $U(\cH)$ satisfying (\ref{fgf1a}). 
This is impossible, the best  one can do is
the following improvement of Theorem \ref{bogocar}:

\bet  Let $\dim\cY$ be even.
For any $r\in O(\cY)$ there exists a unique pair
$\{U_r,-U_r\}\subset U(\cH)$ such that
\[U_r \phi(y)U_r^*=\phi(ry),\]
and such that we have a group homomorphism
\beq O(\cY)\ni r\mapsto \pm U_r\in  U(\cH)/\{1,-1\}.\label{bogo2car}\eeq
\label{bogo1car}\eet

(\ref{bogo2car}) is called the {\em  Pin representation of $O(\cY)$}.

 Note that for $\dim\cY>2$, the
homotopy group of $O(\cY)$ is $\zz_2$. Hence the double covering of $O(\cY)$
is its universal covering.
The image of the Pin representation in $U(\cH)$ is isomorphic to this double
covering and is called the Pin group.

In the physics literature the fact that orthogonal transformations can be
unitarily implemented is again associated with the name of Bogolubov
and the transformations
described in Theorems \ref{bogocar} and \ref{bogo1car} are often called (fermionic)
  Bogolubov transformations. They are used e.g. in the theory
of the  superconductivity.

 Theorems  \ref{bogocar} and \ref{bogo1car}  are well known in mathematics
in the context of theory of Clifford algebras.
They are most conveniently proven by using, 
 what we call, the Fock representation,
 where one has simple formulas for $U_r$. 
We will describe these formulas later on (in a more general context of the
infinite number of degrees of freedom).






\section{Fock spaces}
\label{s5}
In this section we fix the notation for 
  bosonic and
fermionic Fock spaces. Even though these concepts are  widely used, there
seem to be  no universally accepted symbols for many concepts in this area.

\subsection{Tensor algebra}
Let $\cZ$ be  a Hilbert space.
Let $\otimes^n\cZ$ denote the $n$-fold tensor product of $\cZ$.
We set\[\otimes \cZ=\loplus\limits_{n=0}^\infty
\otimes^n\cZ.\]
Here $\oplus$ denotes the  direct sum in the sense of Hilbert spaces, that
is the completion of the algebraic direct sum.
$\otimes\cZ$ is sometimes called the {\em full Fock space}. 
The element $1\in\cc=\otimes^0\cZ\subset\otimes\cZ$ is often called
 the {\em vacuum }and denoted $\Omega$.

Sometimes we will need
$\otimes^\fin\cZ$ which is the subspace of $\otimes \cZ$ with a finite number
of particles, that means
the algebraic direct sum of $\otimes^n\cZ$.

$\otimes\cZ$ and $\otimes^\fin\cZ$ are
 associative
 algebras with the operation $\otimes$ and the identity $\Omega$.

\subsection{Operators $\d\Gamma$ and $\Gamma$
in the full Fock space}
\label{5.3}

If $p$ is a closed operator from $\cZ_1$ to $\cZ_2$, then we define  the closed
operator $\Gamma^n(p)$
 from $\otimes^n\cZ_1 $ to $\otimes^n\cZ_2$ and $\Gamma(p)$
from  $\otimes\cZ_1
$ to $\otimes\cZ_2$:
\[\Gamma^n(p):=p^{\otimes n},     \]
\[\Gamma(p):=\loplus_{n=0}^\infty \Gamma^n(p).\]
$\Gamma(p)$ is  bounded iff $\|p\|\leq1$.
$\Gamma(p)$ is unitary iff $p$ is.

Likewise, if $h$ is a closed operator on $\cZ$,
 then we define the closed operator $\d\Gamma^n(h)$
 on $\otimes^n\cZ $  and $\d\Gamma(h)$ on  $\otimes\cZ$:
\[\d\Gamma^n(h)=\sum_{j=1}^n1_\cZ^{\otimes (j-1)}\otimes h\otimes
1_\cZ^{\otimes(n-j)},\]
\[\d\Gamma (h):=\loplus_{n=0}^\infty\d \Gamma^n(h)
.\] $\d\Gamma(h)$ is self-adjoint iff $h$ is.

The {\em number operator} is defined as $N=\d\Gamma(1)$.
The {\em parity operator }is 
\beq I:=(-1)^N=\Gamma(-1).\label{para0}\eeq

Let us give a sample of properties of operators on full Fock spaces.
\bet \ben
\item
 Let $h,h_1,h_2\in B(\cZ)$, 
$p_1\in B(\cZ,\cZ_1)$, $p_2\in B(\cZ_1,\cZ_2)$, $\|p_i\|\leq1$. We then have
\beq\Gamma(\e^{\i h})=\exp(\d\Gamma(\i h)),\label{piy}\eeq
\[\Gamma(p_2)\Gamma(p_1)=\Gamma(p_2 p_1),\]
\[[\d\Gamma(h_1),\d\Gamma(h_2)]=\d\Gamma([h_1,h_2]).\]
\item
 Let $\Phi,\Psi\in\otimes^\fin\cZ$,
 $h\in B(\cZ)$, $p\in B(\cZ,\cZ_1)$. Then
\[\Gamma(p)\ (\Phi\otimes \Psi)=(\Gamma(p)\Phi)\otimes(\Gamma(p)\Psi),\]
\[\d\Gamma(h)\ (\Phi{\otimes} \Psi)=(\d\Gamma(h)\Phi)\otimes \Psi
+\Phi\otimes(\d\Gamma(h)\Psi).\]
\een\eet

 Of
course, under appropriate technical conditions, similar statements are true
for unbounded operators. In particular (\ref{piy}) is true for any
self-adjoint $h$.




 \subsection{Symmetric and antisymmetric Fock spaces}
\label{s.sym} 
 

Let
$S^n\ni\sigma\mapsto
 \Theta(\sigma)\in U(\otimes^n\cZ)$
be the natural representation of the permutation
 group $S^n$ given by
\[\Theta(\sigma)z_1\otimes\cdots\otimes z_n:=z_{\sigma^{-1}(1)}\otimes
\cdots\otimes
z_{\sigma^{-1}(n)}.\]
 We define
 \[\Theta_\s^n:=\frac{1}{n!}\sum_{\sigma\in S^n}\Theta(\sigma)
,\]
 \[\Theta_\a^n:=\frac{1}{n!}\sum_{\sigma\in S^n}(\sgn\,\sigma)\:\Theta(\sigma)
.\]
 $\Theta_\s^n$ and $\Theta_\a^n $ are orthogonal
 projections in $\otimes^n \cZ$.

We will write $\sa$ as a subscript that can mean either $\s$ or $\a$.
We set
\[\Theta_\sa:=\loplus_{n=0}^\infty\Theta_\sa^n.\]
Clearly, $\Theta_\sa$ is an orthogonal projection in $\otimes\cZ$.

Define
\[ \begin{array}{rl}
\Gamma_\sa^n(\cZ)&:=\Theta_\sa^n\bigl(\otimes^n\cZ\bigr),\\[3mm]
\Gamma_\sa(\cZ)&:=\Theta_\sa\bigl(\otimes\cZ\bigr)=
\loplus_{n=0}^\infty\Gamma_\sa^n(\cZ).\end{array}\]
$\Gamma_\sa(\cZ)$ is often called the {\em bosonic/fermionic} or
{\em symmetric/antisymmetric  Fock space}.

We also introduce the finite particle Fock spaces
\[\Gamma_\sa^\fin(\cZ)=\bigl(\otimes^\fin\cZ\bigr)\cap
\Gamma_\sa(\cZ).\]

$\Gamma_\sa(\cZ)$ is a Hilbert space
(as a closed 
subspace of $\otimes\cZ$).

Note that $\Gamma_\sa^0(\cZ)=\cc$ and $\Gamma_\sa^1(\cZ)=\cZ$. $\cZ$ is often
called the {\em 1-particle space }and 
$\Gamma_\sa(\cZ)$ the second quantization of $\cZ$.

The following property of bosonic Fock spaces is often useful:
\bet 
The span of elements of 
the form $z^{\otimes n}$, $z\in\cZ$, is dense in $\Gamma_\s^n(\cZ)$.
\eet

\subsection{Symmetric and  antisymmetric tensor product}
If $\Psi,\Phi\in\Gamma_\sa^\fin(\cZ)$, then we define
 \[\Psi\otimes_\sa \Phi:=\Theta_\sa \bigl(\Psi\otimes \Phi\bigr)
\in \Gamma^\fin_\sa(\cZ).\]
$\Gamma_\sa^\fin(\cZ)$ 
 is an associative
 algebra with the operation $\otimes_\sa$ and the identity $\Omega$.

Note that
$z^{\otimes n}=z^{\otimes_\s n}$.

Instead of $\otimes_\a$ one often uses the wedge product, which for
$\Psi\in\Gamma_\a^p(\cZ)$, $\Phi\in\Gamma_\a^r(\cZ)$ is defined as
\[\Psi\wedge \Phi:=\frac{(p+r)!}{p!r!} \Psi\otimes_\a \Phi\in \Gamma_\a^\fin(\cZ).\]
It is also an associative operation.
Its advantage over $\otimes_\a$
is visible if we compare the following identities:
\[\begin{array}{lr}
v_1\wedge\cdots\wedge v_p=\sum\limits_{\sigma\in S^p}(\sgn\sigma)\ 
v_{\sigma( 1)}\otimes
\cdots\otimes v_{\sigma (p)},&\\[5mm]
v_1\otimes_\a\cdots\otimes_\a v_p=\frac{1}{p!}
\sum\limits_{\sigma\in S^p}(\sgn\sigma)\ v_{\sigma (1)}\otimes
\cdots\otimes v_{\sigma( p)},&\ \ \ v_1,\cdots,v_p\in\cZ.\end{array}\]

The advantage of $\otimes_\a$ is that it is fully analogous to $\otimes_\s$.


\subsection{$\d\Gamma$ and $\Gamma$ operations}

If $p$ is a closed operator from $\cZ$ to $\cW$,
 then $\Gamma^n(p)$, defined
in Subsect. \ref{5.3}, maps $\Gamma_\sa^n(\cZ)$
into $\Gamma_\sa^n(\cW)$. Hence $\Gamma(p)$ maps
$\Gamma_\sa(\cZ)$
 into  $\Gamma_\sa(\cW)$. We will use the same symbols 
 $\Gamma^n(p)$ and  $\Gamma(p)$ to denote the corresponding 
restricted operators.

If $h$ is a closed operator on $\cZ$,
 then $\d\Gamma^n(h)$ maps $\Gamma_\sa^n(\cZ)$
 into itself. Hence, $\d\Gamma(h)$ maps $\Gamma_\sa(\cZ)$
 into itself. We will use the same symbols 
 $\d\Gamma^n(h)$ and  $\d\Gamma(h)$ to denote the corresponding 
restricted operators.

$\Gamma(p)$ is called the 2nd quantization of $p$. Similarly, 
$\d\Gamma(h)$ is 
sometimes  called
the  2nd quantization of $h$.

\bet Let $p\in B(\cZ,\cZ_1)$, $h\in B(\cZ)$, 
$\Psi,\Phi\in\Gamma_\sa^\fin(\cZ)$. Then
\[\Gamma(p)\ \bigl(\Psi\otimes_\sa \Phi\bigr)
=(\Gamma(p)\Psi)\otimes_\sa(\Gamma(p)\Phi),\]
\[\d\Gamma(h)\ \bigl(\Psi\otimes_\sa \Phi\bigr)=(\d\Gamma(h)\Psi)\otimes_\sa \Phi
+\Psi\otimes_\sa(\d\Gamma(h)\Phi).\]
\eet

\subsection{Tensor product of Fock spaces}

In this subsection we describe the so-called {\em 
exponential law for Fock spaces}.

Let $\cZ_1$ and $\cZ_2$ be Hilbert spaces.
We introduce the identification
\[  U:\Gamma_\sa^\fin(\cZ_1)\otimes
\Gamma_\sa^\fin(\cZ_2)\to  \Gamma_\sa^\fin(\cZ_1\oplus\cZ_2)\]
as follows. Let $\Psi_1\in \Gamma_\sa^{n}
(\cZ_1)$, $\Psi_2\in \Gamma_\sa^{m}
(\cZ_2)$. Let $j_i$ be the imbedding
of $\cZ_i$ in $\cZ_1\oplus\cZ_2$. Then 
\beq \begin{array}{l}
U(\Psi_1\otimes
\Psi_2):=\sqrt{\frac{(n+m)!}{n!m!}}
(\Gamma(j_1)\Psi_1)\otimes_\sa(\Gamma(j_2)\Psi_2)
\end{array}.\label{idi}\eeq

\bet
\ben
\item
 $
 U(\Omega_1\otimes\Omega_2)=\Omega$.   
\item  $U$ extends
to a unitary operator 
 $\Gamma_\sa(\cZ_1)\otimes\Gamma_\sa(\cZ_2)\to\Gamma_\sa(\cZ_1\oplus\cZ_2)$.
\item If $h_i\in B(\cZ_i)$, then
\[\begin{array}{l}

U\big(\d\Gamma(h_{1})\otimes 1 + 1 \otimes \d\Gamma(h_{2})\big)=
\d\Gamma(h_1\oplus h_2)U.
\end{array}\]
\item If $p_i\in B(\cZ_i)$, then\[\begin{array}{l}
U\left(\Gamma(p_1)\otimes\Gamma(p_2)\right)=\Gamma(p_1\oplus p_2)U.\end{array}
\]
\een\label{tensor.1}
\eet

\subsection{Creation and annihilation operators}

Let $\cZ$ be a Hilbert space and $w\in\cZ$. 
We consider the bosonic/fermionic Fock space $\Gamma_\sa(\cZ)$.

Let $w\in\cZ$.
We define two operators with the domain $\Gamma_\sa^\fin(\cZ)$.
The {\em creation operator }is defined as
\[a^*(w)\Psi:=\sqrt{n+1}w\otimes_\sa\Psi,\ \ \Psi\in\Gamma^n_\sa(\cZ)\]
In the fermionic case, $a^*(w)$ is bounded. In the bosonic case, $a^*(w)$ is
densely defined and closable. In both cases we 
define denote the closure of $a^*(w)$ by the same symbol.
Likewise, in both cases we define
 the {\em annihilation operator} by
\[ a(w):=a^*(w)^*.\]
Note that
\[a(w)\Psi=\sqrt{n}\bigl((w|\otimes 1\bigr)\Psi,\ \ \
\Psi  \in\Gamma^n_\sa(\cZ).\]

\bet\ben  \item In the bosonic case we have
 \[\begin{array}{l}[a^*(w_1),a^*(w_2)]=0,\ \ 
[a( w_1),a( w_2)]=0,\\[3mm]
[a( w_1),a^*(w_2)]=(w_1|w_2).
\end{array}\]
\item
  In the fermionic case we have
\[\begin{array}{l}[a^*(w_1),a^*(w_2)]_+=0,\ \ 
[a( w_1),a( w_2)]_+=0,\\[3mm]
[a( w_1),a^*(w_2)]_+=(w_1|w_2).
\end{array}\]
\een\eet

In both bosonic and fermionic cases the following is true:
\bet Let $p,h\in B(\cZ)$ and $w\in\cZ$. Then
\ben\item
$\Gamma(p)a( w)=a(p^{*-1}w)\Gamma(p)$,
\item
$[\d\Gamma(h),a( w)]=-a(h^* w)$,
\item $\Gamma(p) a^*(w)=a^*(pw)\Gamma(p)$,
\item
$[\d\Gamma(h),a^*(w)]=a^*(hw).$\een\eet

The exponential law for creation/annihilation operators is slightly different
in the bosonic and fermionic cases:
\bet Let $\cZ_1$ and $\cZ_2$ be Hilbert spaces and
$(w_1,w_2)\in\cZ_1\oplus\cZ_2$.
 Let
$U$ be the  defined for these spaces as in Theorem \ref{tensor.1}.
\ben
\item
In the bosonic
case we have
\[
a^*(w_1, w_2) U= U(a^*(w_1)\otimes1+
1\otimes a^*(w_2)),\]
\[
a( w_1,  w_2) U=
 U(a( w_1)\otimes1+
1\otimes a(w_2)).\]
\item
In the fermionic
case, if $I_1$ denotes the parity operator for 
$\Gamma_\a(\cZ_1)$ (see (\ref{para0})), then
\[
a^*(w_1, w_2) U= U(a^*(w_1)\otimes1+
I_1\otimes a^*(w_2)),\]
\[
a( w_1,  w_2) U=
 U(a( w_1)\otimes1+
I_1\otimes a(w_2)).\]
\een\label{tensor.2}
\eet

Set \beq\Lambda:=(-1)^{N(N-1)/2}. \label{ksl}\eeq
The following property is valid both in the bosonic and fermionic cases:
\beq\begin{array}{rl}
\Lambda a^*(z)\Lambda=-I a^*(z)=a^*(z)I,\\[3mm]
\Lambda a(z)\Lambda=I
a(z)=-a(z)I.\end{array}\label{gads}\eeq 
In the fermionic case, (\ref{gads}) allows to convert the anticommutation
relations into commutation relations
\[\begin{array}{rll}
[\Lambda a^*(z)\Lambda, a^*(w)]=[\Lambda a(z)\Lambda, a(w)]&=&0,\\[3mm]
[\Lambda a^*(z)\Lambda, a(w)]&=&I(w|z).
\end{array}\]

\bet Let the assumptions of Theorem
\ref{tensor.2} be satisfied. 
Let $N_i$, $I_i$, $\Lambda_i$ be the operators defined as above 
corresponding to $\cZ_i$, $i=1,2$. Then
\ben
\item $\Lambda U=U(\Lambda_1\otimes\Lambda_2)(-1)^{N_1\otimes N_2}$.
\item
In the fermionic case,
\[\begin{array}{l}
\Lambda a^*(w_1,w_2)\Lambda\ U=
U(a^*(w_1) I_1\otimes I_2+1\otimes a^*(w_2) I_2),\\[3mm]
\Lambda a(w_1,w_2)\Lambda\ U=
U(-a(w_1) I_1\otimes I_2-1\otimes a(w_2) I_2).
\end{array}\]\een\label{qrq}\eet

\proof
To prove  2) we use 1) and 
$(-1)^{N_1\otimes N_2}a(w)\otimes 1
(-1)^{N_1\otimes N_2}=a(w)\otimes I_2.$
\qed

\subsection{Multiple creation and annihilation operators}
\label{faf}
Let $\Phi\in\Gamma_\sa^m(\cZ)$. We 
define the operator of creation of $\Phi$ with the domain
$\Gamma_\sa^\fin(\cZ)$ 
as follows:
\[a^*(\Phi)\Psi:=\sqrt{(n+1)\cdots(n+m)}\Phi\otimes_\sa\Psi,\ \ \ 
\Psi\in\Gamma_\sa^n.\]
$a^*(\Phi)$ is a densely defined closable operator. We denote its closure by
the same symbol.
We  set
\[a(\Phi):=\left(a^*(\Phi)\right)^*.\]
$a(\Phi)$ is called the operator of annihilation of $\Phi$.
For $w_1,\dots,w_m\in\cZ$ we have
\[a^*(w_1\otimes_\sa\cdots\otimes_\sa w_m)=a^*(w_1)\cdots a^*( w_m),\]
\[a(w_1\otimes_\sa\cdots\otimes_\sa w_m)=a(w_m)\cdots a( w_1).\]

Recall from Subsect. \ref{s3.6}
that we can identify the space $B^2(\bar\cZ,\cZ)$ with $\otimes^2\cZ$.
Hence, we have an identification of
 $B_\sa^2(\bar\cZ,\cZ)$  with $\Gamma_\sa^2(\cZ)$.

Thus if $c\in B_\sa^2(\bar\cZ,\cZ)$, then by interpreting $c$ as an element of
$\Gamma_\sa^2(\cZ)$, we can use 
the notation $a^*(c)$ /  $a(c)$ for
 the corresponding
two-particle creation/annihilation operators.

\section{Representations of the CCR in Fock spaces}

\subsection{Field operators}

Let $\cZ$ be  a (complex) Hilbert space. 
Define the real vector space
\beq\Re(\cZ\oplus\bar\cZ):=
\{(z,\bar z)\ :\ z\in\cZ\}.\label{split}\eeq
Clearly, $\Re(\cZ\oplus\bar\cZ)$ is a real subspace  of $\cZ\oplus\bar\cZ$.
For shortness, we will usually write $\cY$ for $\Re(\cZ\oplus\bar\cZ)$.
In this section we will treat $\cY$ as a symplectic space equipped
 with  the symplectic
form
\[(z,\bar z)\omega (w,\bar w)=2\Im(z|w).\]

Consider the creation/annihilation operators
$a^*(z)$ and $a(z)$ acting on the bosonic Fock space $\Gamma_\s(\cZ)$.
 For $y=(w,\bar w)\in\cY$ we define 
\[\phi(y):=a^*(w)+a(w).\]

Note that $\phi(y)$ is essentially self-adjoint on $\Gamma_\s^\fin(\cZ)$.
We use the same symbol $\phi(y)$ for its self-adjoint extension.

We have the following commutation relations
\[[\phi(y_1),\phi(y_2)]=\i y_1\omega y_2,\ \ y_1,y_2\in\cY,\]
as an identity on $\Gamma_\s^\fin(\cZ)$.

It is well known that in every bosonic Fock space we have a natural
representation of the CCR:
\bet
 The map \beq
\cY\ni y\mapsto W(y):=\e^{\i\phi(y)}\in U(\Gamma_\s(\cZ))\label{pio}
\eeq
is 
a regular irreducible
representation of the CCR.
\eet (\ref{pio})
 is called {\em the Fock representation of the CCR}. 


One often identifies the spaces $\cY$ and $\cZ$ through
\beq \cZ\ni z\mapsto\frac1{\sqrt2}(z,\bar z)\in\cY.\label{identi1}\eeq
With this identification, one introduces the field operators for $w\in\cZ$ as
\[\phi(w):=\frac{1}{\sqrt2}\left(a^*(w)+a(w)\right).\]
The converse identities are
\[\begin{array}{l}
a^*(w)=\frac{1}{\sqrt2}\left(\phi(w)-\i\phi(\i w)\right),\\[3mm]
a(w)=\frac{1}{\sqrt2}\left(\phi(w)+\i\phi(\i w)\right).
\end{array}\]
Note
 that in the fermionic case a different identification seems more convenient
(see (\ref{identi2})).
In this section we will avoid to  identify $\cZ$ with $\cY$. 

Note that the
physical meaning of $\cZ$ and $\cY$ is different: $\cZ$ is the one-particle
Hilbert space of the system, $\cY$ is its classical phase space and
$\cZ\oplus\bar\cZ$ can be identified with the complexification of the
classical phase space, that is $\cc\cY$.
For instance, if we are interested in  a real 
scalar quantum field theory, then
$\cZ$ is the space of positive energy solutions of the Klein-Gordon equation,
 $\cY$ is the space of real solutions and $\cZ\oplus\bar\cZ$ is the space of
complex solutions.
See e.g. \cite{DB} in this collection of lecture notes, where this point is
dicussed in more detail.

\subsection{Bosonic Gaussian vectors}
\label{squee1}

Let $c\in B_\s^2(\bar\cZ,\cZ)$. Recall that $c$ can be identified with an
 element of $\Gamma_\s^2(\cZ)$. Recall from Subsect \ref{faf} that we 
 defined an unbounded  operator $a^*(c)$ on $\Gamma_\s(\cZ)$ such that
for $\Psi_n\in\Gamma_\s^n(\cZ)$
\begin{equation} a^*(c)\Psi_n:=\sqrt{(n+2)(n+1)}\:c\otimes_\s\Psi_n
\in\Gamma_\s^{n+2}(\cZ). 
\label{ccca}\end{equation}

\bet Assume that
 $\|c\|<1$.
\ben
\item $\e^{\frac12 a^*(c)}$ is 
 a closable operator on $\Gamma_\s^\fin(\cZ)$.
\item  
$\det(1-cc^*)>0$, so that we can define the vector
\beq\begin{array}{l}
\Omega_c:=\left(\det(1-cc^*)\right)^\frac{1}{4}\exp(\12
a^*(c))\Omega\end{array}\label{sq1}\eeq
 It is  
the unique vector in $\Gamma_\s(\cZ)$ satisfying
\[\|\Omega_c\|=1,\ \ \ (\Omega_c|\Omega)>0,\ \ \ \
(a(z)-a^*(c\bar z))\Omega_c=0,\ \ z\in\cZ.\]
\een\label{sqo}\eet

In the Schr\"odinger representation the vectors $\Omega_c$ are normalized
 Gaussians with
an arbitrary dispersion --- hence they are often called
 {\em squeezed states}.

\subsection{Complex structures compatible with a symplectic form}

Before analyzing Bogolubov transformations on a Fock space it is natural to
start with a little  linear algebra of symplectic vector spaces.

We can treat $\cY$ as a  real 
Hilbert space.
In fact, we have a natural scalar product
\[(z,\bar z)\alpha(w,\bar w):=\Re (z|w).\]
This scalar product will have a fundamental importance in the next section,
when we will discuss fermions. In this section we need it only to
define 
 bounded and trace class operators.

 We define $Sp(\cY)$ to be the set of
all bounded invertible linear maps on $\cY$ preserving $\omega$.
(This extends the definition of $Sp(\cY)$ from the case of a
finite dimensional
symplectic space $\cY$ to the present context).

A linear map $j$ is called a {\em complex structure} (or an {\em
  antiinvolution})  iff
$j^2=-1$. We say that it is {\em compatible with a symplectic form $\omega$} 
 iff
$j\in Sp(\cY)$ and the symmetric form $y_1\omega jy_2$, where $y_1,y_2\in\cY$,
 is positive definite.
(One also says that $j$ is K\"ahler with respect to $\omega$).

On $\cY$ we  introduce the linear map
\[j(z,\bar z):=(\i z,-\i \bar z).\]
It is easy to see that $j$ is a complex structure compatible  with $\omega$.

Note that fixing the complex structure $j$ on the symplectic space
$\cY$ compatible with the symplectic form $\omega$
is equivalent to identifying $\cY$ with $\Re(\cZ\oplus\bar\cZ)$
for some complex Hilbert space $\cZ$.

 Let $r\in
B(\cY)$. We can extend $r$ to $\cZ\oplus\bar\cZ$ by complex linearity.
On $\cZ\oplus\bar\cZ$
we can write $r$ as a 2 by 2 matrix
\[r=\left[\begin{array}{cc}p&q\\ \bar q&\bar p\end{array}\right],\]
where $p\in B(\cZ,\cZ)$, $q\in B(\bar\cZ,\cZ)$.
Now $r\in Sp(\cY)$ iff
\[\begin{array}{ll}
p^*p- q^\t\bar q=1,&\ 
 p^\t\bar q-q^*p=0,\\[3mm]
pp^*-qq^*=1,&
p q^\t-q p^\t=0.
\end{array}\]

We have
\[pp^*\geq1,\ \ \ \ p^*p\geq1.\]
Hence $p^{-1}$ exists and $\|p^{-1}\|\leq1$.

We define the operators  $c,d\in B(\bar\cZ,\cZ)$
\[c:=p^{-1} q=q^\t(p^\t)^{-1},\ \ \ \ d:=q\bar p^{-1}=(p^*)^{-1}q^\t.\]
Note that $d,c$ are symmetric (in the sense defined in Sect. \ref{2}), $\|d\|\leq1$, $\|c\|\leq1$,
\beq r=\left[\begin{array}{cc}1&d\\0&1\end{array}\right]
\left[\begin{array}{cc}(p^*)^{-1}&0\\0&\bar p\end{array}\right]
\left[\begin{array}{cc}1&0\\\bar c&1\end{array}\right],\label{adad}\eeq
\[1-cc^*=(p^*p)^{-1},\ \ \ 1-dd^*=(\bar p^*\bar p)^{-1}.\]
The decomposition (\ref{adad}) plays an important role in the 
description of Bogolubov
transformations.

In the following theorem we introduce a certain subgroup of $Sp(\cY)$,
which will play an important role in Shale's Theorem on the implementabilty
of Bogolubov transformations.
\bet Let $r\in Sp(\cY)$. The following conditions are equivalent:
\[\begin{array}{ll} 0)\ \ii-r\ii r^{-1}\in B^2(\cY),\ \ &
1)\  r\ii-\ii r\in B^2(\cY),\\[3mm] 2)\  \Tr q^*q<\infty,& 3)\ 
\Tr(pp^*-\one)<\infty,\\[3mm]
4) \ d\in B^2(\bar\cZ,\cZ),& 5) \ c\in B^2(\bar\cZ,\cZ).
\end{array}\]
Define $Sp_2(\cY)$ 
to be the set of  $r\in Sp(\cY)$ satisfying the
above conditions. Then $Sp_2(\cY)$ is a group.
\label{rdr}\eet

\subsection{Bosonic Bogolubov transformations in the Fock
representation}

Consider now the Fock representation $\cY\ni y\mapsto W(y)\in
U(\Gamma_\s(\cZ))$ defined in (\ref{pio}).

The following theorem describes when a symplectic transformation is
implementable by a unitary transformation. Part 1) was originally proven in
\cite{Sh}. Proof of 1) and 2) can be found in \cite{Ber,Ru2}.

\bet[Shale Theorem] \ben\item Let $r\in Sp(\cY)$.
 Then the following conditions are equivalent:
\newline
a) There exists   $U\in U(\Gamma_\s(\cZ))$ such that
\beq UW(y)U^*=W(ry),\ \ \ y\in\cY.\label{imple}\eeq
b) $r\in Sp_2(\cY)$.
\item
If the above conditions are satisfied, then $U$ is defined uniquely up to a
phase factor. Moreover, if we
define
\beq U_r^\ii=|\det
pp^*|^{-\frac{1}{4}}\e^{-\12a^*(d)}\Gamma((p^*)^{-1})\e^{\12a(
  c)},\label{shale}
\eeq
then $U_r^\ii$   is the unique unitary operator 
satisfying (\ref{imple}) and
\beq (\Omega|U_r^\ii\Omega)>0.\label{posi}\eeq
\item If $\cU_r=\{\lambda U_r^\ii\ :\ \lambda\in\cc,\ |\lambda|=1\}$, then
\[Sp_2(\cY)\ni r\mapsto \cU_r\in U(\Gamma_\s(\cZ))/U(1)\]
is a homomorphism of groups.
\een
\label{omeg}\eet

\subsection{Metaplectic group in the Fock representation}

 $r\mapsto U_r^j$ is not a representation of $Sp(\cY)$, it is only a
projective representation. By taking a certain subgroup of $Sp_2(\cY)$ we can
obtain a representation analogous to the metaplectic representation described
in Theorem \ref{bogo1}.

 Define $Sp_1(\cY):=\{r\in Sp(\cY)\     :\ r-1\in B^1(\cY)\}$.
(Recall that $B^1(\cY)$ are  trace class
 operators). 

\bet \ben \item $Sp_1(\cY)$ is a subgroup of $Sp_2(\cY)$.
\item $r\in Sp_1(\cY)$ iff
$  p-1\in B^1(\cZ)$.
\een
\eet

 For $r\in Sp_1(\cY)$, define
\beq \pm U_r=\pm(\det
p^*)^{-\12}\e^{-\12a^*(d)}\Gamma((p^*)^{-1})\e^{\12a( c)}.\label{dete}\eeq
(We take both signes of the square root, thus
 $\pm U_r$ denotes a pair of operators differing by a sign).

\bet\ben
\item $\pm U_r\in U(\Gamma_\s(\cZ))/\{1-1\}$;
\item
$U_rW(y)U_r^*=W(ry)$.
\item
 The following map is a group homomorphism.
 \beq Sp_1(\cY)\ni r\mapsto \pm U_r\in U(\cH)/\{1,-1\}\label{hho}\eeq
\een\label{impli}\eet

Clearly, 
the operators $\pm U_r$ differ by a phase factor from
$U_r^\ii$ from Theorem  \ref{omeg}.

\subsection{Positive symplectic transformations}

Special role is played by 
positive symplectic transformations.
It is easy to show that
 $r\in Sp_2(\cY)$ is a positive self-adjoint operator on $\cY$ iff
it is of the
form
\beq r=\left[\begin{array}{cc}(1-cc^*)^{-1/2}&(1-cc^*)^{-1/2}c
\\(1-c^*c)^{-1/2}c^*
&(1-c^*c)^{-1/2}\end{array}\right],\label{sqo1}\eeq
for some $c\in B_\s^2(\bar\cZ,\cZ)$.

The following theorem describes Bogolubov transformations associated with
positive symplectic transformations.

\bet\ben
\item The formula
\begin{equation}\begin{array}{l}
R_c:=\det(1-cc^*)^\frac{1}{4}\exp(-\12 a^*(c))\Gamma\left((1-cc^*)^\12\right)
\exp(\12 a(c))\end{array}\label{squ}\end{equation}
defines a unitary operator on $\Gamma_\s(\cZ)$.
\item If $\Omega_c$ is defined in (\ref{sq1}), then $\Omega=R_c\Omega_c$,
\item
\[\begin{split}
R_ca^*(z)R_c^*=a^*\left((1-cc^*)^{-1/2}z\right)+a\left(
(1-cc^*)^{-1/2}c\bar z\right)\\[3mm]
R_ca( z)R_c^*=a^*
\left(
(1-cc^*)^{-1/2}c\bar z\right)+
a\left((1-cc^*)^{-1/2} z\right).\\[3mm]
\end{split}\]
\item
  If $r$ is related to $c$ by
 (\ref{sqo1}), then
$R_c$  coincides with  $U_r^j$ defined in  (\ref{shale}).
\item
$R_c$
 coincides with 
$U_r$ defined in (\ref{dete}), where we take the plus sign and the positive
 square root.
\een
\eet

\section{Representations of the CAR in  Fock spaces}

\subsection{Field operators}

Let $\cZ$ be  a Hilbert space. 
As in the previous section, let  $\cY:=\Re(\cZ\oplus\bar\cZ)
$. This time, however,  we treat it as a real Hilbert space
 equipped with  the scalar product
\[(z,\bar z)\alpha (w,\bar w)=\Re(z|w).\]

For $w\in\cZ$, consider the creation/annihilation operators
$a^*(w)$ and $a(w)$ acting on the fermionic Fock space $\Gamma_\a(\cZ)$.
 For $y=(w,\bar w)\in\cY$ we define 
\[\phi(y):=a^*(w)+a(w).\]

Note that $\phi(y)$ are bounded and self-adjoint for any $y\in \cY$. Besides,
\[[\phi(y_1),\phi(y_2)]_+=2 y_1\alpha y_2,\ \ y_1,y_2\in\cY.\]
Thus we have
\bet
\beq\cY\ni y\mapsto\phi(y)\in B_\h(\Gamma_\a(\cZ))\label{focco}\eeq
is an irreducible representation of the CAR over the space $(\cY,\alpha)$.
\eet
(\ref{focco}) is called the {\em  Fock representation of the CAR}.

Let $w_1,\dots,w_m$ be an orthonormal basis of the complex Hilberts space
$\cZ$. Then 
\beq(w_1,\bar w_1),(-\i w_1,\i\bar w_1),\dots
(w_m,\bar w_m),(-\i w_m,\i\bar w_m)\label{bas}\eeq
is an orthonormal basis of the real Hilbert space
$\cY=\Re(\cZ\oplus\bar\cZ)$. It is easy to see that the orientation of
(\ref{bas}) does not depend on the choice of
 $w_1,\dots,w_m$.

The operator $Q$ defined as in Theorem \ref{qu} for this orientation equals
the parity operator $I=\Gamma(-1)=(-1)^N$. In fact, using Theorem \ref{qu} 4),
we can compute
\[\begin{array}{rl}
Q&=\i^m
\prod\limits_{j=1}^m\phi(w_j,\bar w_j)\phi(-\i w_j,\i\bar w_j)\\[3mm]
&=\prod\limits_{j=1}^m\left(-a^*(w_j)a(w_j)+a(w_j)a^*(w_j)\right)=\Gamma(-1)
.\end{array}\]

In the fermionic case, one
 often identifies the spaces $\cY$ and $\cZ$ through
\beq \cZ\ni w\mapsto(w,\bar w)\in\cY.\label{identi2}\eeq
With this identification, one introduces the field operators for $w\in\cZ$ as
\[\phi(w):=a^*(w)+a(w).\]
The converse identities are
\[\begin{array}{l}
a^*(w)=\frac{1}{2}\left(\phi(w)-\i\phi(\i w)\right),\\[3mm]
a(w)=\frac{1}{2}\left(\phi(w)+\i\phi(\i w)\right).
\end{array}\]

Using these identifications, we have for $z,w\in\cZ$
the identities
\[\begin{array}{l}[\phi(z),\phi(w)]_+=2\Re(w|z),
\\[3mm]
\Lambda\phi(w)\Lambda=-\i\phi(\i w)I=\i I\phi(\i w),
\\[3mm]
[\Lambda \phi(z)\Lambda,\phi(w)]=2\Im(w|z) I,
\end{array}\]
where $I$ is the parity operator and $\Lambda$ was introduced in (\ref{ksl}).

Note that the identification (\ref{identi2})
is different from the one used in the bosonic
 case (\ref{identi1}).
In this section we will avoid identifying  $\cZ$ with $\cY$.

\subsection{Fermionic Gaussian vectors}

Let $c\in\Gamma_\a^2(\cZ)$. Note that it can be identified with an
element of $ B_\a^2(\bar\cZ,\cZ)$. 
 $cc^*$ is trace class, so
$\det(1+cc^*)$ is well defined. 

\bet 
 Define a vector in $\Gamma_\a(\cZ)$ by
\beq\begin{array}{l}
\Omega_c:=\left(\det(1+cc^*)\right)^{-\frac{1}{4}}\exp(\12
a^*(c))\Omega.\end{array}\label{dsa}\eeq 
It is  
the unique vector satisfying
\[\|\Omega_c\|=1,\ \ \ (\Omega_c|\Omega)>0,\ \ \ \
(a(z)+a^*(c\bar z))\Omega_c=0,\ \ z\in\cZ.\]
\label{sqoa}\eet

Vectors of the form $\Omega_c$ are often used in the many body quantum
theory. In particular, they appear as convenient variational states in theory
of
 superconductivity that goes back to the work of
Bardeen-Cooper-Schrieffer, see e.g. \cite{FW}.

\subsection{Complex structures compatible with a scalar product}

Similarly as for bosons, it is convenient to study some abstract properties of
orthogonal transformations on a real Hilbert space as a preparation for the
analysis of fermionic Bogolubov transformations. 

Let $O(\cY)$ denote the group of orthogonal transformations on $\cY$.

We say that a complex structure $j$ is {\em compatible with the scalar product
$\alpha$} (or is K\"ahler with respect to $\alpha$) if
$j\in O(\cY)$.

Recall that on $\cY$ we have a distinguished
complex structure
\[j(z,\bar z):=(\i z,-\i \bar z).\]
It is easy to see that $j$ is compatible with $\alpha$.

Note that fixing the complex structure $j$ on a real Hilbert space $\cY$
compatible with the scalar product
 $\alpha$ is equivalent with identifying $\cY$ with
$\Re(\cZ\oplus\bar\cZ)$ for some complex Hilbert space $\cZ$.

 Let $r\in
B(\cY)$. Recall that we
 can extend $r$ to $\cZ\oplus\bar\cZ$ by complex linearity
and write it as
\[r=\left[\begin{array}{cc}p&q\\ \bar q&\bar p\end{array}\right],\]
where $p\in B(\cZ,\cZ)$, $q\in B(\bar\cZ,\cZ)$. Now $r\in O(\cY)$  iff
\[\begin{array}{ll}
p^*p+ q^\t\bar q=1,&\ 
 p^\t\bar q+q^*p=0,\\[3mm]
pp^*+qq^*=1,&
p q^\t+q p^\t=0.
\end{array}\]

It is convenient to distinguish a certain class of orthogonal
transformations given by the following theorem:

\bet Let $r\in O(\cY)$. Then the following conditions are equivalent:
\ben
\item
 $\Ker(rj+jr)=\{0\}$;
\item $\Ker(r^*j+jr^*)=\{0\}$;
\item $\Ker\: p=\{0\}$;
\item $\Ker \:p^*=\{0\}$. 
\een
\label{nono}\eet

If the conditions of Theorem \ref{nono} are satisfied, then we say
that $r$ is {\em $j$-nondegenerate.}
Let us assume that this is the case.
Then
$p^{-1}$ and $p^{*-1}$ are  densely defined operators.
Set
\[d=q\bar p^{-1}=-(p^*)^{-1}q^\t ,
\ \ \
c= p^{-1}q=-q^\t (p^\t)^{-1}.\]
and assume that they are bounded. Then $d,c\in B_\a(\bar\cZ,\cZ)$.
The following factorization of $r$ plays an important role in the description
of fermionic
Bogolubov transformations:
\[r=\left[\begin{array}{cc}1&d\\0&1\end{array}\right]
\left[\begin{array}{cc}(p^*)^{-1}&0\\0&\bar p\end{array}\right]
\left[\begin{array}{cc}1&0\\\bar c&1\end{array}\right].\]
We also have
\[1+cc^*=(p^*p)^{-1},\ \ \ 1+d^*d=(\bar p\bar p^*)^{-1}.\]

The following group will play an important role in the Shale-Stinespring
Theorem on the implementability of fermionic Bogolubov transformations:

\bep Let $r\in O(\cY)$. The following conditions are equivalent:
\ben
\item $\ii-r\ii r^{-1}\in B^2(\cY)$,\item
$ r\ii-\ii r\in B^2(\cY)$,\item $q\in B^2(\bar\cZ,\cZ)$
.\een
Define $O_2(\cY)$ 
to be the set of  $r\in O(\cY)$ satisfying the
above conditions. Then $O_2(\cY)$ is a group.
\eep

Note that if $r$ is $j$-nondegenerate, then it belongs to $O_2(\cY)$ iff $c\in
B^2(\bar\cZ,\cZ)$, or equivalently,  $d\in
B^2(\bar\cZ,\cZ)$.

\subsection{Fermionic Bogolubov transformations in the Fock
representation}

Consider now the Fock representation of the CAR, $\cY\ni y\mapsto \phi(y)\in
B_\h(\Gamma_\a(\cZ))$.

\bet \ben\item Let $r\in O(\cY)$.
 Then the following conditions are equivalent:
\newline
a) There exists   $U\in U(\Gamma_\a(\cZ))$
 such that
\beq U\phi(y)U^*=\phi(ry),\ \ \ y\in\cY.\label{imple1}\eeq
b) $r\in O_2(\cY)$
\item For $r\in O_2(\cY)$, the unitary operator 
$U$  satisfying (\ref{imple1})
 is defined uniquely up to a phase factor.
Let $\cU_r$ denote the class of these
operators. Then
\[O_2(\cY)\ni r\mapsto \cU_r\in U(\Gamma_\a(\cZ))/U(1)\]
is a homomorphism of groups.
\item Let $r\in O_2(\cY)$ be $j$-nondegenerate. Let $p,c,d$ be defined as
in the previous subsection.
Set
\beq U_r^\ii=|\det
pp^*|^{\frac{1}{4}}\e^{\12a^*(d)}\Gamma((p^*)^{-1})\e^{-\12a(
  c)}.\label{shale1}\eeq 
Then $U_r^\ii$  is the unique unitary operator 
satisfying (\ref{imple1}) and
\beq (\Omega|U_r^\ii\Omega)>0.\label{pos1}\eeq
\een
\label{omeg1}\eet

\subsection{Pin group in the Fock representation}

Define $O_1(\cY):=\{r\in O(\cY)\ :\ r-1\in B^1(\cY)\}$.

\bet\ben
\item
$O_1(\cY)$ is a subgroup of $ O_2(\cY)$.
\item
$r\in O_1(\cY)$ iff $p-1\in B^1(\cZ)$.
\een
\eet

The following theorem describes the Pin representation for an arbitrary number
of degrees of freedom:
\bet There exists a group homomorphism
 \beq O_1(\cY)\ni r\mapsto \pm U_r\in U(\Gamma_\a(\cZ))
/\{1,-1\}\label{hho3}\eeq
satisfying
$U_r\phi(y)U_r^*=\phi(ry)$.
\label{impli3}\eet

In order to give a formula for $\pm U_r$, which is analogous to the bosonic
formula
(\ref{dete}),
 we have to restrict ourselves to $j$-nondegenerate transformations.

\bet
Suppose  that $r\in O_1(\cY)$ is $j$-nondegenerate.
Then
\beq \pm U_r=\pm(\det
p^*)^{\12}\e^{\12a^*(d)}\Gamma((p^*)^{-1})\e^{-\12a( c)}.\eeq
\label{dete1}\eet

Similarly as in the bosonic case,
 it is easy to see that 
the operators $\pm U_r$ differ by a phase factor from
$U_r^\ii$ from Theorem  \ref{omeg1}.

\subsection{$j$-self-adjoint orthogonal transformations}

Special role is played  by
$r\in O_2(\cY)$ satisfying $rj=j^*r$.
Such transformations will be called {\em $j$-self-adjoint}.

One can easily show that $r\in O_2(\cY)$ is    $j$-self-adjoint if
\beq r=\left[\begin{array}{cc}(1+cc^*)^{-1/2}&(1+cc^*)^{-1/2}c
\\-(1+c^*c)^{-1/2}c^*
&(1+c^*c)^{-1/2}\end{array}\right].\label{qxa}\eeq
for some $c\in B_\a^2(\cY)$.

\bet\ben
\item The formula
\begin{equation}\begin{array}{l}
R_c:=\det(1+cc^*)^{-\frac{1}{4}}\exp(\12 a^*(c))\Gamma\left((1+cc^*)^{-\12}\right)
\exp(-\12 a(c))\end{array}\label{squa}\end{equation}
defines a unitary operator on $\Gamma_\a(\cZ)$.
\item If $\Omega_c$ is defined in (\ref{dsa}), then
$\Omega=R_c\Omega_c$.
\item
\[\begin{split}
R_ca^*(z)R_c^*=a^*\left((1+cc^*)^{-1/2}z\right)+a\left(
(1+cc^*)^{-1/2}c\bar z\right),\\[3mm]
R_ca(z)R_c^*=
a^*\left((1+cc^*)^{-1/2}c\bar z\right)+a\left((1+cc^*)^{-1/2}z\right).\\[3mm]
\end{split}\]
\item
If $r$ and $c$ are related by
(\ref{qxa}), then the operator $R_c$ 
coincides with the operator $U_r^j$
defined in (\ref{shale1}).
\item
$R_c$ coincides with
$U_r$  defined in (\ref{dete1}) with the plus sign and the positive square
  root. 
\een\eet

\section{$W^*$-algebras}     

In this section we review some elements of the theory of
$W^*$-algebras needed in our paper. For more details we refer the
reader to \cite{DJP}, and also  \cite{BR1, BR2, St,Ta1,Ta2}.

$\fM$ is a
 {\em $W^*$-algebra}
 if it is a $C^*$-algebra, possessing a {\em predual}. (This means
 that there exists a Banach space $\cY$ such that $\fM$ is isomorphic as a
 Banach  space to the dual of $\cY$. This Banach space $\cY$ is called a
 predual of $\fM$).

One can show that a  predual of a $W^*$-algebra
 is defined uniquely up to an isomorphism.
The topology on $\fM$ 
given by the functionals in the predual (the $*$-weak topology in
the terminology of theory of Banach spaces) will be called the {\em 
$\sigma$-weak
topology}. The set $\sigma$-weakly continuous linear functionals coincides with
 the predual of $\fM$.

$\rr\ni t\mapsto \tau^t$ is called
a {\em  $W^*$-dynamics} if it is a 1-parameter group with values in
$*$-automorphisms of $\fM$ and,
 for any $A\in\fM$,
$t\mapsto\tau^t(A)$ is $\sigma$-weakly continuous.
The pair $(\fM,\tau)$ is called a {\em $W^*$-dynamical system}.

\[
\fM\cap\fM':=\{B\in\fM\ :\ AB=BA,\ A\in\fM\}\]
 is called the {\em center of the algebra $\fM$}.
A $W^*$-algebra with a trivial center is called a {\em factor}.

If $\fA$ is a subset of $B(\cH)$ for some Hilbert space $\cH$, then
\[\fA':=\{B\ :\ AB=BA,\ A\in\fA\}\]
is called the {\em commutant of $\fA$}.

\subsection{Standard representations}

We say that $\cH^+$ is a {\em self-dual cone} in a Hilbert space $\cH$ if
\[\cH^+=\{\Phi\in\cH\ :\ (\Phi|\Psi)\geq0,\ \Psi\in\cH^+\}.\]

 We say that 
 a quadruple $(\pi,\cH,J,\cH^+)$ is a {\em standard representation of a
$W^*$-algebra} $\fM$ if $\pi:\fM\to \cB(\cH)$ is an injective
 $\sigma$-weakly continuous
 $*$-representation, $J$ is an antiunitary involution on $\cH$ and
 $\cH^+$ is a self-dual cone in $\cH$ satisfying the following
conditions:
\ben\item $J\pi(\fM) J=\pi(\fM)'$;
 \item $J\pi(A)J=\pi(A)^*$ for $A$ in the center
 of $\fM$;
\item $J\Psi=\Psi$ for $\Psi\in\cH^+$;
 \item $\pi(A)J\pi(A)\cH^+\subset\cH^+$ for $A\in\fM$.
\een Every $W^*$-algebra has  a unique (up to the unitary 
equivalence) standard representation,  \cite{Ha}
(see also \cite{Ar2, BR2, Co, DJP,Ta2}).

The standard representation has several important properties.
First,  every $\sigma$-weakly continuous state $\omega$  has a unique 
vector representative  in $ \cH^+$ (in other words, there is a unique
normalized  vector $\Omega\in\cH^+$ such that
$\omega(A)=(\Omega|
\pi(A)\Omega)$). Secondly, for 
every 
$*$-automorphism $\tau$ of $\fM$ there exists a unique unitary operator
 $U\in B(\cH)$ such that \[\pi(\tau(A))=U\pi(A)U^*,\ \ \ 
U\cH^+\subset\cH^+.\]
Finally, for every $W^*$-dynamics 
$\rr\ni t\mapsto\tau^t$ on $\fM$ there is a unique self-adjoint operator $L$ 
on $\cH$ such that
\begin{equation}
\begin{split}
\pi(\tau^t(A))=\e^{\i tL}\pi(A)\e^{-\i tL},\ \ \ \ 
\e^{\i t L} \cH^+ =\cH^+.
\end{split}
\label{modrel}
\end{equation}
The operator $L$  is called
the {\em standard
 Liouvillean of  
 $t\mapsto\tau^t$}.

Given a standard representation $(\pi,\cH,J,\cH^+)$
we  also have
 the {\em right
representation} $\pi_\r:\bar\fM\to B(\cH)$ given by
$\pi_\r(\bar A):=J\pi(A)J$. Note that the image of $\pi_\r$ is $\pi(\fM)'$.
We  will often
 write $\pi_\l$ for $\pi$ and call it the {\em left representation}.

\subsection{Tomita-Takesaki theory}

Let $\pi:\fM\to B(\cH)$ be an injective
$\sigma$-weakly continuous $*$-representation 
and
$\Omega$  a cyclic and separating vector for $\pi(\fM)$. One then 
proves that the formula
\[S\pi(A)\Omega=\pi(A)^*
\Omega\] defines a closable antilinear operator $S$ on $\cH$.
The{\em  modular
 operator} $\Delta$ and the {\em modular conjugation } $J$ are defined by the
polar decomposition of $S$:
\[S=J\Delta^{1/2}.\]
If we set
\[\cH^+=\{\pi(A)J\pi(A)\Omega\ :\ A\in\fM\}^\cl,\]
then  $(\pi,\cH,J,\cH^+)$ is a standard representation of $\fM$. Given
$\fM$, $\pi$ and $\cH$, it is the unique standard
representation with the property
$\Omega\in\cH^+$.

\subsection{KMS states}

Let $(\fM,\tau)$ be a $W^*$-dynamical system. 
 Let $\beta$ be a positive number
(having the physical interpretation of  the inverse temperature).
  A $\sigma$-weakly continuous
state $\omega$ on  $\fM$
 is called a {\em $(\tau, \beta)$-KMS state} (or a {\em $\beta$-KMS
state for $\tau$}) 
iff for all $A, B \in \fM$ there 
exists a function $F_{A, B}$,  analytic inside the strip $\{ z : 0 < \Im z <\beta\}$,
 bounded and 
continuous on its closure, and satisfying the KMS boundary conditions
\[F_{A, B}(t)= \omega(A\tau^t(B)), \qquad F_{A, B}(t + \i \beta)= 
\omega(\tau^t(B)A).
\]
A KMS-state is $\tau$-invariant. If $\fM$ is a
factor, then $(\fM, \tau)$ can have at most one 
$\beta$-KMS state.

If $\fM\subset B(\cH)$ and 
$\Phi\in\cH$, we will say that $\Phi$ is a {\em  $(\tau,\beta)$-KMS vector} iff
$(\Phi|\cdot\Phi)$ is a $(\tau,\beta)$-KMS state.

The acronym KMS stands for Kubo-Martin-Schwinger.
\subsection{Type I factors---irreducible representation}

The most elementary example of a factor is the so-called type I factor -- this
means the algebra of all bounded
operators on a given Hilbert space. In this and the next two subsections we
describe various concepts of theory of $W^*$-algebras on this example.

The space of $\sigma$-weakly continuous functionals on
 $B(\cH)$ (the predual of $B(\cH)$) can be identified with $B^1(\cH)$
(trace class operators) by the formula
\begin{equation} \psi(A)=\Tr\gamma A,\ \  \ \ \gamma\in
B^1(\cH),\ \ A\in B(\cH).\label{stat}\end{equation} 
In particular, $\sigma$-weakly continuous
states are determined by positive trace one operators,
called density matrices. A state given by a density matrix $\gamma$ is
faithful iff $\Ker\gamma=\{0\}$.

If $\tau$ is a $*$-automorphism of $B(\cH)$, then there exists  $W\in U(\cH)$
such that
\begin{equation} \tau(A)=WAW^*,\ \ A\in B(\cH).\label{dynna}\end{equation}
If $ t\mapsto \tau^t$ is a
$W^*$-dynamics, then there 
exists a self-adjoint operator $H$ on $\cH$ such that
\[\tau^t(A)=\e^{\i tH}A\e^{-\i tH},\ \  A\in B(\cH).\]
See e.g. \cite{BR1}.

A state given by (\ref{stat})
 is invariant with respect to the
$W^*$-dynamics (\ref{dynna}) iff $H$ commutes
with $\gamma$. 

There exists a $(\beta,\tau)$--KMS state iff $\Tr\,\e^{-\beta
H}<\infty$, and then it has the density matrix $\e^{-\beta
H}/\Tr\,\e^{-\beta H}$.

\subsection{Type I factor---representation in Hilbert-Schmidt
 operators}
\label{I.I}
Clearly, the representation of $B(\cH)$ in $\cH$ is not in a standard
form.
To construct a standard form of $B(\cH)$, 
consider the Hilbert space of Hilbert-Schmidt operators on $\cH$,
denoted $B^2(\cH)$, and two injective representations:
\begin{equation}\begin{array}{l}
B(\cH)\ni A\mapsto\pi_\l(A)\in B(B^2(\cH)),\ \ \ \pi_\l(A)B:=AB,
\ \ B\in B^2(\cH);\\[3mm]
\bar{B(\cH)}\ni \bar A\mapsto\pi_\r(\bar A)\in B(B^2(\cH)),
\ \ \ \pi_\r(\bar A)B 
:=B A^*,
\ \ B\in B^2(\cH).\end{array}\label{repr}\end{equation}
Set $J_\cH B:=B^*$, $B\in B^2(\cH)$. Then
$J_\cH \pi_\l(A) J_\cH =\pi_\r(\bar A)$ and
\[(\pi_l,B^2(\cH),J_\cH,B^2_+(\cH))\] is a standard representation of $B(\cH)$.

If a state on $B(\cH)$ is given by a density matrix $\gamma\in B_+^1(\cH)$,
then its standard vector representative is $\gamma^{\12}\in B^2_+(\cH)$. 
The standard implementation of the $*$-authomorphism
$\tau(A)=WAW^*$ equals $\pi_\l(W)\pi_r(\bar W)$.
If the $W^*$-dynamics $t\mapsto
\tau^t$ is given by a self-adjoint operator
$H$, then its standard Liouvillean is $\pi_\l(H)-\pi_\r(\bar H)$.

\subsection{Type I factors---representation in $\cH\otimes\bar\cH$}
\label{I.II}

An alternative formalism, which can be used to describe a
standard form of type I factors, uses the notion of a conjugate Hilbert space.

Recall  that $B^2(\cH)$ has a natural identification with
$\cH\otimes\bar\cH$. 
Under the identification 
 the representations (\ref{repr}) become
\begin{equation}\begin{array}{l}
B(\cH)\ni A\mapsto A\otimes 1_{\bar\cH}\in B(\cH\otimes\bar\cH);\\[3mm]
\bar{B(\cH)}\ni\bar A\mapsto1_{\cH}\otimes \bar A\in  B(\cH\otimes\bar\cH)
.\end{array}\label{repr1}\end{equation}
(Abusing the notation, sometimes the above representations will also be
denoted by $\pi_\l$ and $\pi_\r$).

 Note that the standard unitary implementation of the
automorphism $\tau(A)=WAW^*$ is then equal to  $W\otimes\bar W$. The
standard Liouvillean for $\tau^t(A)=\e^{\i tH}A\e^{-\i tH}$ equals
$L=H\otimes\one-\one\otimes\bar H$.
The modular conjugation  is $J_\cH$ defined by
\begin{equation} J_\cH\bigl(\Psi_1\otimes\bar{\Psi_2}\bigr):=
\Psi_2\otimes\bar{\Psi_1}.\label{jotk}\end{equation}
The positive cone is then equal to
\[(\cH\otimes\bar\cH)_+:=\Conv\{\Psi\otimes\bar\Psi\ :\ \Psi\in\cH\}^\cl
,\]
where $\Conv$ denotes the convex hull. 


\subsection{Perturbations of $W^*$-dynamics and Liouvilleans}

The material of this subsection will be needed only in the last section
devoted to Pauli-Fierz systems.

Let  $\tau$  be a $W^\ast$-dynamics on a  $W^*$-algebra $\fM$ and let
  $(\pi, \cH, J, \cH_+)$ be a standard representation of $\fM$. Let
 $L$ be  the standard Liouvillean of $\tau$.

The following theorem  is proven in  \cite{DJP}:

\bet   Let $V$ be a self-adjoint
operator on $\cH$
  affiliated to $\fM$. (That means that all spectral projections of
$V$ belong to $\pi(\fM)$). 
Let $L + V$ be essentially self-adjoint on $\Dom(L) 
\cap \Dom(V)$ and  
$L_V:= L + V - JVJ$ be essentially 
self-adjoint on $\Dom(L)\cap \Dom(V)\cap \Dom(
JVJ)$.
Set
\[\tau_V^t(A) := \pi^{-1} \left(\e^{\i t (L +V)}\pi(A)\e^{-\i t (L +
V)}\right).
\]
Then $t\mapsto\tau_V^t$ is a $W^\ast$-dynamics
 on $\fM$ and  $L_V$ is  its standard
Liouvillean.
\label{pertdyn}
\eet

\section{Quasi-free representations of the CCR}

\subsection{Bosonic quasi-free vectors}

Let $(\cY,\omega)$ be a real vector space with an antisymmetric form. Let
\beq \cY\ni y\mapsto W(y)\in U(\cH)\label{ccc5}\eeq
be  a representation of the CCR.
We say that $\Psi\in\cH$ is a {\em quasi-free vector} for (\ref{ccc5})
iff there exists a  quadratic form $\eta$ such that
\beq\begin{array}{l}
(\Psi|W(y)\Psi)=\exp(-\frac14y\eta y).\end{array}\label{bete}\eeq
Note that $\eta$ is necessarily positive, that is
$y\eta y\geq0$ for $y\in\cY$.

A representation (\ref{ccc5})
 is called {\em quasi-free} if there exists a cyclic
quasi-free vector in $\cH$.

The following fact is easy to see:

\bet A quasi-free representation is regular. \eet

Therefore, in  a quasi-free representation we can  define 
the
  corresponding field operators, denoted $\phi(y)$.

If $m$ is an integer, we say that   $\sigma$ is a {\em pairing of
$\{1,\dots,2m\} $} if it is a permutation  of
$\{1,\dots,2m\}$
satisfying
\[\sigma(1)<\sigma(3)<\cdots<\sigma(2m-1),\ \ \ \sigma(2j-1)<\sigma(2j),\
j=1,\dots,m.\]
$P(2m)$ will denote the set of pairings of $\{1,\dots,2m\}$.

\bet
Suppose we are given a regular 
representation of the CCR 
\[\cY\ni y\mapsto \e^{\i\phi(y)}\in U(\cH).\] Let $\Psi\in\cH$.
 Then the following
statements are equivalent:
\ben
\item For any $n=1,2,\dots$,  $y_1,\dots y_n\in\cY$, $\Psi\in
  \Dom\left(\phi(y_1)\cdots\phi(y_n)\right)$, and

\[\begin{array}{ll}
(\Psi|\phi(y_1)\cdots\phi(y_{2m-1})\Psi)&=0,
\\[3mm]
(\Psi|\phi(y_1)\cdots\phi(y_{2m})\Psi)&=
\sum\limits_{\sigma\in P(2m)}
\prod\limits_{j=1}^m
(\Psi|\phi(y_{\sigma(2j-1)})\phi(y_{\sigma(2j)})\Psi).
\end{array}\]
\item
$\Psi$ is a quasi-free vector.\een
\eet

\bet
Suppose that $\Psi$ is a quasi-free vector with
 $\eta$ satisfying
 (\ref{bete}). Then
\ben\item $y_1(\eta+\frac\i2\omega) y_2=(\Psi|\phi(y_1)\phi(y_2)\Psi)$;
\item $
|y_1 \omega y_2|\leq2|y_1\eta y_1|^{1/2}|y_2\eta y_2|^{1/2}
,\ \ y_1,y_2\in\cY$.\een\label{kaz}\eet

\proof Note that $(\Psi|\phi(y)^2\Psi)=y\eta y$.
This implies 
\[\frac12
\Big((\Psi|\phi(y_1)\phi(y_2)\Psi)+(\Psi|\phi(y_2)\phi(y_1)\Psi)\Big) 
=y_1\eta y_2.\]
From the canonical commutation relations we get
\[\frac12
\Big((\Psi|\phi(y_1)\phi(y_2)\Psi)-(\Psi|\phi(y_2)\phi(y_1)\Psi)\Big)
=\frac\i2 y_1\omega y_2.\]
This yields 1).

From
\[\|(\phi(y_1)\pm\i\phi(y_2))\Psi\|^2\geq0.\]
we get
\beq |y_2\omega y_1|\leq y_1\eta y_1+y_2\eta y_2.\label{pst}\eeq
This implies 2). \qed

\subsection{Classical quasi-free representations of the CCR}

Let us briefly discuss  quasi-free representations for the trivial
antisymmetric form. In this case the fields commute and can be interpreted as
classical random variables, hence we will call such representations {\em
  classical}. 

Consider a real vector space $\cY$
 equipped with a positive scalar product $\eta$.
 Consider the probabilistic Gaussian measure given by the covariance $\eta$.
That means, if $\dim\cY=n<\infty$, then it is the measure $\d\mu=
(\det\eta)^{1/2}(2\pi)^{-n/2}\e^{-y\eta y/2}\d y$, where $\d y$ denotes the
  Lebesgue measure on $\cY$. If $\dim\cY=\infty$, see e.g. \cite{Si}.

Consider the Hilbert space $L^2(\mu)$. Note that a dense subspace
of of $L^2(\mu)$ can be
treated as functions on $\cY$. For $y\in\cY$, let $\phi(y)$ denote
the function $\cY\ni v\mapsto y\eta v\in\rr$.
 $\phi(y)$ can be treated as a self-adjoint operator on $L^2(\mu)$.

We equip $\cY$ with the antisymmetric form $\omega=0$.
 Then  \beq\cY\ni y\mapsto\e^{\i\phi(y)}\in U(L^2(\mu))\label{cls}\eeq
 is a  representation of the CCR.

Let $\Psi\in L^2(\mu)$ be the constant function equal to
$1$. Then $\Psi$ is a cyclic quasi-free vector for (\ref{cls}).

In the remaining part of this section we will discuss quasi-free
representations that are fully ``quantum'' -- whose CCR
are given by a non-degenerate antisymmetric form $\omega$.

\subsection{Araki-Woods representation of the CCR}
\label{sAW}

In this subsection we describe the {\em Araki-Woods representations of the
 CCR}
 and the 
corresponding
$W^*$-algebras. These representations were introduced in \cite{AW}.
They are examples of quasi-free representations. 
In our presentation  we follow \cite{DJ2}.

Let $\cZ$ be a Hilbert space and  consider  the Fock space
$\Gamma_\s(\cZ\oplus\bar\cZ)$. 
We will identify the symplectic space
$\Re\bigl((\cZ\oplus\bar\cZ)\oplus\bar{(\cZ\oplus\bar\cZ)}\bigr)$ with
$\cZ\oplus\bar\cZ$, as in (\ref{identi1}).
Therefore,
for $(z_1,\bar
z_2)\in\cZ\oplus\bar\cZ$, 
the operator \[
\phi(z_1,\bar z_2):=\frac1{\sqrt2}\bigl(a^*(z_1,\bar z_2)+a(z_1,\bar
z_2)\bigr)\]
is the corresponding field operator and  $W(z_1,\bar z_2)=\e^{\i\phi
(z_1,\bar z_2)}$ is the corresponding Weyl operator.

We will parametrize the Araki-Woods representation by a self-adjoint
 operator $\gamma $ on
$\cZ$ satisfying   $0\leq\gamma\leq1$, $\Ker(\gamma-1)=\{0\}$. 
Another important object associated to the Araki-Woods representation is a
positive operator $\rho$ on $\cZ$ called the
``1-particle density''. It   is related to $\gamma$ by
\beq
\gamma:=\rho(1+\rho)^{-1},  \ \ \ \rho=\gamma(1-\gamma)^{-1}.\label{casz}
\eeq
(Note  that in 
\cite{DJ2} we used $\rho$ to parametrize Araki-Woods representations).

 For $z\in  \Dom(\rho^\12)$ we define two unitary operators  on
 $\Gamma_\s(\cZ\oplus\bar\cZ)$ as:
\[\begin{array}{l}
 W_{\gamma,\l}^\AW(z):=W\bigl((\rho+1)^{\12}z,\bar\rho^{\12}
\bar z\bigr),\\[3mm]
 W_{\gamma,\r}^\AW(\bar z):=W\bigl(\rho^{\12}z,(\bar\rho+1)^{\12}
\bar z\bigr).\end{array}\]
We denote by $\fM_{\gamma,\l}^\AW$ and $\fM_{\gamma,\r}^\AW$ 
the von Neumann algebras
generated by $\{W_{\gamma,\l}^\AW(z) : z\in \Dom(\rho^\12)\}$ and  
$\{W_{\gamma,\r}^\AW(\bar z) : z\in \Dom(\rho^\12)\}$ respectively.
We  will be call them respectively the {\em 
left} and the {\em right  Araki-Woods algebra}.
We  drop the superscript 
${\AW}$ until the end of the section.

The operators
$\tau$ and $\epsilon$, 
defined by 
\begin{equation}\cZ\oplus\bar\cZ\ni(z_1,\bar{z}_2)\mapsto
\tau(z_1,\bar{z}_2):=(\bar{z}_2,z_1)
\in \bar\cZ\oplus\cZ,\label{tau}\end{equation}
\begin{equation}\cZ\oplus\bar\cZ\ni(z_1,\bar{z}_2)\mapsto
\epsilon(z_1,\bar{z}_2):=(z_2,\bar{z}_1)
\in \cZ\oplus\bar\cZ,\label{epsi}\end{equation}
will be useful.
Note that $\tau$ is linear, $\epsilon$ antilinear, and
\begin{equation} \epsilon(z_1,\bar{z}_2)=\bar{\tau(z_1,\bar{z}_2)}.\end{equation}

In the following theorem  we will describe some basic 
properties of the Araki-Woods algebras.  

\bet \ben
\item  $\cZ\supset \Dom(\rho^\12)\ni z\mapsto  W_{\gamma,\l}(z)\in
U(\Gamma_\s(\cZ\oplus\bar\cZ))$ 
is a regular representation of the CCR. In particular,
\[W_{\gamma,\l}(z_1)W_{\gamma,\l}(z_2)=\e^{-\frac{\i}{2}\Im(z_1|z_2)}
W_{\gamma,\l}(z_1+z_2).\]
It will be  called 
 the left Araki-Woods representation of the CCR associated to 
the pair $(\cZ, \gamma)$. 
The corresponding field, creation
 and annihilation operators are affiliated to $\fM_{\gamma,\l}$ and are given by
\[\begin{split}
\phi_{\gamma,\l}(z)&
=\phi\Big((\rho+1)^{\12}z,\bar\rho^{\12}\bar z\Big),
\\[3mm]
 a_{\gamma, \l}^\ast(z)&
=a^*\Big((\rho+1)^{\12}
z,0\Big)+a\Big(0,\bar\rho^{\12} \bar z\Big),\\[3mm]
a_{\gamma, \l}(z)&
=a\Big((\rho+1)^{\12}\ z,0\Big)+a^*\Big(0,\bar\rho^{\12}\bar 
z\Big).\end{split}\]
\item $\bar\cZ\supset{\Dom(\bar \rho^\12)}\ni \bar z\mapsto  W_{\gamma,\r}(\bar
z)
\in U(\Gamma_\s(\cZ\oplus\bar\cZ))$
is a regular representation of the CCR. In particular
\[W_{\gamma,\r}(\bar z_1)W_{\gamma,\r}(\bar z_2)=\e^{-\frac{\i}{2}\Im(\bar z_1|\bar z_2)}
W_{\gamma,\r}(\bar z_1+\bar z_2)= \e^{\frac{\i}{2}\Im( z_1|z_2)}
W_{\gamma,\r}(\bar z_1+\bar z_2).\]
It  will be called
 the   right Araki-Woods representation of the CCR associated to 
the pair $(\cZ, \gamma)$. 
The corresponding field, creation
 and annihilation operators
 are affiliated to $\fM_{\gamma, \r}$ and are given by 
\[\begin{split}
\phi_{\gamma,\r}(\bar z)
&
=\phi\Big(\rho^{\12}z,(\bar\rho+1)^{\12}\bar z\Big),\\[3mm]
 a_{\gamma,\r}^*(\bar z)&
=a\Big(\rho^{\12} z,0\Big)+a^*\Big(0,(\bar\rho+1)^{\12}
\bar z\Big)
,\\[3mm]
a_{\gamma, \r}(\bar z)&
=a^*\Big(\rho^{\12} z,0\Big)+a\Big(0,(\bar\rho+1)^{\12}
 \bar z\Big).
\end{split}\]
\item  Set
\beq J_\s=\Gamma(\epsilon)\label{js}\eeq Then we have
\[ 
\begin{split}
 J_\s W_{\gamma,\l}(z) J_\s &=W_{\gamma,\r}(\bar z),\\[3mm]
 J_\s \phi_{\gamma,\l}(z) J_\s &=\phi_{\gamma,\r}(\bar z),\\[3mm]
 J_\s 
a_{\gamma,\l}^*(z) J_\s &=a_{\gamma,\r}^*(\bar z),\\[3mm] 
 J_\s  a_{\gamma,\l}( z) J_\s &=a_{\gamma,\r}(\bar  z).
\end{split}\]
\item The vacuum $\Omega$
 is a bosonic quasi-free vector for $W_{\gamma,\l}$, its
 expectation value for the Weyl operators (the 
``generating function'') is equal to
\[\begin{array}{rl}
\bigl(\Omega|W_{\gamma,\l}(z)\Omega\bigr)
&
=\exp\Big(-\frac{1}{4}(z|z)-\frac12(z|\rho z)\Big)
=\exp\Big(-\frac{1}{4}\Big(z|\frac{1+\gamma}{1-\gamma}z\Big)\Big)
\end{array}\]
and  the ``two-point functions'' are equal to
\[\begin{array}{rl}
\bigl(\Omega|\phi_{\gamma,\l}(z_1)\phi_{\gamma,\l}(z_2)\Omega\bigr)&
=\frac{1}{2}(z_1|z_2)+\Re(z_1|\rho z_2),\\[4mm]
\bigl(\Omega|a_{\gamma,\l}(z_1)a_{\gamma,\l}^*(z_2)\Omega\bigr)
&=(z_1|(1+\rho)z_2)=(z_1|(1-\gamma)^{-1}z_2),\\[3mm]
\bigl(\Omega|a_{\gamma,\l}^*(z_1)a_{\gamma,\l}(z_2)\Omega\bigr)
&=(z_2|\rho z_1)=(z_2|\gamma(1-\gamma)^{-1} z_1),\\[3mm]
\bigl(\Omega|a_{\gamma,\l}^*(z_1)a_{\gamma,\l}^*(z_2)\Omega\bigr)&=0,\\[3mm]
\bigl(\Omega|a_{\gamma,\l}(z_1)a_{\gamma,\l}(z_2)\Omega\bigr)&=0.
\end{array}\]
\item $\fM_{\gamma,\l}$ is a factor.
\item  $\Ker\gamma=\{0\}$ iff $\Omega$ is separating  for $\fM_{\gamma,\l}$ 
iff $\Omega$ is cyclic for $\fM_{\gamma,\l}$. If this is the case, then
 the modular conjugation for $\Omega$ is given by (\ref{js}) 
and the modular operator for $\Omega$  is given by
\beq\begin{array}{rl}
\Delta&
=\Gamma\left(\gamma\oplus\bar\gamma^{-1}\right).\end{array}\label{kmsa}\eeq 
\item We have  \beq
\fM_{\gamma,\l}'=\fM_{\gamma,\r}.\label{qoq}\eeq
\item Define
\beq\Gamma_{\s,\gamma}^+(\cZ\oplus\bar\cZ):=\{A
 J_\s A\Omega\ :\
A\in\fM_{\gamma,\l}\}^\cl.\label{cone}\eeq
Then
$(\fM_{\gamma,\l},\Gamma_\s(\cZ\oplus\bar\cZ),
 J_\s ,\Gamma_{\s,\gamma}^+(\cZ\oplus\bar\cZ))$ 
is a $W^*$-algebra in the standard form.
\item If $\gamma$ has some continuous spectrum, then
 $\fM_{\gamma, \l}$ is a factor of type ${\rm III}_1$ \cite{Ta1}.  
\item  If $\gamma=0$,
 then  $\fM_{\gamma, \l}$ is a factor of type {\rm I}.  
\item Let $h$ be a self-adjoint
operator on $\cZ$ commuting with $\gamma$  and 
\[\tau^t(W_{\gamma,\l}(z)):
=W_{\gamma,\l}(\e^{\i th}z).\]
Then $t\mapsto\tau^t$ extends to a $W^*$-dynamics on $\fM_{\gamma,\l}$ and 
 \[L=\d\Gamma(h\oplus(- \bar h))\] is its  standard
Liouvillean. \item $\Omega$
 is a $(\tau, \beta)$-KMS vector
iff $\gamma=\e^{-\beta h}$.\een \label{ArakiWoods}\eet

\proof 1)--4) follow by straightforward computations.

Let us prove 5). We have
\[W_{\gamma,\l}(z_1)W_{\gamma,\r}(\bar z_2)=W_{\gamma,\r}(\bar
z_2)W_{\gamma,\l}(z_1),\ \ \ z_1,z_2\in\Dom\rho^{\frac12}.\]
 Consequently,
 $\fM_{\gamma,\l}$ and $\fM_{\gamma,\r}$ commute with one another.

Now
\[\begin{array}{l}
\bigl(\fM_{\gamma,\l}\cup\fM_{\gamma,\l}'\bigr)'
\subset
\bigl(\fM_{\gamma,\l}\cup\fM_{\gamma,\r}\bigr)'\\[3mm]
=\bigl\{W((\rho+1)^{\frac12} z_1+\rho^{\frac12}z_2,
\bar\rho^{\frac12}\bar z_1+
(\bar\rho+1)^{\frac12} \bar z_2\bigr)\ :\ z_1,z_2\in\cZ
\bigr\}'\\[3mm]
=\{W(w_1,\bar w_2)\ :\ w_1,w_2\in\cZ\}'=\cc\one,\end{array}\]
because
\[\{\bigl((\rho+1)^{\frac12} z_1+\rho^{\frac12}z_2,
\bar\rho^{\frac12}\bar z_1
+(\bar\rho+1)^{\frac12} \bar z_2\bigr)\ :\ z_1,z_2\in\cZ
\bigr\}\]
is dense in $\cZ\oplus\bar\cZ$, and  Weyl operators depend 
strongly continuously on their parameters and act irreducibly on 
$\Gamma_\s(\cZ\oplus\bar\cZ)$. Therefore,
\[\bigl(\fM_{\gamma,\l}\cup\fM_{\gamma,\l}'\bigr)'=\cc\one,\]
which means that $\fM_{\gamma,\l}$ is a factor and proves 5).

Let us prove the $\Rightarrow$ part of 
6). Assume first that $\Ker\gamma=\{0\}$.
Set $\tau^t(A):=\Gamma(\gamma,\bar\gamma^{-1})^{\i t}A
\Gamma(\gamma,\bar\gamma^{-1})^{-\i t}$. We first
check that $\tau^t$ preseves $\fM_{\gamma,\l}$. Therefore, it is a
$W^*$-dynamics on $\fM_{\gamma,\l}$. 

Next we check that  $(\Omega|\cdot\Omega)$ satisfies
  the $(\tau, -1)$-KMS condition.
 This is
straightforward for the Weyl operators $W_{\gamma,\l}(z)$.
 Therefore, it holds for
the $*-$algebra $\fM_{\gamma,\l,0}$ of finite linear combinations of
 $W_{\gamma,\l}(z)$. By the Kaplansky Theorem,
 the unit ball of 
 $\fM_{\gamma,\l,0}$ is $\sigma$-weakly
 dense in the unit ball of
 $\fM_{\gamma,\l}$. Using this we extend the KMS condition to
 $\fM_{\gamma,\l}$.

A KMS state on a factor is always faithful. By 5), $\fM_{\gamma,\l}$ is a
factor. 
Hence
$\Omega$ is separating.

Let $\cH$ be the closure of $\fM_{\gamma,\l}\Omega$.  $\cH$ is an invariant
subspace for $\fM_{\gamma,\l}$, moreover $\Omega$ 
is cyclic and  separating for $\fM_{\gamma,\l}$ on $\cH$.
 Let us compute  the operators $S$, $\Delta$ and $J$ of the modular theory
for $\Omega$ on $\cH$.

Clearly $\cH$ is spanned by vectors of the form $\Psi_z:=
\e^{\i a^*\left((1+\rho)^\12
  z,\bar\rho^{\12}\bar z\right) }\Omega$.
Let us compute:
\[\begin{array}{rl}
\Gamma(\gamma,\bar\gamma^{-1})^{\12}\Psi_z&=
\e^{\i a^*\left(\rho^\12
  z,(1+\bar\rho)^{\12}\bar z\right) }\Omega,\\[3mm]
S\Psi_z&=
\e^{-\i a^*\left((1+\rho)^\12
  z,\bar\rho^{\12}\bar z\right) }\Omega\\[3mm]
&=J_\s \Gamma(\gamma,\bar\gamma^{-1})^{\12}
\Psi_z.
\end{array}\]

We have
\[\Gamma(\gamma,\bar\gamma^{-1})^{\i t}
\Psi_z=
\Psi_{\gamma^{\i t}z}.\]
Hence  $\Gamma(\gamma,\bar\gamma^{-1})$ preserves $\cH$ and vectors $\Psi_z$
form an essential 
domain for its restriction to $\cH$. Besides,
 $\Gamma(\gamma,\bar\gamma^{-1})\cH$
is dense in $\cH$ and
 $S$ preserves $\cH$ as well.
Therefore, $J_\s$ preserves $\cH$.
Thus 
\[S=J_\s\Big|_\cH\ 
\Gamma(\gamma,\bar\gamma^{-1})\Big|_\cH\]
is the the polar decomposition of $S$ and defines the
modular operator and the modular conjugation.
Next we see that
\[\begin{array}{rl}
W_{\gamma,\l}(z_1)J_\s W_{\gamma,\l}(z_2)\Omega&
=W((\rho+1)^{\frac12} z_1+\rho^{\frac12}z_2,
\bar\rho^{\frac12}\bar z_1+
(\bar\rho+1)^{\frac12} \bar z_2\bigr)\Omega
.\end{array}\]
Therefore,
 $\fM_{\gamma,\l}J_\s\fM_{\gamma,\l}\Omega$ is dense 
in $\Gamma_\s(\cZ\oplus\bar\cZ)$.
But $\fM_{\gamma,\l}J_\s\fM_{\gamma,\l}\Omega\subset\cH$.
 Hence 
$\cH=\Gamma_\s(\cZ\oplus\bar\cZ)$ and  $\Omega$ is cyclic.
This proves the $\Rightarrow$ part of 6).

To prove 7), 
we first assume that $\Ker\gamma=\{0\}$. By 6), we can apply
 the modular theory, which   gives
$\fM_{\gamma,\l}'=J_\s\fM_{\gamma,\l}J_\s$.
By 3) we have
 $J_\s\fM_{\gamma,\l}J_\s=\fM_{\gamma,\r}$. 

For a general $\gamma$,
 we decompose $\cZ=\cZ_0\oplus\cZ_1$, where $\cZ_0=\Ker\gamma$
and $\cZ_1$ is equipped 
with a nondegenerate $\gamma_1:=\gamma\big|_{\cZ_1}$.  We
then have $\fM_{\gamma,\l}\simeq B(\Gamma_\s(\cZ_0))\otimes\fM_{\gamma_1,\l}$
and $\fM_{\gamma,\r}\simeq B(\Gamma_\s(\bar\cZ_0))\otimes\fM_{\gamma_1,\r}$.
This implies that $\fM_{\gamma,\l}'=\fM_{\gamma,\r}$ and ends the proof of 7).

From the decomposition
 $\fM_{\gamma,\l}\simeq B(\Gamma_\s(\cZ_0))\otimes\fM_{\gamma_1,\l}$
 we see that if $\Ker\gamma=\cZ_0\neq\{0\}$, then $\Omega$ is neither cyclic
 nor separating. This completes the proof  of 6) \cite{EO}. \qed


\subsection{Quasi-free representations of the CCR as the  Araki-Woods
  representations} 

A large class of quasi-free representation is unitarily
equivalent to the Araki-Woods representation for some $\gamma$.

\bet Suppose that we are given a representation of the CCR 
\beq\cY_0\ni y\mapsto W(y)\in U(\cH),\label{subs}\eeq
with a cyclic quasi-free vector $\Psi$ satisfying
$(\Psi|W(y)\Psi)=\e^{-\frac14y\eta y}$. Suppose that
the symmetric form $\eta$ is nondegenerate. Let $\cY$ be the
 completion of
$\cY_0$ to a real Hilbert space with the scalar product given by $\eta$.
By Theorem \ref{kaz} 2), $\omega$ extends to a bounded antisymmetric form on
$\cY$, which we denote alsoby $\omega$.
Assume that $\omega$ is  nondegenerate
 on $\cY$. Then there exists a Hilbert space $\cZ$ and
a positive operator $\gamma$ on $\cZ$,
a linear injection of $\cY_0$ onto a dense subspace of
 $\cZ$ and an isometric
operator $U:\cH\to\Gamma_\s(\cZ\oplus\bar\cZ)$ such that
\[\begin{array}{rl}
U\Psi&=\Omega,\\[3mm]
U W(y)&=W_{\gamma,\l}(y)U,\ \ \ y\in\cY_0.\end{array}\]
\eet

\proof
Without loss of generality we can assume that $\cY_0=\cY$.

 Working in the real Hilbert space $\cY$ equipped with the scalar
 product $\eta$ and using
 Theorem \ref{kaz} 2), we see that 
 $\omega$ is a bilinear form bounded by $2$. Therefore
there exist a 
 bounded antisymmetric
 operator $\mu$ with a trivial kernel, $\|\mu\|\leq1$,  such that
\[y_1\omega y_2=2y_1\eta\mu y_2.\]
Consider the polar decomposition \[\mu=|\mu|j=j|\mu|.\]
Then $j$ is an orthogonal operator satisfying $j^2=-1$.

Let $\cZ$ be the completion of $\cY$ with respect to the scalar product $\eta|\mu|$.
Then $j$ maps $\cZ$ into itself and is an orthogonal operator for
the scalar product $\eta|\mu|$ satisfying $j^2=-1$.
  We can treat  $\cZ$ as a complex space, identifying $-j$ with the
imaginary unit. We equip it with the (sesquilinear) scalar product
\[(y_1|y_2):=y_1\eta|\mu|y_2+\i y_1\eta\mu y_2
=y_1\omega jy_2+\i y_1\omega y_2.\] 
$\rho:=|\mu|^{-1}-1$ defines a positive operator on $\cZ$ such that
$\cY=\Dom\rho^{\frac12}$. 
Now
\[\begin{array}{rl}
(\Psi|\phi(y_1)\phi(y_2)\Psi)&=
y_1\eta y_2+\frac\i2 y_1\omega y_2\\[3mm]
&=
y_1\eta|\mu|y_2+\i y_1\eta\mu y_2
+y_1\eta|\mu|(|\mu|^{-1}-1)y_2
\\[3mm]
&=(y_1|y_2)+\Re(y_1|\rho y_2)\\[3mm]
&=(\Omega|\phi_{\gamma,\l}(y_1)\phi_{\gamma,\l}(y_2)\Omega),
\end{array}\]
for $\gamma$ as in (\ref{casz}).
Therefore,
\[UW(y)\Psi:=W_{\gamma,\l}(y)\Omega\]
extends to an isometric map from $\cH$ to $\Gamma_\s(\cZ\oplus\bar\cZ)$ that
intertwines 
 the representation (\ref{subs}) with the Araki-Woods representation for
 $(\cZ,\gamma)$. 
\qed

\section{Quasi-free representations of the CAR}

\subsection{Fermionic quasi-free vectors}

Let $(\cY,\alpha)$ be a real Hilbert space. Let
\beq \cY\ni y\mapsto \phi(y)\in B_\h(\cH)\label{ccc5a}\eeq
be  a representation of the CAR.
We say that $\Psi\in\cH$ is a {\em quasi-free vector }for (\ref{ccc5a}) iff 
\[\begin{array}{rl}
(\Psi|\phi(y_1)\cdots\phi(y_{2m-1})\Psi)&=0,\\[3mm]
(\Psi|\phi(y_1)\cdots\phi(y_{2m})\Psi)&=
\sum\limits_{\sigma\in P(2m)}\sgn\sigma
\prod\limits_{j=1}^m (\Psi|\phi(y_{\sigma(2j-1)})\phi(y_{\sigma(2j)})\Psi).
\end{array}\]

We say that (\ref{ccc5a}) is a
{\em quasi-free representation } if there exists a cyclic quasi-free vector $\Psi$
in $\cH$.

Define the antisymmetric form $\omega$
\beq
y_1\omega y_2:=\frac{1}{\i}(\Psi|[\phi(y_1),\phi(y_2)]\Psi).\label{ome}\eeq

\bet\ben\item
$(\Psi|\phi(y_1)\phi(y_2)\Psi)=y_1\alpha y_2+\frac{\i}{2} y_1\omega y_2$;
\item
$|y_1\omega y_2|\leq 2|y_1\alpha y_1|^{\frac12}|y_2\alpha y_2|^{\frac12}$.
\een\label{muu}\eet

\subsection{Araki-Wyss representation of the CAR}
\label{sAWy}

In this subsection we describe {\em  Araki-Wyss representations of the 
CAR} \cite{AWy},
see also \cite{JP}. They are
examples of quasi-free representations of  the CAR.

Let $\cZ$ be a Hilbert space and  consider  the Fock space
$\Gamma_\a(\cZ\oplus\bar\cZ)$.
We will identify the real Hilbert space
$\Re\bigl((\cZ\oplus\bar\cZ)\oplus\bar{(\cZ\oplus\bar\cZ)}\bigr)$ with
$\cZ\oplus\bar\cZ$, as in (\ref{identi2}).
 Therefore,
for $(z_1,\bar
z_2)\in\cZ\oplus\bar\cZ$, \[
\phi(z_1,\bar z_2):=a^*(z_1,\bar z_2)+a(z_1,\bar
z_2)\]
are the corresponding field operators.

We will parametrize Araki-Wyss representation by a positive operator $\gamma$
on $\cZ$,
possibly with a non-dense domain. We will also use the operator  $\chi$, called the
``1-particle density'', which
satisfies $0\leq\chi\leq1$.
The two operators are related to one another by
\beq\gamma:=\chi(1-\chi)^{-1},\ \ \ \chi=\gamma(\gamma+1)^{-1}.
\label{rho}\eeq

 For $z\in  \cZ$ we define  the Araki-Wyss field operators  on
 $\Gamma_\a(\cZ\oplus\bar\cZ)$ as:
\[\begin{array}{rl}
 \phi_{\gamma,\l}^\AW(z)&:=\phi\bigl((1-\chi)^{\12}z,\bar\chi^{\12}\bar z\bigr)
,\\[3mm]
 \phi_{\gamma,\r}^\AW(\bar z)&:=\Lambda
\phi\bigl(\chi^{\12}z,
(1-\bar\chi)^{\12}\bar z\bigr)\Lambda=
\i I\phi\bigl(\i\chi^{\12}z,
\i(1-\bar\chi)^{\12}\bar z\bigr)
,\end{array}\]
where recall that $\Lambda=(-1)^{N(N-1)/2}$ and $I=(-1)^N=\Gamma(-1)$.
The maps $z \mapsto \phi_{\gamma, \l}(z)$,
 $\bar z \mapsto \phi_{\gamma, \r}(\bar z),$ are called 
respectively 
 the left and the right Araki-Wyss representation of the  CAR associated to 
the pair $(\cZ, \gamma)$. 
We denote by $\fM_{\gamma,\l}^\AW$ and $\fM_{\gamma,\r}^\AW$ 
the von Neumann algebras
generated by $\{\phi_{\gamma,\l}(z) : z\in \cZ\}$ and  
$\{\phi_{\gamma,\r}(\bar z) : z\in \cZ\}$.
They will be called respectively the {\em
 left} and the {\em right Araki-Wyss algebra}.

We  drop the superscript 
${\rm AW}$ until the end of the section.

In the following theorem  we will describe some basic 
properties of the Araki-Wyss algebras.  
\bet \ben
\item  $\cZ\ni z\mapsto  \phi_{\gamma,\l}(z)\in
B_\h(\Gamma_\a(\cZ\oplus\bar\cZ))$ 
is a  representation of the CAR. In particular,
\[[\phi_{\gamma,\l}(z_1),\phi_{\gamma,\l}(z_2)]_+=
2\Re(z_1|z_2).\]
The corresponding creation
 and annihilation operators belong to $\fM_{\gamma,\l}$ and are given by
\[\begin{array}{lll}
a_{\gamma,\l}^*(z)&
=a^*\Big((1-\chi)^{\12}z,0\Big)+a\Big(0,\bar\chi^{\12}\bar
z\Big), 
\\[3mm]
a_{\gamma,\l}(z)&
=a\Big((1-\chi)^{\12}z,0\Big)+a^*\Big(0,\bar\chi^{\12}\bar
z\Big)
.\end{array}\]
\item $\bar\cZ\ni \bar z\mapsto  \phi_{\gamma,\r}(\bar
z)
\in B_\h(\Gamma_\a(\cZ\oplus\bar\cZ))$
is a  representation of the CAR. In particular
\[[\phi_{\gamma,\r}(\bar z_1),\phi_{\gamma,\r}(\bar z_2)]_+=
2\Re(z_1|z_2).\]
The corresponding creation
 and annihilation operators belong to $\fM_{\gamma,\r}$ and are given by
\[\begin{array}{lll}
a_{\gamma,\r}^*(z)&
=\Lambda\Big(
a\bigl(\chi^{\12}z,0\bigr)+a^*\bigl(0,(1-\bar\chi)^{\12}\bar
z\bigr)\Big)\Lambda, 
\\[3mm]
a_{\gamma,\r}(z)&
=\Lambda\Big(a^*\bigl(\chi^{\12}z,0\bigr)
+a\bigl(0,(1-\bar\chi)^{\12}\bar z\bigr)\Big)\Lambda
.\end{array}\]
\item Set
\beq J_\a:=\Lambda\Gamma(\epsilon).\label{qa}\eeq
We have
\[
\begin{split}
J_\a \phi_{\gamma,\l}(z)J_\a &=\phi_{\gamma,\r}(\bar z),\\[3mm]
J_\a a_{\gamma,\l}^*(z)J_\a &=a_{\gamma,\r}^*(\bar z),\\[3mm] 
J_\a  a_{\gamma,\l}( z)J_\a &=a_{\gamma,\r}(\bar z).
\end{split}\]
\item
The vacuum $\Omega$
 is a fermionic quasi-free vector,
the ``two-point functions''  are equal
\[\begin{array}{rl}(\Omega|\phi_{\gamma,\l}(z_1)\phi_{\gamma,\l}(z_2)\Omega)&=
(z_1|z_2)-\i2\Im(z_1|\chi z_2),
\\[4mm]
\bigl(\Omega|a_{\gamma,\l}(z_1)a_{\gamma,\l}^*(z_2)\Omega\bigr)
&=(z_1|(1-\chi)z_2)=(z_1|(1+\gamma)^{-1}z_2),\\[3mm]
\bigl(\Omega|a_{\gamma,\l}^*(z_1)a_{\gamma,\l}(z_2)\Omega\bigr)
&=(z_2|\chi z_1)=(z_2|\gamma(\gamma+1)^{-1}z_1),\\[3mm]
\bigl(\Omega|a_{\gamma,\l}^*(z_1)a_{\gamma,\l}^*(z_2)\Omega\bigr)&=0,\\[3mm]
\bigl(\Omega|a_{\gamma,\l}(z_1)a_{\gamma,\l}(z_2)\Omega\bigr)&=0.
\end{array}\]
\item  $\fM_{\gamma,\l}$ is a factor.
\item  $\Ker\gamma=\Ker\gamma^{-1}=\{0\}$ 
(equivalently, $\Ker\chi=\Ker(1-\chi)=\{0\}$) iff 
$\Omega$ is separating for
  $\fM_{\gamma,\l}$ iff $\Omega$ is cyclic for
  $\fM_{\gamma,\l}$. If this is the case, then
the modular conjugation for $\Omega$ is given by (\ref{qa})
and the modular operator  for $\Omega$
is given by
\beq\begin{array}{rl}
\Delta&=\Gamma
\left(\gamma\oplus\bar\gamma^{-1}\right).\end{array}\label{kmsa1}\eeq 
\item We have  \beq
\fM_{\gamma,\l}'=\fM_{\gamma,\r}.\label{qoq1}\eeq
\item Set
 \beq \Gamma_{\a,\gamma}^+(\cZ\oplus\bar\cZ):=\{AJ_\a A\Omega\ :\
A\in\fM_{\gamma,\l}\}^\cl.\label{cone2}\eeq
Then
$(\fM_{\gamma,\l},\Gamma_\a(\cZ\oplus\bar\cZ),
J_\a ,\Gamma_{\a,\gamma}^+(\cZ\oplus\bar\cZ))$ 
is a $W^*$-algebra in the standard form.
\item If $\gamma$ has some continuous spectrum, then
  $\fM_{\gamma, \l}$ is a factor of type ${\rm III}_1$.
\item If $\cZ$ is an infinite dimensional Hilbert
 space and  $\gamma=\lambda$ or $\gamma=\lambda^{-1}$
with
 $\lambda\in]0,1[$,  then 
  $\fM_{\gamma, \l}$  
is a factor of type  ${\rm III}_\lambda$, \cite{Ta2}.  
\item If  $\cZ$ is an infinite dimensional Hilbert space and  $\gamma=1$
  (equivalently,  
$\chi=\frac12$),
 then  $\fM_{\gamma, \l}$ is a factor of type ${\rm II}_1$.  
\item If $\gamma=0$ or $\gamma^{-1}=0$, (equivalently, $\chi=0$   or $\chi=1$),
 then  $\fM_{\gamma, \l}$ is a factor of type {\rm I}.  
\item Let $h$ be a self-adjoint
operator on $\cZ$ commuting with $\gamma$  and 
\[\tau^t(\phi_{\gamma,\l}(z))
:=\phi_{\gamma,\l}(\e^{\i th}z).\]
Then $t\mapsto
\tau^t$ extends uniquely  to a $W^*$-dynamics on $\fM_{\gamma,\l}$ and 
 \[L=\d\Gamma(h\oplus(- \bar h))\] is its  standard
Liouvillean. 
\item $\Omega$
 is a $ (\tau,\beta)$-KMS vector 
iff 
$\gamma=\e^{-\beta h}$.
\een\label{ArakiWyss}\eet

\proof 1)-4) follow by direct computations.

The proof of 5) will be
divided into a number of steps.

\noindent{\bf Step 1.}
 We have
\[\phi_{\gamma,\l}(z_1)\phi_{\gamma,\r}(\bar z_2)=\phi_{\gamma,\r}(\bar
z_2)\phi_{\gamma,\l}(z_1).\]
 Consequently,
  $\fM_{\gamma,\l}$ and $\fM_{\gamma,\r}$ commute with one another.

\noindent{\bf Step 2.}
For simplicity, in this step 
we assume that $\cZ$ is separable and $0\leq \chi\leq\frac12$;
 the generalization of the
proof to the general case is easy. By a well known theorem (see
e.g. \cite{Ta1}, Ex. II.1.4), for any $\epsilon>0$, we can find a self-adjoint
operator $\nu$ such that
$\Tr(\chi^{\frac12}-\nu^{\frac12})^*(\chi^{\frac12}-\nu^{\frac12})
<\epsilon^2$ and there
exists an orthonormal basis $w_1,w_2,\dots$ of eigenvectors of $\nu$. Let
$\nu w_i=\nu_{i}w_i$, $\nu_{i}\in\rr$.

Introduce the operators
\[\begin{array}{rll}
A_j&:=&
\phi\left((1-\nu_j)^{\frac12} w_j, \bar\nu_j^{\frac12}\bar w_j\right)
\phi\left(\i(1-\nu_j)^{\frac12} w_j, -\i\bar\nu_j^{\frac12}\bar w_j\right)
\\[3mm]&&\times\Lambda
\phi\left(\nu_j^{\frac12} w_j, (1-\bar\nu_j)^{\frac12}\bar w_j\right)
\phi\left(\i\nu_j^{\frac12} w_j, -\i(1-\bar\nu_j)^{\frac12}\bar w_j\right)
\Lambda
\\[3mm]
&:=&
\phi\left((1-\nu_j)^{\frac12} w_j, \bar\nu_j^{\frac12}\bar w_j\right)
\phi\left(\i(1-\nu_j)^{\frac12} w_j, -\i\bar\nu_j^{\frac12}\bar w_j\right)
\\[3mm]
&&\times
\phi\left(\i\nu_j^{\frac12} w_j, \i(1-\bar\nu_j)^{\frac12}\bar w_j\right)
\phi\left(-\nu_j^{\frac12} w_j, (1-\bar\nu_j)^{\frac12}\bar w_j\right)
\\[3mm]
&=&\left(2a^*(w_j,0)a(w_j,0)-1
\right)
\bigl(2a^*(0,\bar w_j)a(0,\bar w_j)-1
\bigr).
\end{array}\]
Note that $A_j$ commute with one another and $\prod\limits_{j=1}^\infty A_j$
converges in the $\sigma$-weak topology to $I$.

Introduce also
\[\begin{array}{rll}
B_j&:=&
\phi_{\gamma,\l}(w_j)\phi_{\gamma,\l}(\i w_j)
\phi_{\gamma,\r}(\bar w_j)\phi_{\gamma,\r}(\i\bar w_j)\\[3mm]
&=&\left(
2a^*\left((1-\chi)^{\frac12}w_j,\bar\chi^{\frac12}\bar w_j\right)
a\left((1-\chi)^{\frac12}w_j,\bar\chi^{\frac12}\bar w_j\right)-1\right)
\\[3mm]
&&\times\left(
2a^*\left(-\chi^{\frac12}w_j,(1-\bar\chi)^{\frac12}\bar w_j\right)
a\left(-\chi^{\frac12}w_j,(1-\bar\chi)^{\frac12}\bar w_j\right)-1\right)
\end{array}\]
Note that $B_j$ belongs to the algebra generated by
$\fM_{\gamma,\l}$ and $\fM_{\gamma,\r}$ and
\[\left\|\prod\limits_{j=1}^n A_j-\prod\limits_{j=1}^n B_j
\right\|\leq
c\epsilon.\] This proves that 
\beq
I\in \bigl(\fM_{\gamma,\l}\cup\fM_{\gamma,\r}\bigr)''.\label{step}\eeq

\noindent{\bf Step 3.}
We have
\[\begin{array}{l}
\bigl(\fM_{\gamma,\l}\cup\fM_{\gamma,\l}'\bigr)'
\subset
\bigl(\fM_{\gamma,\l}\cup\fM_{\gamma,\r}\bigr)'
=\bigl(\fM_{\gamma,\l}\cup\fM_{\gamma,\r}\cup\{I\}\bigr)'\\[3mm]
=\{\phi((1-\chi)^{\frac12} w_1-\chi^{\frac12}w_2,
\bar\chi^{\frac12}\bar w_1+(1-\chi)^{\frac12}\bar w_2)\ 
:\ w_1,w_2\in\cZ\}'
\\[3mm]
=\{\phi(w_1,\bar w_2)\ :\ w_1,w_2\in\cZ\}'=\cc\one,\end{array}\]
where at the beginning we used Step 1,  then
 (\ref{step}), next we used
\[
\phi_{\gamma,\r}(\bar z)=\i \phi(-\chi^{\frac12} \i z,-(1-\chi)^{\frac12}
\bar{\i z})I,\]
\[
\{(1-\chi)^{\frac12} w_1-\chi^{\frac12}w_2,
\bar\chi^{\frac12}\bar w_1+(1-\chi)^{\frac12}\bar w_2)\ 
:\ w_1,w_2\in\cZ\}=\cZ\oplus\bar\cZ,\]
and finally the irreducibility of fermionic fields.
This shows that 
$\bigl(\fM_{\gamma,\l}\cup\fM_{\gamma,\l}'\bigr)'=\cc\one$, which means
 that
 $\fM_{\gamma,\l}$ is a factor and ends the proof of 5).

The proof of the $\Rightarrow$ part of
 6) is similar to its bosonic analog. Assume that $\Ker\gamma=\{0\}$. 
Set \[\tau^t(A):=\Gamma(\gamma,\bar\gamma^{-1})^{\i t}A
\Gamma(\gamma,\bar\gamma^{-1})^{-\i t}.\] We first
check that $\tau^t$ preserves $\fM_{\gamma,\l}$. Therefore, it is a
$W^*$-dynamics on $\fM_{\gamma,\l}$. 

Next we check that  $(\Omega|\cdot\Omega)$ satisfies
  the $(\tau, -1)$-KMS condition.
 This is
straightforward for the Weyl operators $\phi_{\gamma,\l}(z)$.
 Therefore, it holds for
the $*-$algebra $\fM_{\gamma,\l,0}$  of polynomials
in 
 $\phi_{\gamma,\l}(z)$. By the Kaplansky Theorem,
 the unit ball of 
 $\fM_{\gamma,\l,0}$ is $\sigma$-weakly dense in the unit ball of
 $\fM_{\gamma,\l}$. Using this we extend the KMS condition to
 $\fM_{\gamma,\l}$.

A KMS state on a factor is always faithful. By 5),  $\fM_{\gamma,\l}$ is a
factor. Hence
$\Omega$ is separating.

Let $\cH$ be the closure of $\fM_{\gamma,\l}\Omega$. $\cH$ is invariant for
$\fM_{\gamma,\l}$, moreover $\Omega$ is cyclic and separating 
 for $\fM_{\gamma,\l}$ on $\cH$. 
Computing on  polynomials in
 $\phi_{\gamma,\l}(z)$ acting on $\Omega$,  we
check that  $\Gamma(\gamma,\bar\gamma^{-1})$ and $J_\a$ preserve $\cH$,
the modular
conjugation for $\Omega$
is given by $J_\a\Big|_\cH $
 and  the modular operator equals
 $\Gamma(\gamma,\bar\gamma^{-1})\Big|_\cH$. 
Now\[\begin{array}{rl}&
\phi_{\gamma,\l}(z_1)\cdots \phi_{\gamma,\l}(z_n)J_\a
\phi_{\gamma,\l}(z_1')\cdots \phi_{\gamma,\l}(z_m')
 \Omega\\[3mm]=&
\phi_{\gamma,\l}(z_1)\cdots \phi_{\gamma,\l}( z_n)
\phi_{\gamma,\r}(\bar z_1')\cdots \phi_{\gamma,\r}(\bar z_m')
 \Omega
\end{array}\]
Thus,
 $\fM_{\gamma,\l}J_\a\fM_{\gamma,\l}\Omega$ is dense 
in $\Gamma_\a(\cZ\oplus\bar\cZ)$. But
 $\fM_{\gamma,\l}J_\a\fM_{\gamma,\l}\Omega\subset\cH$. 
Hence $\Omega$ is cyclic and
$\cH=\Gamma_\a(\cZ\oplus\bar\cZ)$.
This proves 6).

In this section, 
we  will  prove  7) only under the assumption
 $\Ker\gamma=\Ker\gamma^{-1}=\{0\}$. By 5) this implies that
 $(\Omega|\cdot\Omega)$ is faithful and we can apply the modular theory, which
 gives 
$J_\a\fM_{\gamma,\l}J_\a=\fM'_{\gamma,\l}$. By 3) we have
$J_\a\fM_{\gamma,\l}J_\a=\fM_{\gamma,\r}$.

7) for a general $\gamma$ will follow from Theorem \ref{latta}, proven later.
 \qed

\subsection{Quasi-free representations of the CAR as the Araki-Wyss
  representations}

There is a simple condition, which allows to check whether a
given  quasi-free representation of the CAR is unitarily equivalent to
an Araki-Wyss representation:

\bet
Suppose that $(\cY,\alpha)$ is a real Hilbert space,
\beq \cY\ni y\mapsto \phi(y)\in B_\h(\cH)\label{ccc5ab}\eeq
is a  representation of the CAR 
and $\Psi$ a cyclic quasi-free vector for 
(\ref{ccc5ab}). Let $\omega$ be defined by (\ref{ome}).
Suppose that $\Ker\omega$ is even or infinite dimensional.
Then 
there exists a complex Hilbert space $\cZ$,
 an operator $\gamma$ on $\cZ$ satisfying
$0\leq\gamma\leq 1$,
and an isometric operator $U:\cH\to\Gamma_\a(\cZ\oplus\bar\cZ)$ such that
\[\begin{array}{rl}
U\Psi&=\Omega,\\[3mm]
U \phi(y)&=\phi_{\gamma,\l}(y)U,\ \ \ y\in\cY.\end{array}\]
$\cZ$ equipped with the real part of its scalar product
coincides with  
 $(\cY,\alpha)$. 
\eet

\proof
By Theorem \ref{muu} 2),
there exists an antisymmetric operator $\mu$ such that $\|\mu\|\leq1$ and
\[y_1\omega y_2=2y_1\alpha\mu y_2.\]
Let $\cY_0:=\Ker\mu$ and $\cY_1$ be its orthogonal complement.
On $\cY_1$ we can make the polar decomposition 
\[\mu=|\mu|j=j|\mu|,\]
$j$ is an orthogonal operator such that $j^2=-1$. Thus $j$ is a complex
structure. If $\dim\cY_0$ is even or infinite, then we  can extend $j$ to a
complex structure on $\cY$.
Interpreting $-j$ as the complex structure, we convert $\cY$ into a complex
space, which will be denoted by $\cZ$,
 and equip it with the (sesquilinear) scalar product
\[(y_1|y_2):=y_1\alpha y_2+\i y_1\alpha jy_2.\]
Now  $\chi:=\frac12(1-|\mu|)$ 
and $\gamma:=\chi(1-\chi)^{-1}$
define  operators on $\cZ$ such that
$0\leq\chi\leq \frac12$ and $0\leq\gamma\leq1$. We have
\[\begin{array}{rl}
(\Psi|\phi(y_1)\phi(y_2)\Psi)&=
y_1\alpha y_2+\frac{\i}{2} y_1\omega y_2\\[3mm]
&=y_1\alpha y_2+\i y_1\alpha jy_2+\i y_1\alpha j(|\mu|-1) y_2\\[3mm]
&=(y_1|y_2)-2\i\Im (y_1|\chi y_2)\\[3mm]
&=
(\Omega|\phi_{\gamma,\l}(y_1)\phi_{\gamma,\l}(y_2)\Omega).
\end{array}\]
Now we see that 
\[U\phi(y)\Psi:=\phi_{\gamma,\l}(y)\Omega\]
extends to an isometric operator intertwining
 the representation (\ref{ccc5ab}) with the Araki-Wyss representation for 
 $(\cZ,\gamma)$. 
 \qed

\subsection{Tracial quasi-free representations}

{\em Tracial representations of the CAR}
 are the fermionic analogs of classical
quasifree representations of the CCR.  

Consider a real Hilbert space $\cV$. Let $\cW=\cV\oplus\i\cV$ be its
 complexification. Let $\kappa$ denote the natural conjugation on $\cW$,
which means $\kappa(v_1+\i v_2):=v_1-\i v_2$, $v_1,v_2\in\cV$.
 Consider the pair 
of representations 
of the CAR
\beq\begin{array}{l}
\cV\ni v\mapsto\phi_{\cV,\l}(v):=\phi(v)\in B_\h(\Gamma_\a(\cW)),\\[3mm]
\cV\ni v\mapsto\phi_{\cV,\r}(v):=
\Lambda\phi(v)\Lambda\in B_\h(\Gamma_\a(\cW)).\end{array}\label{cas}\eeq
Let $\fM_{\cV,\l}$ and $\fM_{\cV,\r}$ be the 
von Neumann algebras generated by 
$\{\phi_{\cV,\l}(v)\ :\ v\in\cV\}$
 and $\{\phi_{\cV,\r}(v)\ :\ v\in\cV\}$ respectively.

\bet\ben\item (\ref{cas}) are two commuting representations of the CAR:
\[\begin{array}{l}
[\phi_{\cV,\l}(v_1),\phi_{\cV,\l}(v_2)]_+=2(v_1|v_2),\\[3mm]
[\phi_{\cV,\r}(v_1),\phi_{\cV,\r}(v_2)]_+=2(v_1|v_2),\\[3mm]
[\phi_{\cV,\l}(v_1),\phi_{\cV,\r}(v_2)]=0.
\end{array}\]
\item Set 
\beq J_\a:= \Lambda\Gamma(\kappa).\label{ja1}\eeq
We have
\[J_\a\phi_{\cV,\l}(v)J_\a=\phi_{\cV,\r}(v).\]
\item $\Omega$ is a quasi-free vector for  (\ref{cas}) 
with the 2-point function
\[(\Omega|\phi_{\cV,\l}(v_1)\phi_{\cV,\l}(v_2)\Omega)=(v_1|v_2).\]
\item
 $\Omega$ is cyclic
 and separating  on $\fM_{\cV,\l}$.
$(\Omega|\cdot\Omega)$ is  tracial, which means
\[(\Omega|AB\Omega)=(\Omega|BA\Omega),\ \ \ A,B\in\fM_{\cV,\l}.\]
The 
corresponding modular conjugation is given by (\ref{ja1}) 
and the modular operator equals $\Delta=1$.
\item We have
 $\fM_{\cV,\r}=\fM_{\cV,\l}'$.
\item If $\dim\cV$ is even or infinite, then
  the  tracial representations of the CAR are unitarily equivalent to
  the Araki-Wyss representations with $\gamma=1$ (equivalently,
  $\chi=\frac12$). 
\item If $\dim\cV$ is odd, then the center of  $\fM_{\cV,\l}$
is 2-dimensional: it is spanned by $1$ and $Q$ introduced in  Theorem \ref{qu}.
\een\label{trac}\eet

\subsection{Putting together an Araki-Wyss and a tracial representation}

A general quasifree representation of the CAR can be obtained by 
putting together an Araki-Wyss representation and a tracial representation.
Actually, one can restrict oneself to a tracial representation
with just one dimensional $\cV$, but we will consider the general case.

 Let $\cZ$, $\gamma$ 
be as in the subsection on Araki-Wyss representations and $\cV,\cW$ be as
in the subsection on tracial representations.
Define the following operators on $\Gamma_\a(\cZ\oplus\bar\cZ\oplus\cW)$
\beq\begin{array}{l}
 \cZ\oplus\cV\ni (z,v)\mapsto
\phi_{\cV,\gamma,\l}^\AW
(z,v):=\phi\bigl((1-\chi)^{\12}z,\bar\chi^{\12}\bar z,v\bigr)
,\\[3mm]
\bar \cZ\oplus\cV\ni (\bar z,v)\mapsto
 \phi_{\cV,\gamma,\r}^\AW(\bar z,v):=\Lambda
\phi\bigl(\chi^{\12}z,
(1-\bar\chi)^{\12}\bar z,v\bigr)\Lambda
,\end{array}\label{cas1}\eeq
(We drop the superscript $\rm AW$ in what follows).

\bet\ben\item (\ref{cas1}) are two commuting representations of the  CAR:
\[\begin{array}{l}
[\phi_{\cV,\gamma,\l}(z_1,v_1),\phi_{\cV,\gamma,\l}(z_2,v_2)]_+
=2\Re(z_1|z_2)+2(v_1|v_2),\\[3mm]
[\phi_{\cV,\gamma,\r}(\bar z_1,v_1),
\phi_{\cV,\gamma,\r}(\bar z_2,v_2)]_+=2\Re(z_1|z_2)+
2(v_1|v_2),\\[3mm]
[\phi_{\cV,\gamma,\l}( z_1,v_1),\phi_{\cV,\gamma,\r}(\bar z_2,v_2)]=0.
\end{array}\]
\item Set
\beq J_\a:=\Lambda\Gamma(\epsilon\oplus
\kappa). \label{ja2}\eeq 
We have
\[ J_\a\phi_{\cV,\gamma,\l}(v) J_\a=\phi_{\cV,\gamma,\r}(v).\]
\item $\Omega$ is a quasi-free vector for with the 2-point function
\[(\Omega|\phi_{\cV,\gamma,\l}(z_1,v_1)\phi_{\cV,\gamma,l}(z_2,v_2)\Omega)
=(z_1|z_2)-2\i\Im(z_1|\chi z_2)+
(v_1|v_2).\]
\item
$\Ker\gamma=\Ker \gamma^{-1}=\{0\}$ (equivalently, 
$\Ker\chi=\Ker(1-\chi)=\{0\}$) iff $\Omega$ is separating  on
  $\fM_{\cV,\gamma,\l}$
iff $\Omega$ is cyclic  on $\fM_{\cV,\gamma,\l}$. If this is the case,
the 
corresponding modular conjugation is given by (\ref{ja2}) and the
 modular operator
 equals 
$\Delta=\Gamma(\gamma\oplus\bar\gamma^{-1}\oplus1)$.
\item $\fM_{\cV,\gamma,\r}=\fM_{\cV,\gamma,\l}'$.
\item If $\dim\cV$ is even or infinite, then
  the  representations (\ref{cas1}) are unitarily equivalent to
  the Araki-Wyss representations.
\item If $\dim\cZ$ is finite and
$\dim\cV$ is odd, then the center of  $\fM_{\cV,\l}$
is 2-dimensional: it is spanned by $1$ and $Q$ introduced in  Theorem \ref{qu}.
\een\label{tra}\eet

\section{Confined Bose and Fermi gas}

Sometimes the Araki-Woods representation of the CCR and the Araki-Wyss 
representations of the  CAR are equivalent to a multiple of  the Fock
representation and the corresponding $W^*$-algebra is type I. 
This happens e.g.
 in the case of a finite number of degrees of freedom.
More generally, this holds if
\beq\Tr\gamma<\infty.\label{confi}\eeq
Representations satisfying this condition will be called ``confined''.

Let us explain the name ``confined''.
Consider   free Bose or Fermi gas with the
 Hamiltonian  equal to $\d\Gamma(h)$,
where $h$ is the 1-particle Hamiltonian.
One can argue that in the physical description of this system
stationary quasi-free states are
of special importance. They are
 given by  density matrices  of the form
\beq \Gamma(\gamma)/\Tr\Gamma(\gamma),\label{densi}\eeq
 with $\gamma$ commuting with $h$.
In particular, $\gamma$ can have the form $\e^{-\beta h}$, in which case
(\ref{densi}) is the Gibbs state at inverse temperature $\beta$.

For (\ref{densi}) to make sense,
 $\Tr\Gamma(\gamma)$ has to be finite.
As we will see later on,
$\Tr\Gamma(\gamma)<\infty$ is equivalent to (\ref{confi}).

A typical 1-particle Hamiltonian $h$   of free Bose or Fermi gas is the
Laplacian with, say, Dirichlet boundary conditions at the boundary of its
domain. If the domain is unbounded, then, usually, the spectrum of $h$ is
continuous and, therefore,
there are no non-zero operators  $\gamma$ that commute with $h$ and
satisfy (\ref{confi}).
If the domain is bounded (``confined''), then  the spectrum of $h$ is
discrete, and hence many such operators $\gamma$ exist.
In particular,
 $\gamma=\e^{-\beta h}$ has this property.
This is the reason why we call ``confined''
the free Bose or Fermi gas satisfying (\ref{confi}).

In this section
 we will show how Araki-Woods and Araki-Wyss representations
arise in confined systems.
We will construct a natural intertwiner
between the
 Araki-Woods/Araki-Wyss representations 
and the Fock representation. We will treat the bosonic and fermionic case
 parallel. Whenever possible, we will use the same formula to describe both
 the bosonic and fermionic case. Some of the symbols will denote different
 things  in the bosonic/fermionic 
cases (e.g. the fields $\phi(z)$); others will have subscripts $\sa$
 indicating the two possible meanings.
 Sometimes there will be signs
 $\pm$ or $\mp$ indicating the two possible versions of the formula, the upper
 in  the bosonic case, the lower in the fermionic case.

\subsection{Irreducible representation}
\label{a2.1}

In this subsection we consider the $W^*$-algebra $B(\Gamma_\sa(\cZ))$ acting
in the obvious way on the Hilbert space
 $\Gamma_\sa(\cZ)$. Recall that  the $W^*$-algebra 
$B(\Gamma_\s(\cZ))$ is generated by the representation of the CCR
\[\cZ\ni z\mapsto W(z)=\e^{\i\phi(z)}\in U(\Gamma_\s(\cZ))\]
and 
the $W^*$-algebra 
$B(\Gamma_\a(\cZ))$ is generated by the representation of CAR
\[\cZ\ni z\mapsto \phi(z)\in B_\h(\Gamma_\a(\cZ)).\]

In both bosonic and fermionic cases we will also use
a certain operator $\gamma$ on $\cZ$. 

Recall that in the bosonic case, $\gamma$ satisfies $0\leq\gamma\leq1$,
$\Ker(1-\gamma)=\{0\}$, and we  introduce the 1-particle density operator
denoted $\rho$, as in (\ref{casz}).

 In  the fermionic case, $\gamma$ is a positive operator, possibly with a
 non-dense domain, and we introduce the
 1-particle density
denoted $\chi$, as in (\ref{rho}).

Throughout the section we assume that $\gamma$ is trace class.
In the bosonic case it is equivalent to assuming that $\rho$ is trace class.
We have
\[ \Tr\Gamma(\gamma)=\det(1-\gamma)^{-1}=
\det(1+\rho).\]

In the fermionic case, if we assume that $\Ker\gamma^{-1}=\{0\}$
 (or
$\Ker(\chi-1)=\{0\}$), $\gamma$ is trace class iff
 $\chi$ is trace class.
We have
\[ \Tr\Gamma(\gamma)=\det(1+\gamma)=
\det(1-\chi)^{-1}.\]

Define the state $\omega_\gamma$ on the $W^*$-algebra
  $B(\Gamma_\sa(\cZ))$ given by the density matrix
\[\Gamma(\gamma)/\Tr\Gamma(\gamma).\]

Let $h$ be another self-adjoint operator on $\cZ$.
Define the dynamics on $B(\Gamma_\sa(\cZ))$:
\[\tau^t(A):=\e^{\i t\d\Gamma(h)}A\e^{-\i t\d\Gamma(h)}\ \ \ \ 
A\in B(\Gamma_\s(\cZ)).\]
Clearly, $\omega_\gamma$ is $\tau$-invariant iff $h$  commutes  with $\gamma$.

The state $\omega_\gamma$ is  $(\beta,\tau)$-KMS iff 
$\gamma$ is proportional to $\e^{-\beta h}$.

\subsection{Standard representation}  

\label{a2.2}

We need to identify the  complex conjugate of the Fock space
 $\bar{\Gamma_\sa(\cZ)}$ with the Fock space over the complex conjugate
 $\Gamma_\sa(\bar\cZ)$.
 In the bosonic case this is straightforward. In the fermionic case, however,
 we will not use  the naive identification, but
 the identification that ``reverses the order of particles'', consistent with
 the convention adopted in Subsection \ref{s5.1}.
 More precisely,
 if $z_1,\dots,z_n\in\cZ$, then the identification looks as follows:
\beq\begin{array}{rl}
\bar{\Gamma_\a^n(\cZ)}\ni\bar{z_1\otimes_\a\cdots\otimes_\a z_n}&
\mapsto \ V\bar{z_1\otimes_\a\cdots\otimes_\a z_n}\\[3mm]
&:=\ \bar z_n\otimes_\a\cdots\otimes_\a\bar
 z_1\in\Gamma_\a^n(\bar\cZ).\end{array}\label{reve}\eeq
(Thus the identification  $V:\bar{\Gamma_\a(\cZ)}\to\Gamma_\a(\bar\cZ)$ 
equals $\Lambda$ times the naive,
``nonreversing'', identification).

 Using  (\ref{reve}) at the second step and the exponential law
at the 
last step, we
have the identification
\begin{equation}\begin{array}{rl}
 B^2(\Gamma_\sa(\cZ))&\simeq\Gamma_\sa(\cZ)\otimes
\bar{\Gamma_\sa(\cZ)}\\[3mm]
&\simeq
\Gamma_\sa(\cZ)\otimes\Gamma_\sa(\bar\cZ)\simeq
\Gamma_\sa(\cZ\oplus\bar\cZ).\end{array}\label{isom3}\end{equation}

 As before, define
\[J_\s:=\Gamma(\epsilon),\ \ \ J_\a:=\Lambda\Gamma(\epsilon).\]
\bet 
In the bosonic/fermionic case, under the above identification, the hermitian
conjugation $*$  becomes $J_\sa$.
\eet

\proof
 We restrict ourselves to the fermionic case.
Consider
\[B=|z_1\otimes_\a\cdots\otimes_\a z_n)(w_1\otimes_\a\cdots\otimes_\a w_m|
\in B^2(\Gamma_\a(\cZ)).\]
It corresponds to
\[\begin{array}{rl}
&\sqrt{(n+m)!}
z_1\otimes_\a\cdots\otimes_\a z_n\otimes_\a\bar{w_1\otimes_\a\cdots\otimes_\a
  w_m}\\[3mm] =&
\sqrt{(n+m)!}
z_1\otimes_\a\cdots\otimes_\a z_n\otimes_\a\bar w_m\otimes_\a\cdots\otimes_\a
  \bar w_1 \in\Gamma_\a(\cZ\oplus\bar\cZ)
.\end{array}\]

On the other hand,
\[B^*
=|w_1\otimes_\a\cdots\otimes_\a w_m)(z_1\otimes_\a\cdots\otimes_\a z_n|
\]
corresponds to
\[\begin{array}{rl}
&\sqrt{(n+m)!}
w_1\otimes_\a\cdots\otimes_\a w_m\otimes_\a\bar{z_1\otimes_\a\cdots\otimes_\a
  z_n} \\[3mm]=
&\sqrt{(n+m)!}
w_1\otimes_\a\cdots\otimes_\a w_m\otimes_\a\bar z_n\otimes_\a\cdots\otimes_\a
  \bar z_1 \\[3mm]
 =&
(-1)^{\frac{n(n-1)}2+\frac{m(m-1)}2+nm}
\sqrt{(n+m)!}
\bar z_1\otimes_\a\cdots\otimes_\a \bar z_n\otimes_\a w_m\otimes_\a\cdots\otimes_\a
  w_1
\\[3mm]
=&\Lambda\Gamma(\epsilon)
\sqrt{(n+m)!}
z_1\otimes_\a\cdots\otimes_\a z_n\otimes_\a\bar w_m\otimes_\a\cdots\otimes_\a
  \bar w_1
\end{array}\]
where at the last step we used $\Gamma(\epsilon)z_i=\bar z_i$,
$\Gamma(\epsilon)\bar w_i=w_i$ and
\[\begin{array}{l}
\frac{n(n-1)}2+\frac{m(m-1)}2+nm=\frac{(n+m)(n+m-1)}{2}.\end{array}\]
\qed

The $W^*$-algebras $B(\Gamma_\sa(\cZ))$ and 
$\bar{B(\Gamma_\sa(\cZ))}$ have a natural standard
representation 
in the Hilbert space $B^2(\Gamma_\sa(\cZ))$, as described in Subsection
\ref{I.I}. They have also a natural
  representation in
the Hilbert space $\Gamma_\sa(\cZ){\otimes}\bar{\Gamma_\sa(\cZ)}$,
 as described in  Subsection \ref{I.II}. Using the identification
(\ref{isom3}) we obtain the representation
$\theta_\l$  of $B(\Gamma_\sa(\cZ))$
and 
$\theta_\r$ of  $\bar{B(\Gamma_\sa(\cZ))}$
in the space $ \Gamma_\sa(\cZ{\oplus}\bar\cZ)$.

Let us describe the last representation in detail.
Let 
 \begin{equation} U: \Gamma_\sa(\cZ)\otimes\Gamma_\sa(\bar\cZ)\to
\Gamma_\sa(\cZ\oplus\bar\cZ)\label{vonn}\end{equation}
be the unitary map defined as in (\ref{idi}).
Let $V$ be defined 
in (\ref{reve}).
Then
\begin{equation}\begin{array}{rl}
B(\Gamma_\sa(\cZ)) \ni A&\mapsto\theta_\l(A):
=U\:A{\otimes} 
1_{\Gamma_\sa(\bar \cZ)}\:U^*\in B(\Gamma_\sa(\cZ\oplus\bar\cZ)),\\[3mm]
\bar{B(\Gamma_\s(\cZ))}\ni \bar A&\mapsto
\theta_\r(\bar A):=U\: 1_{\Gamma_\s( \cZ)}{\otimes}\bar A
\: U^*\in B(\Gamma_\s(\cZ\oplus\bar\cZ)),\\[3mm]
\bar{B(\Gamma_\a(\cZ))}\ni \bar A&\mapsto
\theta_\r(\bar A):=U\: 1_{\Gamma_\a( \cZ)}{\otimes}(V\bar AV^*)
\: U^*\in B(\Gamma_\a(\cZ\oplus\bar\cZ)).
\end{array} 
\label{repo1a}\end{equation}

We have 2 commuting representations of the CCR
\begin{equation}\cZ\ni z\mapsto W(z,0)=\theta_\l(W(z))\in
U(\Gamma_\s(\cZ\oplus\bar\cZ)), \label{spl1}\end{equation} 
\begin{equation}\bar\cZ\ni \bar z\mapsto W(0,\bar
z)=\theta_\r(\bar{W( z)})\in
U(\Gamma_\s(\cZ\oplus\bar\cZ)), \label{spl2}\end{equation} 
The algebra $\theta_\l(B(\Gamma_\s(\cZ)))$ is generated by the image of
(\ref{spl1}) and the algebra
 $\theta_\r(B(\Gamma_\s(\bar\cZ)))$ is generated by
the image of 
(\ref{spl2}).

We have also  2 commuting representations of the CAR
\begin{equation}\cZ\ni z\mapsto \phi(z,0)=\theta_\l(\phi(z))\in
B_\h(\Gamma_\a(\cZ\oplus\bar\cZ)), \label{spl1a}\end{equation} 
\begin{equation}\bar\cZ\ni \bar z\mapsto\Lambda \phi(0,\bar
z)\Lambda=\theta_\r(\bar{\phi( z)})\in
B_\h(\Gamma_\a(\cZ\oplus\bar\cZ)). \label{spl2a}\end{equation} 
The algebra $\theta_\l(B(\Gamma_\a(\cZ)))$ is generated by the image of
(\ref{spl1a}) and the algebra
 $\theta_\r(B(\Gamma_\a(\bar\cZ)))$ is generated by
the image of 
(\ref{spl2a}).


Let $\Gamma_{\sa}^+(\cZ\oplus\bar\cZ)$ 
be the image of $B_+^2(\Gamma_{\sa}(\cZ))$ under the identification
 (\ref{isom3}).

\bet \ben\item
$\Big(\theta_\l,\Gamma_\sa(\cZ\oplus\bar\cZ),J_\sa,
\Gamma_{\sa}^+(\cZ\oplus\bar\cZ)\Big)$ is a standard representation of
$B(\Gamma_\sa(\cZ))$. \item
$
J_\sa\theta_\l(A)J_\sa=\theta_\r(\bar A).$
\item $\d\Gamma(h\oplus(-\bar h))$
 is the standard Liouvillean of $t\mapsto\tau^t$ in this representation
\item The standard vector
representative of $\omega_\gamma$ in this representation is
\begin{equation}\begin{array}{l}
\Omega_\gamma:=
\det(1\mp\gamma)^{\pm\12}\exp\left(\12 a^*\left(\left[\begin{array}{cc}0&\gamma^\12\\
\pm\bar\gamma^\12&0\end{array}\right]\right)\right)
\Omega.\end{array}\label{squee}\end{equation}
\een
\eet

\proof 1), 2) and 3) are straightforward. Let us prove
4), which is a little involved, since we have to use various identifications
we have introduced.

In the representation of Subsection \ref{I.I}, the standard vector
representative of $\omega_\gamma$ equals
\[\begin{array}{rl}&
(\Tr\Gamma(\gamma))^{-\frac12}\Gamma(\gamma^{\frac12})\\[4mm]
=&(\Tr\Gamma(\gamma))^{-\frac12}\sum\limits_{n=0}^\infty
\Theta_\sa^n(\gamma^{\frac12})^{\otimes n}
\Theta_\sa^n
\in B^2(\Gamma_\sa(\cZ)).\end{array}\]
(Recall that $\Theta_\sa^n$ denotes the orthogonal projection onto
$\Gamma_\sa^n(\cZ)$).

Clearly, $\gamma^{1/2}\in B^2(\cZ)$ corresponds to a certain vector
$\Psi\in\cZ\otimes\bar\cZ$.

Let $\sigma$ be the permutation of $(1,\dots,2n)$
 given by
\[\sigma(2j-1)=j,\ \ \sigma(2j)=2n-j+1,\ \ \ j=1,\dots,n.\]
This permutation defines the unitary transformation
\[\Theta(\sigma):(\cZ\otimes\bar\cZ)^{ \otimes n}\to
\left(\otimes^n\cZ\right)\otimes\left(\otimes^n\bar\cZ\right).\]

Now $(\gamma^{1/2})^{\otimes n}$ can be interpreted in two fashions.
It can be interpreted as an element of $\otimes^n B^2(\cZ)$, and then it
corresponds to the vector $\Psi^{\otimes n}\in
(\cZ\otimes\bar\cZ)^{ \otimes n}$. It can be also interpreted as an element of
$ B^2(\cZ^{\otimes n})$
 and then it corresponds to
\[\Theta(\sigma)\Psi^{\otimes
  n}\in \left(\otimes^n\cZ\right)\otimes\left(\otimes^n\bar\cZ\right)
 \simeq \otimes^n\cZ\otimes\left(\bar{\otimes^n\cZ}\right).\]
(Note that we have taken into account the convention about the complex 
conjugate of
the tensor product adopted in Subsect. \ref{s5.1}).

Now $\Theta_\sa^n(\gamma^{1/2})^{\otimes n}\Theta_\sa^n\in
B^2(\Gamma_\sa^n(\cZ))$ corresponds to
\beq\left(\Theta_\sa^n\otimes\bar{\Theta_\sa^n}\right)\ 
\Theta(\sigma)\ \Psi^{\otimes
  n}\in\Gamma_\sa^n(\cZ)\otimes\Gamma_\sa^n(\bar\cZ).\label{lal}\eeq

The identification \[U:
 \Gamma_\sa^n(\cZ)\otimes\Gamma_\sa^n(\bar\cZ)\to
 \Gamma_\sa^{2n}(\cZ\oplus\bar\cZ)\]
 is obtained 
  by first treating $\Gamma_\sa^n(\cZ)\otimes\Gamma_\sa^n(\bar\cZ)$ 
 as a subspace of
$\otimes^{2n}(\cZ\oplus\bar\cZ)$ and then applying
$\frac{\sqrt{(2n)!}}{n!}\Theta_\sa^{2n}$. Therefore, (\ref{lal}) is identified
 with
\beq\begin{array}{rll}&\frac{\sqrt{(2n)!}}{n!}\Theta_\sa^{2n}\ 
\left(\Theta_\sa^n{\otimes}\bar{\Theta_\sa^n}\right)\ 
\Theta(\sigma)\ \Psi^{\otimes
  n}&\\[3mm]=&
\frac{\sqrt{(2n)!}}{n!}\Theta_\sa^{2n}
\left(\Theta_\sa^2\Psi\right)^{\otimes
  n}=
\frac{\sqrt{(2n)!}}{n!}\left(\Theta_\sa^2\Psi\right)^{\otimes_\sa
  n}
&
\in\Gamma_\sa^n(\cZ\oplus\bar\cZ),\end{array}\label{qfq}\eeq
where we used the fact that
\beq\begin{array}{rl}
\Theta_\sa^{2n}\ \left(\Theta_\sa^n\otimes\bar{\Theta_\sa^n}\right)\
\Theta(\sigma)&=\Theta_\sa^{2n} \\[3mm]
&=\Theta_\sa^{2n}\ \left(\Theta_\sa^2{\otimes}\cdots{\otimes}\Theta_\sa^2\right)
.\end{array}\label{qfq1}\eeq
(In the fermionic case, to see the first identity of
(\ref{qfq1}) we need to note that the permutation
$\sigma$ is even).
Now, if $\tau$ denotes the transposition, then
 $\Theta_\sa^2\Psi=\frac12(\Psi\pm\Theta(\tau)\Psi)$. 
Recall  that $\Psi\in\cZ\otimes\bar\cZ$  corresponds to
$\gamma^{1/2}\in B^2(\cZ)$. Therefore, $\Theta(\tau)\Psi$ corresponds to
$(\gamma^{1/2})^\t=\bar\gamma^{1/2}$.
Hence,
$\Theta_\sa^2\Psi\in\Gamma_\sa^2(\cZ\oplus\bar\cZ)$  is identified with
\begin{equation}
{\displaystyle\frac12}c={\displaystyle\frac12}
\left[\begin{array}{cc}0&\gamma^\12\\ 
\pm\bar\gamma^\12&0\end{array}\right]\in
B_\sa^2(\bar\cZ\oplus\cZ,\cZ\oplus\bar\cZ). 
\label{rer}\end{equation}
Thus, (\ref{qfq}) corresponds to
\[(n!)^{-1}
\left({\displaystyle\frac12}
a^*\left(c\right)\right)^n\Omega\in\Gamma_\sa^{2n}(\cZ\oplus\bar\cZ).\] 
So, finally,
 $\Gamma(\gamma^{\frac12})$ corresponds to
\[\begin{array}{l}
\exp
\left(\frac12a^*\left(c\right)\right)
\Omega.\end{array}\]

Clearly,
\[cc^*=\left[\begin{array}{cc}0&\gamma^\12\\
\pm\bar\gamma^\12&0\end{array}\right]
\left[\begin{array}{cc}0&\pm\bar\gamma^\12\\
\gamma^\12&0\end{array}\right]
=\left[\begin{array}{cc}\gamma&0\\
0&\bar\gamma\end{array}\right].\]
Therefore,
\[\det(1\mp cc^*)^{\mp\12}=\det(1\mp\gamma)^{\mp1}=\Tr\Gamma(\gamma)
.\]
\qed

Note that the vector $\Omega_\gamma$ is an example of a bosonic/fermionic
 Gaussian state 
considered in (\ref{sq1}) and (\ref{dsa}), where it was denoted $\Omega_c$:
\[\begin{array}{l}
\Omega_\gamma=
\det(1\mp cc^*)^{\pm\frac{1}{4}}\exp(\12 a^*(c)
)\Omega.
\end{array}\]

\subsection{Standard representation in the
Araki-Woods/Araki-Wyss
 form}
\label{a2.3}

Define the following transformation on $\Gamma_\sa(\cZ\oplus\bar\cZ)$:
\beq\begin{array}{rl}
R_\gamma &:=\det(1\mp\gamma)^{\pm\12}
\exp\left(\mp\12 a^*\left(\left[\begin{array}{cc}0&\gamma^\12\\
\pm\bar\gamma^\12&0\end{array}\right]\right)\right)\\[4mm]&\times
\Gamma\bigl((1\mp\gamma)\oplus(1\mp\bar\gamma)\bigr)^{\pm\12}
\exp\left(\pm\12 a\left(\left[\begin{array}{cc}0&\gamma^\12\\
\pm\bar\gamma^\12&0\end{array}\right]\right)\right)
.\end{array}\label{rrr}\eeq

\bet
 $R_\gamma $ is a unitary operator satisfying
\begin{equation}\begin{array}{l}
R_\gamma \phi(z_1,\bar{z_2})R_\gamma^*\\[3mm]
=\phi((1\mp\gamma)^{\pm\12}z_1\pm(\gamma\mp1)^{\pm\12} z_2,
(\bar\gamma\mp1)^{\pm\12} \bar{z_1}+
(1\mp\bar\gamma)^{\pm\12}\bar{z_2}).\end{array}\label{qq5}\end{equation}
\eet

\proof Let $c$ be defined as in (\ref{rer}).
Using \[\Gamma(1\mp cc^*)=
\Gamma((1\mp\gamma)\oplus(1\mp\bar\gamma)),\]
we see that \[\begin{array}{l}
R_\gamma:=\det(1\mp cc^*)^{\pm\frac{1}{4}}\exp(\mp\12  a^*(c))
\Gamma(1\mp cc^*)^{\pm\12}
\exp(\pm\12 a(c)).\end{array}
\]Thus $R_\gamma$ is in fact the transformation $R_c$ considered in 
(\ref{squ}) and (\ref{squa}).
\qed

Let $\phi_{\gamma,\l}(z)$, $\phi_{\gamma,\r}(z)$, 
 $\Gamma_{\sa,\gamma}^+(\cZ\oplus\bar\cZ)$, etc.  be defined as in Theorems
 \ref{ArakiWoods} and \ref{ArakiWyss}.

 $R_\gamma$ intertwines between the usual left/right
representations  and the left/right Araki-Woods/Araki-Wyss representations,
which is expressed by the following identities:
\[\begin{array}{rl}
R_\gamma \phi(z,0))R_\gamma^*&=\phi_{\gamma,\l}(z),\\[3mm]
R_\gamma  \phi(0,\bar z)R_\gamma^*&=\phi_{\gamma,\r}(\bar z), \hbox{ in the bosonic
  case},\\[3mm]
R_\gamma \Lambda \phi(0,\bar z)\Lambda R_\gamma^*&=\phi_{\gamma,\r}(\bar z),
 \hbox{ in the fermionic
  case},\\[3mm]
R_\gamma\theta_\l(B(\Gamma_\sa(\cZ)))R_\gamma^*&=\fM_{\gamma,\l},\\[3mm]
R_\gamma\theta_\r(\bar{B(\Gamma_\sa(\cZ))})R_\gamma^*&=\fM_{\gamma,\r},\\[3mm]
R_\gamma J_\sa R_\gamma^*&=J_\sa,\\[3mm]
 R_\gamma \Gamma_{\sa}^+(\cZ\oplus\bar\cZ)&=
\Gamma_{\sa,\gamma}^+(\cZ\oplus\bar\cZ),\\[3mm]
R_\gamma\Omega_\gamma&=\Omega,\\[3mm]
R_\gamma\d\Gamma(h,-\bar h)R_\gamma^*&=\d\Gamma(h,-\bar h).\end{array}\]

For $A\in B(\Gamma_\sa(\cZ))$, set
\begin{equation}\begin{array}{l}
\theta_{\gamma,\l}(A):=R_\gamma\theta_\l(A) R_\gamma^*
\in
B(\Gamma_\sa(\cZ\oplus\bar\cZ)) ,\\[3mm]
\theta_{\gamma,\r}(\bar A):=R_\gamma\theta_\r(\bar A) R_\gamma^*
\in
B(\Gamma_\sa(\cZ\oplus\bar\cZ)). \end{array}\label{aqa}\end{equation}

Finally, we see that in the confined case the algebra $\fM_{\gamma,\l}$ is
 isomorphic to $B(\Gamma_\sa(\cZ))$:

\bet\ben
\item $\Big(\theta_{\gamma,\l},\Gamma_\sa(\cZ\oplus\bar\cZ),J_\sa,
\Gamma_{\sa,\gamma}^+(\cZ\oplus\bar\cZ)\Big)$ is a standard representation of
$B(\Gamma_\sa(\cZ))$. 
\item
  $J_\sa\theta_{\gamma,\l}(A)J_\sa=\theta_{\gamma,\r}(
\bar
  A)$. 
\item $\d\Gamma(h\oplus(-\bar h))$
is the standard Liouvillean of $t\mapsto\tau^t$ in this representation.
\item $\Omega$ is the standard vector representative of $\omega_\gamma$
in this representation.
\een\eet

\section{Lattice of von Neumann algebras in a Fock space}

Let $\cZ$ be a Hilbert space. With every real closed subspace of $\cZ$ we can
naturally associate a certain von Neumann 
subalgebra of $B(\Gamma_\sa(\cZ))$,
both in the bosonic and fermionic case. These von Neumann subalgebras form a
 complete lattice. Properties of this lattice are studied in
this section. They have important applications in quantum field theory.

\subsection{Real subspaces in a complex Hilbert space}

In this subsection
we analyze  real subspaces in a complex Hilbert space.
(For a similar analysis of two complex subspaces in a complex Hilbert space
see \cite{Di,Hal}.)
This analysis
will be then used both in the bosonic and fermionic case.
We start with a simple fact which is true both in the complex and real case.
 
\bel Let $\cV_1,\cV_2$ be  closed subspaces
 of a (real or complex) Hilbert space.
 Let $p_1,p_2$ be the corresponding orthogonal projections.
Then \[
\cV_1\cap\cV_2+\cV_1^\perp\cap\cV_2^\perp=\Ker(p_1-p_2).\]
\eel

\proof
 Let $(p_1-p_2)z=0$.
Then $z=p_1z+(1-p_1)z$, where
$p_1z=p_2z\in\cV_1\cap\cV_2$ and
$(1-p_1)z=(1-p_2)z\in \cV_1^\perp\cap\cV_2^\perp$. \qed

Next suppose that $\cW$ is a complex Hilbert space. Then it is at the same
time a real Hilbert space with the scalar product given by the real part of
the original scalar product. If $K\subset\cW$, then $K^\perp$ will denote the
orthogonal complement of $K$ in the sense of the complex scalar product and
$K^\perpr$ will denote the orthogonal complement with respect to the real
scalar product. That means
\[ K^{\perpr}:=\{z\in\cZ\ :\ \Re(v| z)=0,\ v\in K\}.\]
Moreover, 
\[\i K^{\perpr}=\{z\in\cZ\ :\ \Im(v| z)=0,\ v\in K\},\]
so $\i K^{\perpr}$ can be called the symplectic  complement of
$K$.
Note that if $\cV$ is a closed real subspace of $\cW$, then
 $(\cV^{\perpr})^{\perpr}=\cV$ and
 $\i(\i\cV^{\perpr})^{\perpr}=\cV$.

\bet Let $\cV$ be a closed
real subspace of a complex Hilbert space $
\cW$.
Let $p, q$ be the orthogonal projections onto $\cV$ and $\i\cV$ respectively. Then
the following conditions holds:
\ben\item
 $\cV\cap\i\cV=\cV^\perpr\cap\i\cV^\perpr=\{0\}\ \Leftrightarrow
\ \Ker (p-q)=\{0\}$;
\item $\cV^\perpr\cap\i\cV=\{0\}\Leftrightarrow\ 
\Ker(p+q-1)=\{0\}$.\een
\eet

\proof By the previous lemma applied to the real Hilbert space $\cW$ and its
subspaces $\cV$, $\i\cV$ we get
\[\cV\cap\i\cV+\cV^\perpr\cap\i\cV^\perpr=\Ker(p-q).\]
This gives 1). Applying this lemma to $\cV$, $\i\cV^\perpr$ yields
\[
\cV\cap\i\cV^\perpr+\cV^\perpr\cap\i\cV=\Ker(p+q-1).
\]
Using $\cV\cap\i\cV^\perpr=\i(\cV^\perpr\cap\i\cV)$ we obtain 2). \qed

We will say that a real subspace $\cV$
of a complex Hilbert space is {\em in a general
position }if it satisfies both conditions of the previous theorem.
The following fact is immediate:

\bet Let $\cV$ be a closed real subspace of a complex Hilbert space $\cW$.
Set
\[\begin{array}{ll}
\cW_+:=\cV\cap\i\cV,&\cW_-:=\cV^\perpr\cap\i\cV^\perpr,\\[3mm]
\cW_1:=\cV\cap\i\cV^\perpr+\i\cV\cap\cV^\perpr,\ \ &
\cW_0:=(\cW_++\cW_-+\cW_0)^\perp,
\end{array}\]
Then $\cW_-,\cW_+,\cW_0,\cW_1$ are complex subspaces of $\cW$ and
\[\cW=\cW_+\oplus\cW_0\oplus\cW_1\oplus\cW_-.\]
We have 
\[\cW_+=\cV\cap\cW_+,\ \ \ \{0\}=\cV\cap\cW_-.\]
Set
\[\cV_0:=\cV\cap\cW_0,\ \ \ \ \cV_1:=\cV\cap\cW_1=\cV\cap\i\cV^\perpr.\]
Then $\cV_0$ is a subspace  of $\cW_0$ in a general position and
\[\begin{array}{rl}
\cV&=\cW_+\oplus\cV_0\oplus\cV_1\oplus\{0\},\\[3mm]
\i\cV^\perpr&=\{0\}\oplus\i\cV_0^\perpr\oplus\cV_1\oplus\cW_-,\end{array}\] 
(where $\cV_0^\perpr$ is the real orthogonal complement of $\cV_0$ taken
inside $\cW_0$).
\label{popo}\eet

\bet Let $\cV$ be a closed real subspace of a complex Hilbert space $\cW$
in a general position.
Then the following is true:
\ben\item There exists a closed complex subspace $\cZ$ of $\cW$, a unitary
 antilinear
operator $\epsilon$ on  $\cW$ and a self-adjoint operator $\chi$ such that
$\epsilon^2=1$, $\epsilon\cZ=\cZ^\perp$,
 $0\leq \chi\leq\frac12$, $\Ker\chi=\Ker(\chi-\frac12)=\{0\}$ and
\[\begin{array}{rl}
 \{(1-\chi)^\frac12z+\epsilon\chi^\frac12 z\ : \
z\in\cZ\}&=\cV,\\[3mm]
 \{\chi^\frac12z+\epsilon(1-\chi)^\frac12 z\ : \
z\in\cZ\}&=\i\cV^\perpr.
\end{array}\]
\item
Set $\rho:=\chi(1-2\chi)^{-1}$.  Then $\rho$ is
 a positive operator on $\cZ$ with $\Ker\rho=\{0\}$ and
\[\begin{array}{rl}\{(1+\rho)^\frac12z+\epsilon\rho^\frac12
 z\ : \ z\in\Dom \rho^{1/2}\}&\ \hbox{ is dense in
}\ \cV,\\[3mm]
\{\rho^\frac12z+\epsilon(1+\rho)^\frac12
 z\ : \ z\in\Dom\rho^{1/2}\}&\ \hbox{ is dense in
}\ \i\cV^\perpr.\end{array}\]
\een
\label{deco}\eet

\proof
 Let $p$, $q$ be defined as above. Clearly, $q=\i
p\i^{-1}$.

Define the self-adjoint real-linear operators
 $m:=p+q-1$, $n:=p-q$. Note that
\[\begin{array}{rll}
  n^2&=1-m^2&=p+q-pq-qp,\\[3mm] mn&=-nm&=qp-pq,
\end{array}\]
\[\begin{array}{ll}
\i m=m\i,\ \ &\i n=-n\i,\\[3mm]
\Ker\: m=\{0\},\ \ \ &\Ker\: n=\{0\},\\[3mm]
\Ker(m\pm1)=\{0\},\ \ \ &\Ker(n\pm1)=\{0\},\\[3mm]
-1\leq m\leq1,\ \ \ &-1\leq n\leq1.
\end{array}
\]
We can introduce their polar decompositions
\[n=|n|\epsilon=\epsilon|n|,\ \ \ m=w|m|=|m|w.\]Clearly,
 $\epsilon$ and $w$ are orthogonal operators satisfying
\[\begin{array}{l}
w^2=\epsilon^2=1,\ \ \ \ 
w\epsilon=-\epsilon w,\\[3mm]
 w\i=\i w,\ \ \ \ \ \i\epsilon=-\i\epsilon.
\end{array}
\]
Set \[\cZ:=\Ker(w-1)=\Ran 1_{]0,1[}(m).\]
Let $1_\cZ$ denote the orthogonal projection from $\cW$ onto
$\cZ$.

We have \[\epsilon\cZ=\Ker(w+1)=\Ran 1_{]-\infty,0[}(m),\]
Clearly, we have the orthogonal direct sum $\cW=\cZ\oplus\epsilon\cZ$.

Using $p=\frac{m+n+1}{2}$ 
we get
\[\begin{array}{rl}
1_\cZ p1_\cZ&=\frac{m+1}{2}1_\cZ,\\[3mm]
\epsilon1_{\cZ }\epsilon p1_{\cZ}&=\epsilon 1_{\cZ}\epsilon
\frac{n}{2}1_{\cZ}
=\epsilon 
\frac{\sqrt{1-m^2}}{2}1_{\cZ}.\end{array}\]
Therefore,\[\begin{array}{l}
p1_\cZ=\frac{m+1}{2}1_\cZ+\epsilon 
\frac{\sqrt{1-m^2}}{2}1_{\cZ}.\end{array}\]

Set 
$\chi:=\frac121_\cZ(1- m)$.
Then
\beq \{(1-\chi)^\frac12z+\epsilon\chi^\frac12 z\ : \
z\in\cZ\}=\{pz\ :\ z\in\cZ\}\subset \cV.\label{spa1}\eeq

Suppose now that $v\in\cV\cap\{pz\ :\ z\in\cZ\}^\perpr$.
Then
\[0=\Re(v|p1_\cZ v)=\Re(v|1_\cZ v)=\|1_\cZ v\|^2.\]
Hence, $v\in\cZ^\perp=\epsilon\cZ$. 
Therefore, using $q=m+1-p$ we obtain
\[\Re(v|qv)=\Re(v|m v)\leq0.\]
Hence $qv=0$. Thus $v\in\i\cV^\perpr\cap\cV$, which means that
$v=0$. Therefore, the left hand side of
(\ref{spa1}) is dense in $\cV$.

To see that we have an equality in (\ref{spa1}), we note that the operator
\beq(1-\chi)^{1/2}1_\cZ+\epsilon\chi^{1/2}1_\cZ\label{daf}\eeq
 is an isometry from $\cZ$ to
$\cW$. Hence the  range of (\ref{daf}) is closed.
 This ends the proof of 1).

2)  follows easily from 1).
\qed

Note that $\epsilon\cZ$ can be identified with $\bar\cZ$. Thus $\cW$ can be
identified with $\cZ\oplus\bar\cZ$. Under this identification, the operator
$\epsilon$ coincides with $\epsilon$ defined in (\ref{epsi}). 

Theorem \ref{deco} gives
 2 descriptions of a real subspace $\cV$. The description 2) is
 used in the Araki-Woods representations of the CCR
 and 1) in the Araki-Wyss representations of the CAR.

\subsection{Complete lattices}

In this subsection we recall some definitions concerning abstract lattices
(see e.g. \cite{Pe}). They provide a convenient language that can be used to
express some  properties of a class of von Neumann algebras acting on a
Fock space.

Suppose that $(X,\leq)$ is an ordered set.
Let
$\{x_i\ :\ i\in I\}$ be a nonempty subset of $X$. 

We say that $u$ is a largest minorant of $\{x_i\ :\ i\in I\}$
if
\ben \item $i\in I$ implies $u\leq x_i$;
\item $u_1\leq x_i$ for all $i\in I$ implies $u_1\leq u$.
\een

If $\{x_i\ :\ i\in I\}$
 possesses a largest minorant, then it is uniquely defined. The largest
minorant of a set $\{x_i\ :\ i\in I\}$ is usually denoted
\[\mathop{\wedge}\limits_{i\in I}x_i.\]

Analogously we define the smallest majorant of
$\{x_i\ :\ i\in I\}$, which is usually denoted by
 \[\mathop{\vee}\limits_{i\in I}x_i.\]

We say that $(X,\leq)$
 is a complete lattice, if every nonempty subset of $X$ possesses
the largest minorant and the smallest majorant. It is then equipped with the
 operations $\wedge$  and $\vee$.

We will say that the complete lattice is complemented if it is equipped with
the operation $X\ni x\mapsto \sim x\in X$ such that
\ben\item $\sim(\sim x)=x$;
\item $x_1\leq x_2$ implies $\sim x_2\leq \sim x_1$;
\item $\sim
\mathop{\wedge}\limits_{i\in I}x_i=
\mathop{\vee}\limits_{i\in I}(\sim x_i)$.
\een
The operation $\sim$ will be
 called the complementation.

Let $\cW$ be a topological vector space.
For a family $\{\cV_i\}_{i\in I}$ of closed subspaces of $\cW$  we define
\[\begin{array}{l}
\mathop{\vee}\limits_{i\in I}\cV_i:=\left(\sum\limits_{i\in
  I}\cV_i\right)^\cl.\end{array}
\]
Closed subspaces of 
$\cW$ form a complete lattice with the order relation
$\subset$ and the operations $\cap$
and $\vee$.
If in addition $\cW$ is a (real or complex) Hilbert space, then taking the
orthogonal complement is an example of  a complementation. 
In the case of a complex
Hilbert space (or a finite dimensional symplectic space),
 taking the symplectic complement is also
an  example of a complementation.

Let $\cH$ be a Hilbert space.
For a family of von Neumann algebras $\fM_i\subset B(\cH)$, $i\in I$,
 we set
\[\begin{array}{l}
\mathop{\vee}\limits_{i\in I}\fM_i:=\left(\bigcup\limits_{i\in
  I}\fM_i\right)''.\end{array}
\](Recall that the prime denotes the commutant).
Von Neumann algebras in $B(\cH)$ form a complete lattice with the order
relation $\subset$
and the operations
$\cap$ and $\vee$.
Taking the commutant is an example of 
 a complementation.

\subsection{Lattice of von Neumann algebras in a bosonic Fock space}

In this subsection we describe the result of Araki describing the lattice of von
Neumann algebras naturally associated to a bosonic Fock space \cite{Ar4,EO}.
 In the proof of
this result it is convenient to use the facts about the Araki-Woods
representation derived earlier.

Let $\cW$ be a complex Hilbert space.
 We will identify $\Re(\cW\oplus\bar\cW)$ with $\cW$.
Consider the Hilbert space $\Gamma_\s(\cW)$ and the corresponding Fock
representation $\cW\ni w\mapsto W(w)\in
B(\Gamma_\s(\cW))$.

For a  real
 subspace $\cV\subset\cW$ we define the von Neumann algebra
\[\fM(\cV):=\{W(w)\ :\ w\in\cV\}''\subset B(\Gamma_\s(\cW)).\]
First note that it follows from the strong continuity of $\cW
\ni w\mapsto W(w)$ that $\fM(\cV)=\fM(\cV^\cl)$. Therefore, in what follows it
is enough to restrict ourselves to closed subspaces of $\cW$.

The following theorem was proven by Araki \cite{Ar4},
 and then a simpler proof of the
most difficult statement, the  duality (6),
 was given by Eckmann and Osterwalder
 \cite{EO}: 
\bet
\ben
\item $\fM(\cV_1)=\fM(\cV_2)$ iff $\cV_1=\cV_2$.
\item $\cV_1\subset\cV_2$ implies $\fM(\cV_1)\subset\fM(\cV_2)$.
\item
$\fM(\cW)=B(\Gamma_\s(\cW))$ and $\fM(\{0\})=\cc\one.$
\item
$\fM\Big(\mathop{\vee}\limits_{i\in I}\cV_i\Big)
=\mathop{\vee}\limits_{i\in I}\fM(\cV_i)$.
\item
$\fM\Big(\bigcap\limits_{i\in I}\cV_i\Big)=\bigcap\limits_{i\in I}\fM(\cV_i)$.
\item
$\fM(\cV)'=\fM(\i\cV^{\perpr})$.
\item
$\fM(\cV)$ is a factor iff $\cV\cap\i\cV^{\perpr}=\{0\}$.
\een\item
\label{latt}\eet

\proof 
To prove 1), 
assume that $\cV_1$ and $\cV_2$ are distinct closed subspaces. It
is enough to assume that $\cV_2\not\subset\cV_1$. Then we can find
$w\in\i\cV_1^\perpr\backslash\i\cV_2^\perpr$. Now
$W(w)\in\fM(\cV_1)'\backslash\fM(\cV_2)'$. This implies
$\fM(\cV_1)'\neq\fM(\cV_2)'$, which yields (1).

 2) and 3) are immediate. The inclusion  $\subset$ in 4)
and the inclusion $\supset$ in 5)
are immediate.  The inclusion $\supset$ in 4)
follows easily if we invoke the strong continuity of
 $\cW
\ni w\mapsto W(w)$.

If we know 6), then the remaining inclusion $\subset$ in 5) follows from
$\supset$ in 4).
 7) follows from 1), 5) and 6).

Thus what remains to be shown is (6).
Its original proof 
was surprisingly involved, see 
 \cite{Ar4}. We will give a somewhat
simpler proof  \cite{EO}, which uses properties of the Araki-Woods
 representations, which in turn are based on the Tomita-Takesaki theory.

First, assume that $\cV$ is in general position in $\cW$.
 Then, according to Theorem \ref{deco} 2), and identifying $\epsilon\cZ$ with
 $\bar\cZ$,
we obtain a
decomposition $\cV=\cZ\oplus\bar\cZ$ and a positive operator $\rho$ such
that 
\[\{(1+\rho)^\frac12z+\bar\rho^\frac12 \bar z
\ : \ z\in\cZ\}\ \hbox{ is dense in
}\ \cV.\] 
Then we see that $\fM(\cV)$ is the left Araki-Woods algebra 
$\fM_{\rho,\l}^\AW$. By Theorem \ref{ArakiWoods}, the commutant of 
$\fM_{\rho,\l}^\AW$ is
$\fM_{\rho,\r}^\AW$.
But
\[\{
\rho^\frac12 z+(1+\bar\rho)^\frac12\bar z\ :
 \ z\in\cZ\}\ \hbox{ is dense in
}\ \i\cV^\perpr.\] 
Therefore,
$\fM_{\rho,\r}^\AW$ coincides with
$\fM(\i\cV^\perpr)$.
This ends the proof of 6) in the case of $\cV$ in a general position.

For an arbitrary $\cV$,
we decompose $\cW=\W_+\oplus\cW_0\oplus\cW_1\oplus
\cW_-$ and
$\cV=\cW_+\oplus\cV_0\oplus\cV_1\oplus\{0\}$,
as in  Theorem \ref{popo}.
Then we can write
\[\begin{array}{rl}
B(\Gamma_\s(\cW))&\simeq B(\Gamma_\s(\cW_+))\otimes B(\Gamma_\s(\cW_0))
\otimes B(\Gamma_\s(\cW_1))\otimes B(\Gamma_\s(\cW_-)),\\[3mm]
\fM(\cV)&\simeq B(\Gamma_\s(\cW_+))\otimes \fM(\cV_0)\otimes\fM(\cV_1)\otimes
 \one.
\end{array}\]
Clearly, $\i\cV^\perpr=\{0\}\oplus\i\cV_0^\perpr\oplus\cV_1
\oplus\cW_-$ and the commutant of 
$\fM(\cV)$ equals 
\[\begin{array}{rl}
\fM(\cV)'&\simeq \one\otimes\fM(\cV_0)'\otimes\fM(\cV_1)
\otimes B(\Gamma_\s(\cW_-))\\[3mm]
&= \one\otimes\fM(\i\cV_0^\perpr)\otimes\fM(\cV_1)
\otimes B(\Gamma_\s(\cW_-))\\[3mm]
&\simeq\fM(\i\cV^\perpr).\end{array}\]
\qed

Note that the above theorem can be interpreted as an isomorphism of the
 complete lattices of closed real subspaces of the complex Hilbert space $\cW$
 with the symplectic complement as the complementation, and the lattice of von
 Neumann algebras $\fM(\cV)\subset B(\Gamma_\s(\cW))$, with the
 complementation  given by the commutant.

\subsection{Lattice of von Neumann algebras in a fermionic Fock space}

In this subsection we describe the fermionic analog of Araki's result about
the lattice of von Neumann algebras in a bosonic Fock space.

Again, consider a complex Hilbert space $\cW$.
 We will identify $\Re(\cW\oplus\bar\cW)$ with $\cW$.
Consider the Hilbert space $\Gamma_\a(\cW)$ and the corresponding Fock
representation $\cW\ni w\mapsto \phi(w)\in
B(\Gamma_\a(\cW))$.

Consider the Hilbert space $\Gamma_\a(\cW)$ and the corresponding Fock
representation. We will identify $\Re(\cW\oplus\bar\cW)$ with $\cW$.
For a real subspace $\cV\subset\cW$ we define the von Neumann algebra
\[\fM(\cV):=\{\phi(z)\ :\ z\in\cV\}''\subset B(\Gamma_\a(\cW)).\]
Let  the operator $\Lambda$ defined in (\ref{ksl}).

First note that it follows from the norm continuity of $\cW
\ni w\mapsto \phi(w)$ that 
$\fM(\cV)=\fM(\cV^\cl)$. Therefore, in what follows it
is enough to restrict ourselves to closed real subspaces of $\cW$.

\bet
\ben
\item $\fM(\cV_1)=\fM(\cV_2)$ iff $\cV_1=\cV_2$.
\item $\cV_1\subset\cV_2$ implies $\fM(\cV_1)\subset\fM(\cV_2)$.
\item
$\fM(\cW)=B(\Gamma_\a(\cW))$ and $\fM(\{0\})=\cc\one.$
\item
$\fM\Big(\bigcap\limits_{i\in I}\cV_i\Big)=\bigcap\limits_{i\in I}\fM(\cV_i)$.
\item
$\fM\Big(\mathop{\vee}\limits_{i\in I}\cV_i\Big)
=\mathop{\vee}\limits_{i\in I}\fM(\cV_i)$.
\item
$\fM(\cV)'=\Lambda\fM(\i\cV^{\perpr})\Lambda$.
\een
\label{latta}\eet

The proof of the above theorem is very similar to the proof of Theorem
\ref{latt} from the bosonic case. The main additional difficulty is the
behavior of fermionic fields under the tensor product. They are studied in the
following theorem.

\bet
Let $\cW_i$, $i=1,2$ be two Hilbert spaces and $\cW=\cW_1\oplus\cW_2$
Let $N_i$ be the number
 operators in $\Gamma_\a(\cZ_i)$, $i=1,2$, $I_i:=(-1)^{N_i}$ and
 $\Lambda_i:=(-1)^{N_i(N_i-1)/2}$. We identify the
 operators on $\Gamma_\a(\W)$ with those
 on 
 $\Gamma_\a(\cW_1)\otimes\Gamma_\a(\cW_2)$ using
 $U$ defined in (\ref{idi}). This identification is denoted by $\simeq$.
Let $\cV_i$, $i=1,2$ be real closed subspaces of $\cW_i$, $i=1,2$ resp.
Then
\beq\fM(\cV_1\oplus\cV_2)\simeq\bigl(
\fM(\cV_1)\otimes 1+(-1)^{N_1\otimes N_2}1\otimes\fM(\cV_2)(-1)^{N_1\otimes
  N_2} \bigr)'',\label{fgf}\eeq
\beq\begin{array}{rl}
\fM(\cV_1\oplus\{0\})&\simeq\fM(\cV_1)\otimes\one,\\[3mm]
\fM(\cW_1\oplus\cV_2)&\simeq B(\Gamma_\a(\cW_1))\otimes \fM(\cV_2).
\end{array}\label{fgf1}\eeq
\beq\begin{array}{rl}
\Lambda\fM(\cV_1\oplus\cW_2)\Lambda&\simeq
\Lambda_1\fM(\cV_1)\Lambda_1\otimes B(\Gamma_\a(\cW_2)),\\[3mm]
\Lambda\fM(\{0\}\oplus \cV_2)\Lambda&\simeq\ \one\otimes
\Lambda_2\fM(\cV_2)\Lambda_2\ .
\label{fgf2}\end{array}\eeq
\label{iii}\eet

\proof Let
$v\in\cV_2$. Then, by Theorem  \ref{tensor.2} 2),
 we have the identification
\[\begin{array}{rl}
\phi(0,v)&\simeq (-1)^{N_1}\otimes\phi(v)=
(-1)^{N_1\otimes N_2}\ \one\otimes\phi(v)\ (-1)^{N_1\otimes N_2}.
\end{array}\]
Therefore, the von Neumann
algebra generated by $\phi(0,v)$, $v\in\cV_2$ equals
$(-1)^{N_1\otimes N_2}1\otimes\fM(\cV_2)(-1)^{N_1\otimes
  N_2}$. 

Clearly, the von Neumann algebra generated by $\phi(v,0)$, $v\in\cV_1$ equals
$\fM(\cV_1)\otimes 1$. This implies (\ref{fgf}).

(\ref{fgf}) implies immediately (\ref{fgf1}).
It also implies
\beq\begin{array}{rl}
\fM(\cV_1\oplus\cW_2)&\simeq(-1)^{N_1\otimes N_2}
\fM(\cV_1)\otimes B(\Gamma_\a(\cW_2))(-1)^{N_1\otimes N_2},\\[3mm]
\fM(\{0\}\oplus \cV_2)&\simeq(-1)^{N_1\otimes N_2}\ \one\otimes
\fM(\cV_2)\ (-1)^{N_1\otimes N_2},\end{array}
\label{fgf3}\eeq
from which (\ref{fgf2}) follows by Theorem \ref{qrq} 1).
\qed

\noindent{\bf Proof of Theorem \ref{latta}.}
Let us first prove 1). 
Assume that $\cV_1$ and $\cV_2$ are distinct closed subspaces. It
is enough to assume that $\cV_2\not\subset\cV_1$. Then we can find
$w\in\i\cV_1^\perpr\backslash\i\cV_2^\perpr$. Now
$\Lambda\phi(w)\Lambda\in\fM(\cV_1)'\backslash\fM(\cV_2)'$. 
Thus $\fM(\cV_1)'\neq\fM(\cV_2)'$, which yields 1).

Similarly as in the proof of Theorem \ref{latt}
the only difficult part is a proof of  6).

Assume first that $\cV$ satisffies
 $\cV\cap\i\cV=\cV^\perpr\cap\i\cV^\perpr=\{0\}$. By Theorem \ref{popo}, we can
 write $\cW=\cW_0\oplus\cW_1$ and $\cV=\cV_0\oplus\cV_1$ where $\cV_0$  
is in general position in $\cW_0$ and $\cV_1^\perpr=\i\cV_1$ inside $\cW_1$.
By Theorem \ref{deco}, we can find a complex subspace $\cZ$ of $\cW_0$, an
 antilinear involution $\epsilon$ on $\cW_0$ and a self-adjoint operator $0\leq 
\chi\leq \frac12$ such that $\epsilon\cZ=\cZ^\perp$,
 $\Ker\chi=\Ker(\chi-\frac12)=\{0\}$ and 
\[\begin{array}{rl}
 \{(1-\chi)^\frac12z+\epsilon\chi^\frac12 z\ : \
z\in\cZ\}\oplus\cV_1&=\cV,\\[3mm]
 \{\chi^\frac12z+\epsilon(1-\chi)^\frac12 z\ : \
z\in\cZ\}\oplus\cV_1&=\i\cV^\perpr.
\end{array}\]
We can identify  $\epsilon\cZ$ with $\bar\cZ$, using $\epsilon$ as the
conjugation. Then we are precisely in the framework of Theorem \ref{tra}, which
implies that $\fM(\cV)'=\Lambda  \fM(\i\cV^\perpr)\Lambda$.

For an arbitrary $\cV$, to decompose $\cW=\W_+\oplus\cW_0\oplus\cW_1
\oplus\cW_-$ and
$\cV=\cW_+\oplus\cV_0\oplus\cV_1\oplus\{0\}$ as in
 Theorem \ref{popo}.
Then we can write
\[\begin{array}{rl}
B(\Gamma_\a(\cW))&\simeq B(\Gamma_\a(\cW_+))\otimes
B(\Gamma_\a(\cW_0\oplus\cW_1 ))
\otimes B(\Gamma_\a(\cW_-)),\\[3mm]
\fM(\cV)&\simeq B(\Gamma_\a(\cW_+))\otimes \fM(\cV_0\oplus\cV_1)\otimes\one
\end{array}\]
Let $N_{01}$ be the number operator on 
$\Gamma_\a(\cW_0\oplus\cW_1)$ and $\Lambda_{01}:=(-1)^{N_{01}(N_{01}-1)/2}$.
The commutant of 
$\fM(\cV)$ equals 
\[\begin{array}{rl}
\fM(\cV)'
&\simeq \one\otimes\fM(\cV_0\oplus\cV_1)'\otimes B(\Gamma_\a(\cW_-)),\\[3mm]
&=\one\otimes\Lambda_{01}
\fM(\i(\cV_0\oplus\cV_1)^\perpr)\Lambda_{01}\otimes
B(\Gamma_\a(\cW_-))\\[3mm] 
&\simeq\Lambda\fM(\i\cV^\perpr)\Lambda,\end{array}\]
where in the last step we used Theorem \ref{iii}.
\qed

\section{Pauli-Fierz systems}

In this section
 we would like to illustrate the following phenomenon. We start from a
certain physically well motivated quantum system describing a small system
 interacting with the Bose gas. The Hamiltonian that generates the dynamics
 is a certain self-adjoint operator,  bounded from below,
 partly 
expressed in terms of
 the usual creation and annihillation operators.

Suppose that we want to consider the same system in the thermodynamical limit
corresponding to a nonzero  density $\rho$ (and the corresponding $\gamma$ 
defined by (\ref{casz})).
For instance, we are interested in  the density given by the
Planck law at inverse temperature $\beta$. 
We can do this as follows:
 we change the representation of the CCR from the original Fock representation
to the Araki-Woods representation at  $\gamma$. We still assume that
 the dynamics is  formally generated by the same expression.

To make this idea rigorous, it is convenient to use
 the framework of 
$W^*$-dynamical systems. In fact, what we obtain is a family of
$W^*$-dynamical systems $(\fM_\gamma^\PF,\tau_\gamma)$ depending on $\gamma$,
 in general non-isomorphic to one another.

Even if we fix $\gamma$, then we can consider various unitarily non-equivalent
 representations of the $W^*$-dynamical system  $(\fM_\gamma^\PF,\tau_\gamma)$. In
 fact, 
 in the literature such
 systems are considered in at least  two
different representations.

 The first one is what we call
 the semistandard
representation. It was used mostly in the older literature, e.g. 
 by Davies \cite{Da}. It
is quite simple: the small system is assumed to interact with positive
density Araki-Woods fields. In this representation, the dynamics 
 has a unitary implementation given  by the unitary group
generated
 by, what we call,
the
 semi-Liouvillean.

 The second one  is the standard
representation. It is commonly used
 in the more recent literature \cite{DJ2,BFS}. One can argue that
it is the most natural representation from the point of view of theory of
$W^*$-algebras. In any case, it is a useful tool to study various properties
of $(\fM_\gamma^\PF,\tau_\gamma)$.
 On the other hand, it is more complicated than the
semistandard representation. 
The natural implementation of the dynamics
in this representation  is
generated by the
standard Liouvillean.

If the bosons are confined
in the sense of
Subsections \ref{a2.1}, then   $(\fM_\gamma^\PF,\tau_\gamma)$
 are for various $\gamma$
isomorphic. In this case, the algebra $\fM_\gamma^\PF$ has  also a third useful
representation: the 
 irreducible one, which is not available in the
general case.

The 
 main goal of this section is to illustrate the above 
 ideas with the
 so-called  Pauli-Fierz systems.
We will
 use the name {\em
a Pauli-Fierz operator} to denote a self-adjoint operator
describing bosons interacting with
 a small quantum system with an interaction
linear in fields. 
We 
reserve the name {\em a Pauli-Fierz  Hamiltonian} to 
Pauli-Fierz operators with a
positive dispersion relation. This condition guarantees that they are bounded
 from below.
(Note that Pauli-Fierz
 Liouvilleans and semi-Liouvilleans are in general
 not bounded from below).

Pauli-Fierz Hamiltonians 
 arise in  quantum physics in various contexts
and are known under many names (e.g. the spin-boson Hamiltonian).  
The Hamiltonian
of QED in the dipole approximation is an example of such an operator.

Several aspects of
 Pauli-Fierz operators have
been recently 
studied in mathematical literature, both because of their physical
importance and because of their interesting mathematical properties, see
\cite{DG,DJ,DJ2,BFS} and references therein.

The plan of this section is as follows. First we fix some notation useful in
 describing small quantum systems interacting with Bose gas (following mostly
 \cite{DJ2}). 
Then
 we describe  a Pauli-Fierz Hamiltonian. It is described by a positive 
boson 1-particle energy $h$,
small system Hamiltonian $K$ and a coupling operator $v$. (Essentially the
 only reason to assume that $h$ is positive is the fact that such 1-particle
 energies are typical for physical systems).
The corresponding
$W^*$-algebraic system is just the
 algebra of all bounded operators on  the Hilbert space with the Heisenberg
dynamics generated by the Hamiltonian.

Given the operator $\gamma$ describing the boson
fields (and the corresponding operator $\rho$ 
describing the boson density, related to $\gamma$ by (\ref{casz}))
  we construct the
$W^*$-dynamical system $(\fM_\gamma^\PF,\tau_\gamma)$
 -- the Pauli-Fierz system corresponding to $\gamma$.
The  system $(\fM_\gamma^\PF,\tau_\gamma)$
is described in two representations: the semistandard
and the standard one. 

Parallel to the general case, we  describe the confined case. 
We show, in particular,
that  in the confined case the semi-Liouvilleans and Liouvilleans are
 unitarily equivalent for various densities $\gamma$. 

The constructions presented in this
 section are mostly taken from \cite{DJ2}. The only new  material is the
 discussion of the confined case, which, even if  straightforward, we
 believe to be  quite instructive.

\begin{remark}
In all our considerations about Pauli-Fierz systems
we restrict ourselves to the $W^*$-algebraic
formalism. 
It would be tempting to apply the  $C^*$-algebraic approach
 to describe
Pauli-Fierz systems \cite{BR2}. 
This approach proposes that a quantum system should
be described by a certain  $C^*$-dynamical system (a $C^*$-algebra with a
strongly continuous dynamics). By
considering various representations
of this $C^*$-dynamical system one could
 describe its various thermodynamical behaviors.

Such an approach works usually well in the case of infinitely extended spin or
fermionic systems, because in a finite volume
   typical interactions are bounded \cite{BR2}),
  and in algebraic local quantum field theory, because of the finite speed of
  propagation \cite{Haa}.
  Unfortunately, for Pauli-Fierz systems the $C^*$-approach
  seems to be inappropriate -- we do not know
of a  good choice of a
  $C^*$-algebra with the dynamics generated by a non-trivial Pauli-Fierz
  Hamiltonian. The problem is due to the unboundedness of  bosonic
fields that are involved in  
 Pauli-Fierz Hamiltonians. 
\end{remark}

\subsection{Creation  and annihilation operators in coupled systems}
\label{s4.1}

Suppose that $\cW$ is a Hilbert space. Consider a
 bosonic  system described by the Fock space $\Gamma_\s(\cW)$
 interacting  with a  quantum system
described by a Hilbert space $\cE$. The composite system is
 described by the Hilbert space $\cE\otimes\Gamma_\s(\cW)$. In this
 subsection we discuss the  formalism that we will use  to describe the 
interaction of such coupled systems.

Let $q\in
\cB(\cE,\cE\otimes\cW)$.
The annihilation operator   $a(q)$ is  a densely defined operator on
$\cE\otimes\Gamma_\s(\cW)$ with the domain equal to the finite particle 
subspace of
$\cE\otimes\Gamma_\s(\cW)$. For $\Psi\in\cE\otimes\Gamma_\s^n(\cW)$ we set
\beq
a(q)\Psi:
=\sqrt{n}q^*{\otimes
}1_\cW^{\otimes(n-1)}\,\Psi\in\cE\otimes\Gamma_\s^{n-1}(\cW) .\label{qua}\eeq
($\cE\otimes\Gamma_\s^n(\cW)$  can be viewed as a subspace of
$\cE\otimes\cW^{\otimes n}$. Moreover, $q^*{\otimes
}1_\cW^{\otimes(n-1)}$ is an operator from
$\cE\otimes\cW^{\otimes n}$ to $\cE\otimes\cW^{\otimes (n-1)}$, which maps
$\cE\otimes\Gamma_\s^n(\cW)$ into $\cE\otimes\Gamma_\s^{n-1}(\cW)$. Therefore,
(\ref{qua}) makes sense).

The operator $a(q)$ is closable and we will denote its closure by the same
symbol. The creation operator $a^*(q)$ is defined as
\[a^*(q):=a(q)^*.\]

Note 
that if $q=B\otimes|w)$, for  $B\in \cB(\cE)$ and $w\in\cW$,
then
\[\begin{array}{l}a^*(q)=B{\otimes} a^*(w),\ \ \ 
a(q)=B^*{\otimes}a(w),\end{array}\]
where $a^*(w)$/$a(w)$ are the usual creation/annihilion operators on the Fock
space $\Gamma_\s(\cW)$.

\subsection{Pauli-Fierz Hamiltonians}


Throughout this section we assume that $K$ is a self-adjoint operator on a
finite dimensional 
Hilbert space $\cK$, $h$ is a positive operator on a Hilbert space
$\cZ$ and $v\in \cB(\cK,\cK\otimes\cZ)$. 
The self-adjoint operator  
\[H_\fr:=K\otimes1+1\otimes\d\Gamma(h)\]
on $\cK\otimes\Gamma_\s(\cZ)$
 will be called a {\em free Pauli-Fierz Hamiltonian}. The
interaction is described by  the self-adjoint operator
\[ V= a^*(v) + a(v).\]
The operator
\[H:=H_\fr+ V\]
 is called a Pauli-Fierz Hamiltonian.

If
\beq \label{posi1} 
h^{-\12}v\in \cB(\cK,\cK\otimes\cZ),
\eeq
then $H$ is 
self-adjoint on $\Dom(H_\fr)$ and bounded from below, see e.g \cite{DJ2}.

 $\bigl(B(\cK\otimes\Gamma_\s(\cZ)),\,\e^{\i t H}\cdot\e^{-\i t H}\bigl)$
will be called a {\em Pauli-Fierz $W^*$-dynamical system at zero density}.

\begin{remark}
We will usually drop $\one_\cK\otimes$ in formulas, so that
$h^{-\12}v$ above should be read $(h^{-\12}{\otimes}\one_\cK)v$.
\end{remark}

\begin{remark} Self-adjoint operators of the form of a Pauli-Fierz Hamiltonian,
 but without
 the requirement
 that the boson energy is positive, will be called Pauli-Fierz operators.
\end{remark}

\subsection{More notation}

In order to describe  Pauli-Fierz systems  at a positive density
in a compact and elegant way
 we need more
notation.

Let $\cK$ and $\cZ$ be Hilbert spaces. Remember that we assume $\cK$ to be
 finite dimensional.
 First  we
introduce a certain antilinear map $\star$ from 
 $\cB(\cK,\cK\otimes\cZ)$ to 
$\cB(\cK,\cK\otimes\bar\cZ)$. 

Let $v\in \cB(\cK,\cK\otimes\cZ)$. We define
 $v^\star \in B(\cK,\cK\otimes\bar\cZ)$ such that for 
$\Phi, \Psi \in \cK$ and $w \in \cZ$, 
\[(\Phi{\otimes} w\,|\,v\Psi)_{\cK\otimes\cZ}
=(v^\star\Phi|\Psi\otimes\bar w)_{\cK\otimes\bar\cZ}.\]
It is easy to see that  $v^\star$ is  uniquely defined. 
(Note that $\star$ is different from $*$ denoting the Hermitian 
conjugation).

\begin{remark} Given an orthonormal
 basis $\{w_i\ :\ i\in I\}$ in $\cZ$, any $v\in  B(\cK,\cK\otimes\cZ)$
can be decomposed as 
\begin{equation} v=\sum_{i\in I} B_i\otimes |w_i),\label{cza}\end{equation}
where $B_i\in B(\cK)$, then
\[v^\star:=\sum_{i\in I}B_i^*\otimes|\bar{w_i}).\]
\eer

Next we introduce the operation
 $\check \otimes$, which can be called
 {\em tensoring in the middle}.
Let $\cH_1$, $\cH_2$ be Hilbert spaces.
If $\bar B\in \cB(\bar\cK)$, $A\in \cB(\cK\otimes\cH_1,\cK\otimes\cH_2)$,
 we define
\begin{equation} \bar B\check\otimes A:= (\theta^{-1}{\otimes}1_{\cH_2})\
(\bar B{\otimes} A)  \ (\theta{\otimes}1_{\cH_1})\in
\cB(\cK\otimes\bar\cK\otimes\cH_1,\cK\otimes\bar\cK\otimes\cH_2)
,\label{check}\end{equation}
where $\theta:\cK\otimes\bar\cK\to\bar\cK\otimes\cK$ is defined as 
$\theta\ \Psi_1{\otimes}\bar{\Psi}_2:=\bar{\Psi}_2{\otimes}\Psi_1$. 

\ber If $C\in \cB(\cK)$, $A\in
 \cB(\cH_1,\cH_2)$, then
$\bar B\:\check\otimes\: (C{\otimes} A):= C\otimes \bar B\otimes A$.
\eer

\subsection{Pauli-Fierz systems at a positive density}
\label{s5.4a}

In this subsection we introduce Pauli-Fierz
$W^*$-dynamical systems. They will be the main subject of the remaining
part of this section.

Let $\rho$ be a positive
 operator commuting with $h$ having the interpretation of the {\em radiation
 density}. Let $\gamma$ be the operator related to $\rho$ as in  (\ref{casz}).
 Let
$\fM_{\gamma,\l}^\AW\subset B(\Gamma_\s(\cZ\oplus\bar\cZ))$ 
be the left Araki-Woods algebra introduced in Subsection \ref{sAW}.
The {\em Pauli-Fierz algebra corresponding to $\gamma$} is defined by
\begin{equation}
\fM_{\gamma}^\PF:=B(\cK)\otimes\fM_{\gamma,\l}^\AW.
\label{pau}\end{equation}

The identity map
\begin{equation}
\fM_\gamma^\PF\to B(\cK\otimes\Gamma_\s(\cZ\oplus\bar\cZ))\label{semmi}\end{equation}
will be called {\em  the semistandard representation of $\fM_\gamma^\PF$}.
 (The
bosonic part of (\ref{semmi}) is already standard, the part involving
$\cK$ is not---hence the name).

\bep Assume that  
\begin{equation}(1+\rho)^{1/2}v \in \cB(\cK, \cK \otimes \cZ).\label{nzn}\end{equation}
 Let 
\[\begin{array}{rl}
V_\gamma
&:=a^*\Big((1+\rho)^\12v,\bar\rho^\12 v^{\star}\Big)
+a\Big( (1+\rho)^\12v, \bar\rho^\12v^{\star }\Big)
.\end{array}\]
Then the operator $V_\gamma$ is 
essentially self-adjoint on the space of finite particle
vectors and affiliated to $\fM_{\gamma}^\PF$.
\label{zxz}\eep

The {\em free Pauli-Fierz semi-Liouvillean }is the self-adjoint operator on\\
$\cK\otimes\Gamma_\s(\cZ\oplus\bar\cZ)$ defined as
\[\begin{array}{rl}L_{\fr}^\semi&
:=K\otimes1+1\otimes\d\Gamma(h\oplus(-\bar h)).\end{array}\]
The {\em  full Pauli-Fierz semi-Liouvillean corresponding to
$\gamma$} is
\begin{equation} L_\gamma^\semi:=L_\fr^\semi+  V_\gamma.
\label{lo}\end{equation}

Let us formulate the following assumption:

\beq (1+h)(1 +\rho)^{1/2}v \in 
\cB(\cK,\cK\otimes \cZ).
\label{inter}\eeq

Using  Theorem 3.3 of \cite{DJP}  we obtain
\bet \ben\item
 \[\tau_\fr^t(A):=\e^{\i tL_\fr^\semi}A\e^{-\i tL_\fr^\semi} \]
is a $W^*$-dynamics on $\fM_{\gamma}^\PF$.\item
 Suppose that 
 (\ref{inter})
holds. Then 
 $L_\gamma^\semi$ is essentially self-adjoint on\\ $\Dom(L_{\fr}^\semi)\cap
\Dom(V_\gamma)$ and
\[\tau_\gamma^t(A):=\e^{\i t
L_\gamma^\semi}A
\e^{-\i t L_\gamma^\semi}\] is a $W^*$-dynamics 
on $\fM_{\gamma}^\PF$.
\een\label{dynx}\eet

The pair $\bigl(\fM_{\gamma}^\PF,\tau_\gamma\bigr)$
will be
called the {\em Pauli-Fierz $W^*$-dynamical system corresponding to $\gamma$}.

\subsection{Confined Pauli-Fierz
  systems---semistandard representation}

In this subsection we make the assumption $\Tr\gamma<\infty$. As before, 
we will call it the confined case.

We can use the identity representation for $B(\cK)$ and the
Araki-Woods representation $\theta_{\gamma,\l}$ 
for $B(\Gamma_\s(\cZ))$. Thus we
obtain the faithful  representation
\[\pi_{\gamma}^{\semi}
 :B(\cK\otimes 
\Gamma_\s(\cZ))\to B(\cK\otimes\Gamma_\s(\cZ\oplus\bar\cZ)),\] 
which will be called the semistandard representation of 
$B(\cK\otimes 
\Gamma_\s(\cZ))$.
In other words, $\pi_{\gamma}^\semi$ is defined by 
\[\pi_{\gamma}^\semi(A)=
U_{\gamma}^\semi
A{\otimes} 1_{\Gamma_\s(\bar\cZ)}
U_{\gamma}^{\semi*}
\ \ \ A\in B(\cK\otimes\Gamma_\s(\cZ)),\]
where\[U_{\gamma}^\semi:=
1_\cK{\otimes} R_\gamma U,\]
and $U$ was defined in (\ref{idi}) and $R_\gamma$ in (\ref{rrr}).


\bet
\[\begin{array}{rl}
\pi_{\gamma}^{\semi}
\left(B\bigr(\cK\otimes \Gamma_\s(\cZ)\bigr)\right)&= \fM_{\gamma}^\PF,\\[3mm]
 \pi_{\gamma}^{\semi}\Big(\e^{\i tH}A\e^{-\i
tH}\Big)&=\tau_\gamma^t\left(\pi_{\gamma}^{\semi}(A)\right)
,\ \ \ A\in B(\cK\otimes\Gamma_\s(\cZ)),\\[3mm]
L_\gamma^\semi&=U_{\gamma}^\semi
\bigl(H\otimes
1_{\Gamma_\s(\bar\cZ)}-  
1_{\cK\otimes\Gamma_\s(\cZ)}\otimes\d\Gamma(\bar h)\bigr)
U_{\gamma}^{\semi*}
.\end{array}\]
\eet

 Let us stress that in the confined case the semi-Liouvilleans 
$L_\gamma^\semi$ and the $W^*$-dynamical systems
$(\fM_{\gamma,\l}^\PF,\tau_\gamma)$  are unitarily equivalent for
 different $\gamma$.

\subsection{Standard  representation of Pauli-Fierz systems}
\label{s5.4b}

In this subsection we drop the assumption $\Tr\gamma<\infty$
 about the confinement of
the bosons and we consider the general case again.

Consider the  representation
\[\pi^{}
:\fM_{\gamma}\to 
\cB(\cK\otimes\bar\cK
\otimes\Gamma_\s(\cZ\oplus\bar\cZ))\]
defined by
\[\pi^{}(A):= 1_{\bar\cK}\check\otimes A,\ \ \  A\in
\fM_{\gamma}^\PF,\]
where $\check\otimes$ was introduced in  (\ref{check}).
Clearly, 
\[\pi^{}(\fM_{\gamma}^\PF)=\cB(\cK)\otimes1_{\bar\cK}\otimes
\fM_{\gamma,\l}^\AW.\]
Set $J:=J_\cK\otimes\Gamma(\epsilon)$, where
\begin{equation} J_\cK\Psi_1\otimes\bar{\Psi}_2:=
\Psi_2\otimes\bar{\Psi}_1,\qquad \Psi_1,\Psi_2\in\cK,
\label{jotk1}\end{equation}
 and $\epsilon$ was introduced in (\ref{epsi}).
Note that
\[J\:\cB(\cK){\otimes}1_{\bar\cK}{\otimes}\fM_{\gamma,\l}^\AW\:J=1_\cK{\otimes}
\cB(\bar\cK){\otimes}\fM_{\gamma,\r}^\AW,\]
and
 if $A \in \cB(\cK)\otimes\fM_{ \gamma,\l}^\AW$, then
\[J\pi(A)J=1_{\cK}\otimes\Big(
1_{\bar\cK}{\otimes}\Gamma(\tau)\ \bar{A}\ 1_{\bar\cK}{\otimes} 
\Gamma(\tau)\Big),\]
where $\tau$ was introduced in (\ref{tau}).

\bep
\[\Big(\pi^{},\: \cK{\otimes}\bar\cK{\otimes}
\Gamma_\s(\cZ{\oplus}\bar\cZ),\: J,\: 
\cH_\gamma^+\Big)\]
 is a standard representation
of $\fM_\gamma$, where
\[\cH_\gamma^+:=\{\pi(A)J\pi(A)\  B{\otimes}\Omega\ : B\in B^2_+(\cK),
A\in\fM_\gamma^\PF\}^\cl.\] 
\eep

Set
\[L_{\fr}:=K\otimes1\otimes1-1\otimes\bar
K\otimes1+1\otimes1\otimes\d\Gamma(h\oplus(-\bar h)).\]  

\bep Assume (\ref{nzn}). Then
\[\begin{split}
\pi(V_\gamma)&:=1_{\bar\cK}\check\otimes V_\gamma\\[3mm]
&=1_{\bar\cK}\check\otimes a^*\Big((1+\rho)^\12v,\bar\rho^\12 v^{\star}\Big)
+1_{\bar\cK}\check\otimes a\Big( (1+\rho)^\12 v, \bar\rho^\12v^{\star }\Big)
\end{split}\]
 is essentially self-adjoint on finite particle vectors of
$\cK \otimes \bar \cK 
\otimes \Gamma_{\s}(\cZ \oplus \bar \cZ)$ and is 
affiliated  to the $W^*$-algebra 
$B(\cK)\otimes\one_{\bar\cK}\otimes\fM_{ \gamma,\l}^\AW$. Moreover, 
\[\begin{array}{rl}
J\pi(V_\gamma) J&:=1_{\cK}\otimes \Big(
1_{\bar\cK}{\otimes}\Gamma(\tau)\ \bar{
V_\gamma}\ 1_{\bar\cK}{\otimes}\Gamma(\tau)\Big)\\[3mm]
&=1_{\cK}\otimes a^*\Big(\rho^\12\bar v^{\star},(1+\bar\rho)^\12 \bar v\Big)
+1_{\cK}\otimes
a\Big( \rho^\12\bar v^{\star}, (1+\bar\rho)^\12\bar v\Big)
.\end{array}\]
\eep

Set
\begin{equation} L_\gamma:=L_{\fr}
+\pi(V_\gamma)- J\pi(V_\gamma) J.\label{lst}\end{equation}
\bet
\ben\item
 $L_\fr$ is the standard Liouvillean of the free Pauli-Fierz 
system  $(\fM_\gamma^\PF, \tau_\fr)$.
\item
  Suppose that
(\ref{inter})  holds.
Then 
$L_\gamma$ is essentially self-adjoint on \\
 $\Dom(L_{\fr})\cap
\Dom(\pi(V_\gamma))\cap\Dom(J\pi(V_\gamma) J)$ and is the Liouvillean of 
the Pauli-Fierz system $(\fM_\gamma^\PF, \tau_\gamma)$.  
\een\label{dynn}\eet

\subsection{Confined
 Pauli-Fierz systems---standard representation}

Again we make the assumption $\Tr\gamma<\infty$
 about the confinement of the bosons.

We can use the  standard representation $\pi_\l$ for $B(\cK)$ in
$B(\cK\otimes\bar\cK)$ (in the form of Subsection \ref{I.II})
and the Araki-Woods representation $\theta_{\gamma,\l}$
for
$B(\Gamma_\s(\cZ))$ in $B(\Gamma_\s(\cZ\oplus\bar\cZ))$.
  Thus we obtain the representation
\[\pi_{\gamma,\l}:
B(\cK\otimes 
\Gamma_\s(\cZ))\to
B(\cK\otimes\bar\cK\otimes\Gamma_\s(\cZ\oplus\bar\cZ)).\]
defined by
\[\pi_{\gamma,\l}(A_1\otimes A_2)=A_1\otimes 1_{\bar\cK}\otimes
\theta_{\gamma,\l}(A_2),
\ \ \ A_1\in B(\cK),\ \ A_2\in B(\Gamma_\s(\cZ)).\]
Note that
\[\pi_{\gamma,\l}(A):=1_{\bar\cK}\check\otimes
 \pi_{\gamma}^\semi(A),\ \ \ \ \ A\in
B(\cK\otimes\Gamma_\s(\cZ)).\]
One can put it in a  different way. Introduce the obvious 
unitary identification
\[\tilde U:\cK\otimes\Gamma_\s(\cZ)\otimes\bar\cK\otimes\Gamma_\s(\bar\cZ)\to
\cK\otimes\bar\cK \otimes\Gamma_\s(\cZ\oplus\bar\cZ).\]
Set
\[U_\gamma:=1_{\cK\otimes\bar\cK}{\otimes} R_\gamma\
\tilde U.\]
Then
\[\pi_{\gamma,\l}(A)=U_\gamma\ 
 A{\otimes}1_{\bar\cK\otimes\Gamma_\s(\bar\cZ)}\ 
U_\gamma^*,\ \ \ A\in
B(\cK\otimes\Gamma_\s(\cZ)).\]


\bet 
\[\begin{array}{rl}
\pi_{\gamma,\l}\left(B\bigl(\cK\otimes\Gamma_\s(\cZ)\bigr)\right)
&=\pi(\fM_\gamma^\PF),\\[3mm]
\pi_{\gamma,\l}\Big(\e^{\i tH}A\e^{-\i
tH}\Big)
&=\pi\left(\tau_\gamma^t(\pi_\gamma^\semi(A))\right)
,\ \ \ A\in B(\cK\otimes \Gamma_\s(\cZ)),\\[3mm]
L_\gamma
&=U_\gamma\left(H\otimes1_{\bar\cK\otimes\Gamma_\s(\bar\cZ)}
-1_{\cK\otimes\Gamma_\s(\cZ)}\otimes \bar H\right) U_\gamma^*
.\end{array}\]
\eet

 Let
 us stress that in the confined case
 the Liouvilleans $L_\gamma$ are unitarily equivalent for
 different $\gamma$.

\end{document}